\documentclass{aa}  
\usepackage{graphicx}
\usepackage{txfonts}
\usepackage{natbib}
\bibpunct{(}{)}{;}{a}{}{,}
\begin{document}
   \title{Spatially resolved mid-infrared observations of the triple system T~Tauri             
         \thanks{Based on observations with the Very Large Telescope Interferometer (VLTI, proposal 074. C-0209(A) and 077.C-0176(C)).}}

   \subtitle{}

   \author{Th. Ratzka\inst{1,2,3}
          \and
          A.~A. Schegerer\inst{4,2}
          \and
          Ch. Leinert\inst{2}
          \and
          P. \'Abrah\'am\inst{5}
          \and
          Th. Henning\inst{2}
          \and
          T. M. Herbst\inst{2}
          \and
          R.~K\"ohler\inst{2,6}
          \and 
          S. Wolf\inst{7,2}
          \and
          H. Zinnecker\inst{2}}

   \offprints{ratzka@usm.uni-muenchen.de}

   \institute{Astrophysical Institute Potsdam,
              An der Sternwarte 16, 14482 Potsdam, Germany
	  \and
              Max-Planck-Institute for Astronomy,
              K\"onigstuhl 17, 69117 Heidelberg, Germany
          \and
	      Universit{\"a}ts-Sternwarte M{\"u}nchen,
              Scheinerstra{\ss}e 1, 81679 M{\"unchen}, Germany
	  \and
	      Helmholtz Zentrum M{\"u}nchen, German Research Center for Environmental Health,
	      Ingolst{\"a}dter Landstra{\ss}e 1, 85758 Neuherberg, Germany
	  \and
              Konkoly Observatory of the Hungarian Academy of Sciences,
              P.O.~Box 67, 1525 Budapest, Hungary
          \and
 	      ZAH, Landessternwarte,
 	      K\"onigstuhl, 69117 Heidelberg, Germany
          \and
	      University of Kiel, Institute of Theoretical Physics and Astrophysics, 
	      Leibnizstra{\ss}e 15, 24098 Kiel, Germany
             }

   \date{Received Nov 21, 2008; accepted May 20, 2009}

 
  \abstract
   {}
   {The aim of this study is to enhance our knowledge of the characteristics and distribution of the circumstellar dust associated with the individual components of the young hierarchical triple system T Tau.}
   {To reach this goal, observations in the N-band (8-13 $\mu$m) with the two-telescope interferometric instrument MIDI at the VLTI were performed. For the northern component of the T~Tau system, projected baseline lengths of 43\,m, 62\,m, and 85\,m were used. For the southern binary projected baseline lengths of equivalent resolution could be utilised. Our study is based on both the interferometric and the spectrophotometric measurements and is supplemented by new visual and infrared photometry. Also, the phases were investigated to determine the dominating mid-infrared source in the close southern binary. The data were fit with the help of a sophisticated physical disc model. This model utilises the radiative transfer code MC3D that is based on the Monte-Carlo method.}
   {Extended mid-infrared emission is found around all three components of the system. Simultaneous fits to the photometric and interferometric data confirm the picture of an almost face-on circumstellar disc around T~Tau~N. Towards this star, the silicate band is seen in emission. This emission feature is used to model the dust content of the circumstellar disc. Clear signs of dust processing are found. Towards T~Tau~S, the silicate band is seen in absorption. This absorption is strongly pronounced towards the infrared companion T~Tau~Sa as can be seen from the first individual N-band spectra for the two southern components. Our fits support the previous suggestion that an almost edge-on disc is present around T~Tau~Sa. This disc is thus misaligned with respect to the circumstellar disc around T~Tau~N. The interferometric data indicate that the disc around T~Tau~Sa is oriented in the north-south direction, which favours this source as launching site for the east-western jet. We further determine from the interferometric data the relative positions of the components of the southern binary in the N-band. We find good agreement with recent position measurements in the near-infrared.}
   {}


   \maketitle
%

\section{Introduction}

Our knowledge of the T Tau triple system has grown as observational techniques were improved. \object{T~Tau} was chosen by \cite{joy45} as the prototype for a new class of variable objects. The members of this class are characterised as low-mass ($<3\,{\rm M}_{\odot}$) pre-main sequence stars still surrounded by accretion discs and thus exhibiting large infrared excesses \citep{appenzeller89}. A number of nebulous patches have been detected in the vicinity of T~Tau. They show similar spectral features to Herbig-Haro objects. Surrounding the source on scales of arcseconds, they clearly witness stellar outflows. Today an east-west and a northwest-southeast oriented outflow, as well as many interlocking loops and filaments of H$_2$, are known, e.g., \cite{herbst07}.

A close companion was detected by \cite{dyck82}. It was later named \object{T~Tau~S} because it is located 0.7'' south of \object{T~Tau~N}. T~Tau~S is the source of one of the jets and a prototypical infrared companion (IRC) deeply embedded and not detectable in the visual \citep{stapelfeldt98}. Various explanations for its higher extinction have been discussed. On the one hand, a different evolutionary status of the southern component may be the reason. Already \cite{dyck82} saw in T~Tau~S a star that is less evolved than T~Tau~N. On the other hand, the extinction may be caused by a special geometry of the system like viewing T~Tau~S through an edge-on disc or viewing a background southern component through a foreground screen. Then T~Tau~S can be a normal T~Tauri star coeval with its northern neighbour. \cite{koresko97} concluded T~Tau~S is experiencing episodes of embedding, perhaps because of an enhanced accretion rate at certain orbital phases. 

T~Tau~S was resolved by \cite{koresko00} into two components named \object{T~Tau~Sa} and \object{T~Tau~Sb}, with a separation of about 0.05'' in 1997. While T~Tau~Sb appears to be a ``normal'' active low-mass pre-main-sequence star simply residing behind an absorbing screen with A$_\mathrm{V}\approx15\,\mathrm{mag}$ \citep{duchene05}, T~Tau~Sa remains an enigmatic source. According to the picture drawn by \cite{duchene05}, T~Tau~Sa is the most massive star of the T~Tau system with a stellar mass of $2.5-3\,M_{\odot}$. A large fraction of the visual and near-infrared light emitted by the photosphere of this intermediate-mass young stellar object are scattered away thanks to an edge-on circumstellar disc or lost through its ``polar'' regions. Furthermore, a circumbinary envelope or thick disc may obscure both southern components leading to an additional visual extinction of A$_\mathrm{V}\approx15\,\mathrm{mag}$ as it has been found towards T~Tau~Sb. 

Associated with T~Tau~Sb is a radio source observed by \cite{loinard07} with the VLBA to determine its distance precisely. The derived value of $146.7\pm0.6$\,pc has been adopted in this paper for the whole T~Tau system.

The northern component has shown a significant brightness variation neither in the K- nor in the L'-band \citep{beck04}. Even its visual flux seems to be constant \citep{ghez91}. The southern component shows variability with a total amplitude of $\sim 2$\,mag in the K-, and $\sim 3$\,mag in the L'-band. \cite{beck04} explained the variations by changes in the amount of obscuring material. In its bright state, T~Tau~S dominates the flux of the system longwards of 3\,$\mu$m. According to \cite{ghez91}, the flux ratio between the southern and the northern component in the N-band ranges from $\sim 0.4$ in the minimum to $\sim 2.6$ in the maximum. Both components forming T~Tau~S have been identified as variable. The flux ratio of T~Tau~Sb with respect to T~Tau~Sa varies between $\sim 0.1$ \citep{koresko00} and  $\sim 3.6$ \citep{beck04} in the K-band.

There has been general agreement that T~Tau~N is surrounded by a circumstellar disc of moderate mass \citep{beckwith90} seen not far from face-on, although its proposed size \citep{akeson98} is still under discussion. For T~Tau~S it was not at all obvious that its circumstellar material would be concentrated in a circumstellar disc. Its outburst in 1989 led \cite{ghez91} to model its SED as dominated by a FUOR-type accretion disc. The picture of T~Tau~S then shifted to that of a rather normal young star \citep{hogerheijde97} extincted by a disc seen nearly edge-on \citep{solf99},  but with very low mass according to the limits derived by millimetre interferometry \citep{hogerheijde97, akeson98}. This missing mass was naturally explained by truncation due to the companion T~Tau~Sb  \citep{koresko00}. Ultraviolet studies from HST showed an absorbing screen in front of T~Tau~S \citep{walter03}, possibly associated with a circumbinary disc around T Tau S, which would be responsible for most of the extinction towards T Tau Sb. By mid-infrared adaptive optics imaging in the silicate feature \cite{skemer08} add to the evidence for different disc orientations in the T Tau system and  question the importance of a circumbinary disc. The actual orientations of the circumstellar discs in the close binary T~Tau~S have remained largely unknown.

Summarising the brief description above, the T~Tau system is one of the best-suited laboratories for investigating the dynamics and the formation of young multiple systems. A first step in this direction has been made by monitoring the orbital motions of the southern components around each other and around the northern component. These studies led to reliable masses of the stars and concluded that T Tau~Sb is indeed bound to the system \citep{duchene06, koehler08, koehler08a}. A next step is the investigation of the individual discs with respect to their dust content and their geometry. The information about the geometry will help in addressing the issue of coplanarity of the discs with respect to each other, as well as to the orbits.

In this paper we present observations (Section~\ref{obs}) of the T~Tau system with the ``MID-infared Interferometric instrument'' MIDI, which resides in the interferometric laboratory of the ``Very Large Telescope Interferometer'' (VLTI) on Paranal, Chile. After a brief description of the data reduction (Section~\ref{dr}), we present first the directly derived results (Section~\ref{results}). In Section~\ref{ttaun}, we model the northern component with a radiative transfer model and investigate the dust properties of its circumstellar disc by fitting the correlated fluxes. In Section~\ref{ttaus}, the visibilities of T~Tau~S are ana\-lysed to derive relative positions and separate spectra for the two components. With the help of radiative transfer models, the properties and the relative orientations of the discs are determined. Finally, a picture of the system is drawn in Section~\ref{3d}. In Section~\ref{sum} a summary can be found. 

%

\section{Observations\label{obs}}

\begin{table*}
\caption{Journal of MIDI Observations}
\label{table1}
\centering
\begin{tabular}{cccccccccc}
\hline\hline
\noalign{\smallskip}
Date of      & Universal      & Object             & IRAS & \multicolumn{2}{c}{Proj.~Baseline$^*$\ \ } &  Airmass$^{**}$ & Acquisition & Interferometric & Photometric\\
Observation  & Time           &                    & [Jy]  & \     [m] & [deg]                    &                 & Frames      & Frames	         & Frames\\
\noalign{\smallskip}
\hline
\noalign{\smallskip}
31-10-2004   & 03:53 - 04:06  & \object{HD 20644}  &   14.7 & 28.1 &\ 54.2 & 1.82 & \ \ 1000$\times$4\,ms &  -$^{***}$           & 2$\times$1500$\times$12\,ms \\
31-10-2004   & 06:45 - 06:59  & \object{HD 25604}  &\ \ 5.1 & 42.0 &\ 50.4 & 1.50 & \ \ 1000$\times$4\,ms &  10000$\times$12\,ms & 2$\times$2000$\times$12\,ms \\
31-10-2004   & 06:59 - 07:22  &          T~Tau~N   &      - & 42.9 &\ 49.4 & 1.45 & \ \ 1000$\times$4\,ms &  10000$\times$12\,ms & 2$\times$2000$\times$12\,ms \\
31-10-2004   & 07:22 - 07:41  &          T~Tau~S   &      - & 43.9 &\ 48.9 & 1.49 & \ \ 1000$\times$4\,ms &  10000$\times$12\,ms & 2$\times$2000$\times$12\,ms \\
31-10-2004   & 07:41 - 08:10  & \object{HD 37160}  &\ \ 6.5 & 43.5 &\ 46.2 & 1.22 & \ \ 1000$\times$4\,ms &  10000$\times$12\,ms & 2$\times$2000$\times$12\,ms \\
31-10-2004   & 08:40 - 09:08  &         HD 37160   &\ \ 6.5 & 45.8 &\ 46.7 & 1.31 & \ \ 1000$\times$4\,ms &  10000$\times$12\,ms & 2$\times$2000$\times$12\,ms \\
\noalign{\smallskip}
\hline
\noalign{\smallskip}
02-11-2004   & 05:10 - 05:36  &         T~Tau~N    &	  - & 85.0 &\ 87.9 & 1.42 & \ \ 1000$\times$4\,ms &  16000$\times$12\,ms & 2$\times$2500$\times$12\,ms \\
02-11-2004   & 05:36 - 06:10  &         T~Tau~S    &	  - & 87.6 &\ 85.6 & 1.40 & \ \ 1000$\times$4\,ms &  16000$\times$12\,ms & -$^{***}$  \\
02-11-2004   & 06:10 - 06:37  & \object{HD 31421}  &\ \ 9.3 & 87.7 &\ 84.5 & 1.27 & \ \ 1000$\times$4\,ms &  16000$\times$12\,ms & 2$\times$2500$\times$12\,ms \\
02-11-2004   & 07:08 - 07:41  &         HD 37160   &\ \ 6.5 & 89.1 &\ 82.8 & 1.21 &    10000$\times$4\,ms &  16000$\times$12\,ms & 2$\times$2500$\times$12\,ms \\
02-11-2004   & 08:09 - 08:37  & \object{HD 50778}  &   17.3 & 88.6 &\ 80.2 & 1.03 & \ \ 1000$\times$4\,ms &  16000$\times$12\,ms & 2$\times$2500$\times$12\,ms \\
02-11-2004   & 08:37 - 09:03  & \object{HD 61935}  &\ \ 7.1 & 87.2 &\ 79.6 & 1.05 & \ \ 1000$\times$4\,ms &  16000$\times$12\,ms & 2$\times$2500$\times$12\,ms \\
\noalign{\smallskip}
\hline
\noalign{\smallskip}
04-11-2004   & 00:01 - 00:26  & \object{HD 178345} &\ \ 8.6 & 57.0 & 145.6 & 1.42 & \ \ 1000$\times$4\,ms &  20000$\times$12\,ms & 2$\times$5000$\times$12\,ms \\
04-11-2004   & 02:10 - 02:47  & \object{HD 188603} &   11.4 & 45.5 & 168.6 & 2.47 & \ \ 1000$\times$4\,ms &  -$^{***}$           & 2$\times$5000$\times$12\,ms \\
04-11-2004   & 02:47 - 03:54  &         HD 25604   &\ \ 5.1 & 60.7 & 117.1 & 1.75 & \ \ 1000$\times$4\,ms &  16000$\times$12\,ms & 2$\times$5000$\times$12\,ms \\
04-11-2004   & 03:54 - 04:36  &         T~Tau~N    &      - & 61.6 & 114.0 & 1.56 & \ \ 1000$\times$4\,ms &  16000$\times$12\,ms & -$^{***}$  \\
04-11-2004   & 04:36 - 05:04  &         T~Tau~S    &      - & 62.3 & 111.4 & 1.47 & \ \ 1000$\times$4\,ms &  16000$\times$12\,ms & 2$\times$5000$\times$12\,ms \\
04-11-2004   & 05:04 - 05:29  &         HD 20644   &   14.7 & 59.0 & 101.5 & 1.69 & \ \ 1000$\times$4\,ms &  12000$\times$12\,ms & 2$\times$3000$\times$12\,ms \\
04-11-2004   & 07:18 - 07:36  &         HD 37160   &\ \ 6.5 & 61.0 & 107.4 & 1.21 & \ \ 1000$\times$4\,ms & \ 8000$\times$12\,ms & 2$\times$2000$\times$12\,ms \\
04-11-2004   & 09:00 - 09:23  &         HD 50778   &   17.3 & 61.0 & 112.6 & 1.04 & \ \ 1000$\times$4\,ms & \ 8000$\times$12\,ms & 2$\times$2000$\times$12\,ms \\
\noalign{\smallskip}
\hline
\noalign{\smallskip}
\multicolumn{8}{l}{$^{\ \ *}$ determined for the fringe tracking sequence}\\
\multicolumn{8}{l}{$^{\ **}$ determined for the photometric measurements}\\
\multicolumn{8}{l}{$^{***}$ peculiar instrumental visibility or chopping problems}\\
\noalign{\smallskip}
\hline

\end{tabular}
\end{table*}

We obtained three separate interferometric measurements of both T~Tau~N and T~Tau~S within the ``Guaranteed Time Observations'' (GTO) in October and November 2004. The baselines used were UT2-UT3 (Oct.~30/31), UT2-UT4 (Nov.~1/2), and UT3-UT4 (Nov. 3/4). A journal of observations appears in Table~\ref{table1}. 

\subsection{Observing Sequence}

The standard observing sequence for an interferometric measurement with MIDI on the VLTI produces an image of the object at 8.7\,$\mu$m, a spectrum from 8\,$\mu$m to 13\,$\mu$m, and spectrally resolved visibilities and correlated fluxes over the same wavelength range. For descriptions of the instrument and its operation see \cite{leinert03a, leinert03}, \cite{morel04}, and \cite{ratzka05}.

The chopped image is taken after the coarse acquisition by the telescopes. These images allow the observer to adjust the position of the objects to a predetermined pixel in order to maximise the overlap of both images for the following interferometric measurements. 

Then, the beam combiner, which produces two interferometric outputs of opposite sign, is put into the optical train. Together with a dispersion device -- in our case, the low resolution prism ($\lambda/\Delta\lambda\approx 30$) -- {\it dispersed fringes}, i.e., spectrally resolved interferograms, can be obtained. After locating the zero optical path difference (OPD) by scanning a few millimetres around the expected point of path length equalisation, an interferometric measurement with self-fringe-tracking is started. In this mode, the piezo-mounted mirrors within MIDI are used to scan a range in OPD of 40-80\,$\mu$m in steps of typically 2\,$\mu$m. At each step, with the corresponding fixed OPD, an exposure is taken that gives the instantaneous value of the interferometric signal from 8\,$\mu$m to 13\,$\mu$m. After each scan, the position of the fringe packet in the scan is measured and the VLTI delay lines are adjusted in order to recentre the fringe packet for the next scan. Even for the low resolution of the prism, the coherence length is about $\pm$~300~$\mu$m. This means that imperfect centring of the interferometric scan within a few $\lambda$ has no adverse effect on the visibility determination, i.e., no noticeable spectral smearing occurs. The output after a scan is the spectrally resolved temporal fringe pattern, which gives the fringe amplitude or correlated flux. The scans are repeated typically 100-200 times in saw-tooth manner to increase the statistical accuracy. For typical exposure times of 15\,ms and readout times of 3\,ms this means that the temporal fringe signal is modulated at a frequency of $\approx$~10\,Hz. For these inteferometric measurements no chopping is used, because the two resulting output signals are subtracted from each other as first step of the data reduction. This leads -- when supplemented with high-pass filtering -- to an efficient background subtraction.
 
To derive the (wavelength dependent) visibilities, spectra are measured with the same pixels of the detector that have been used before for measuring the fringe signal. To accomplish this, first the light from one, then from the other telescope is blocked. For the photometric measurements the secondary mirrors of the UTs are chopped to remove the thermal background radiation. 

By definition the visibility is obtained as the ratio of correlated flux (from the interferometric measurement) to the total flux (from the spectrophotometry). This gives the raw visibility, still suffering from atmospheric and instrumental correlation losses. To correct for these losses, the sequence of observations has to be repeated for a source with known diameter, and therefore known visibility.

\subsection{Calibration\label{calib}}

Calibrator stars with known diameter were observed immediately after the object and in the same region of the sky to correct for the reduction in the fringe contrast due to optical imperfections and atmospheric turbulence. However, with the present accuracy of about 5-10\,\% per single visibility measurement, it is also possible to use calibrators observed in the same mode during the same night. The calibrators given in Table~\ref{table1} with their HD number were taken from the ``MIDI Calibrator Catalogue''$^1$ of 509 stars with a flux of at least 5\,Jy at $10\,\mu$m and selected for the absence of circumstellar emission, close companions that could disturb the visibility measurement, or strong variability. A subsample of these calibrators can also be used for absolute flux calibration$^2$.
\footnotetext[1]{http://www.eso.org/$\sim$arichich/download/vlticalibs-ws/}
\footnotetext[2]{http://www.eso.org/sci/facilities/paranal/instruments/midi/tools/}

\subsection{Additional Photometric Datasets}

\begin{table*}
\caption{Photometric Measurements}
\label{table2}
\centering
\begin{tabular}{ccccrrrc}
\hline\hline
\noalign{\smallskip}
Band & $\lambda_{\rm{central}}$ & Date & Instrument & \multicolumn{1}{c}{T Tau N} & \multicolumn{1}{c}{T Tau Sa} & \multicolumn{1}{c}{T Tau Sb} & Reference\\
     & [$\mu$m]       &      &            &                             &                              &                              &          \\
\noalign{\smallskip}
\hline
\noalign{\smallskip}
U  &  0.36 & 21-10-2004 & Phot. / Mt. Maidanak  & $11.74\pm0.04$\,mag &	      - &	            - & $^a$\\
B  &  0.43 & Oct. 2004  & KING / MPIA           & $11.2\pm0.1$\,mag &	      - &	            - & $^b$\\
V  &  0.55 & Oct. 2004  & KING / MPIA           & $ 9.9\pm0.1$\,mag &	      - &	            - & $^b$\\
R  &  0.70 & Oct. 2004  & KING / MPIA           & $ 8.9\pm0.1$\,mag &	      - &	            - & $^b$\\
I  &  0.90 & Oct. 2004  & KING / MPIA           & $ 8.3\pm0.1$\,mag &	      - &	            - & $^b$\\
\noalign{\smallskip}
\hline
\noalign{\smallskip}
J   &  1.25 & 14-12-2002 & NACO / VLT          &  $7.1\pm0.1$\,mag   & $>14.5$\,mag        & $12.9\pm0.1$\,mag & $^c$\\
H   &  1.65 & 14-12-2002 & NACO / VLT          &  $6.2\pm0.1$\,mag   & $12.3\pm0.1$\,mag   & $10.4\pm0.1$\,mag & $^c$\\
Ks  &  2.16 & 14-12-2002 & NACO / VLT          &  $5.7\pm0.1$\,mag   & $ 9.2\pm0.1$\,mag   & $ 8.8\pm0.1$\,mag & $^c$\\
L'  &  3.80 & 13-12-2002 & NIRC-2 / Keck       &  $4.32\pm0.05$\,mag & $ 5.64\pm0.03$\,mag & $ 6.25\pm0.03$\,mag & $^{de}$\\
Ms  &  4.67 & 13-12-2002 & NIRC-2 / Keck       &  $4.13\pm0.10$\,mag & $ 4.96\pm0.06$\,mag & $ 5.77\pm0.06$\,mag & $^{df}$\\
\noalign{\smallskip}
\hline
\noalign{\smallskip}
    &  8.5 & Oct./Nov. 2004 & MIDI/VLTI          & $5.6\pm0.2$\,Jy & $2.6\pm0.2$\,Jy & $1.0\pm0.1$\,Jy & $^b$\\
    &  9.0 & Oct./Nov. 2004 & MIDI/VLTI	 & $6.7\pm0.3$\,Jy & $1.5\pm0.2$\,Jy & $0.6\pm0.1$\,Jy & $^b$\\
    &  9.5 & Oct./Nov. 2004 & MIDI/VLTI	 & $7.4\pm0.5$\,Jy & $1.0\pm0.2$\,Jy & $0.4\pm0.1$\,Jy & $^b$\\
    & 10.0 & Oct./Nov. 2004 & MIDI/VLTI	 & $8.4\pm0.4$\,Jy & $1.1\pm0.2$\,Jy & $0.5\pm0.1$\,Jy & $^b$\\
    & 10.5 & Oct./Nov. 2004 & MIDI/VLTI	 & $8.8\pm0.4$\,Jy & $1.6\pm0.3$\,Jy & $0.7\pm0.2$\,Jy & $^b$\\
    & 11.0 & Oct./Nov. 2004 & MIDI/VLTI	 & $8.8\pm0.4$\,Jy & $2.3\pm0.3$\,Jy & $0.9\pm0.2$\,Jy & $^b$\\
    & 11.5 & Oct./Nov. 2004 & MIDI/VLTI	 & $8.7\pm0.4$\,Jy & $3.2\pm0.4$\,Jy & $1.1\pm0.2$\,Jy & $^b$\\
    & 12.0 & Oct./Nov. 2004 & MIDI/VLTI	 & $8.4\pm0.5$\,Jy & $4.4\pm0.5$\,Jy & $1.3\pm0.2$\,Jy & $^b$\\
    & 12.5 & Oct./Nov. 2004 & MIDI/VLTI	 & $7.8\pm0.5$\,Jy & $5.3\pm0.6$\,Jy & $1.3\pm0.3$\,Jy & $^b$\\
\noalign{\smallskip}
\hline
\noalign{\smallskip}
Q  & 19.91 & 27-08-1996 & MAX / UKIRT         &  \multicolumn{1}{r}{$20\pm 5$\,Jy} &\multicolumn{2}{r}{------------------ \ \ $28\pm7$\,Jy\ \ -----------} & $^g$\\
   &    25 & 25-09-1997 & ISOPHOT             &  \multicolumn{3}{r}{-----------------------$\ \ \ \ 40.88\pm \ \ 6.13$\,Jy\ \  ---------------------} & $^b$\\
   &    60 & 25-09-1997 & ISOPHOT             &  \multicolumn{3}{r}{-----------------------$\ \    117.77\pm	 17.67$\,Jy\ \  ---------------------} & $^b$\\
   &   100 & 25-09-1997 & ISOPHOT             &  \multicolumn{3}{r}{-----------------------$\ \    105.55\pm	 15.83$\,Jy\ \  ---------------------} & $^b$\\
   &   150 & 25-09-1997 & ISOPHOT             &  \multicolumn{3}{r}{-----------------------$\ \    113.39\pm	 17.01$\,Jy\ \  ---------------------} & $^b$\\
   &   170 & 25-09-1997 & ISOPHOT             &  \multicolumn{3}{r}{-----------------------$\ \ \ \ 82.79\pm	 12.42$\,Jy\ \  ---------------------} & $^b$\\
\noalign{\smallskip}
\hline
\noalign{\smallskip}
   &   350 &  2004-2005 & CSO                 & \multicolumn{3}{r}{-----------------------$\ \ \ \ 8.149\pm0.253$\,Jy ----------------------} & $^h$\\
   &   450 &  1997-2002 & SCUBA / JCMT        & \multicolumn{3}{r}{-----------------------$\ \ \ \ 1.655\pm0.218$\,Jy ----------------------} & $^h$\\
   &   840 & 28-10-1994 & JCMT \& CSO         & \multicolumn{3}{r}{-----------------------$\ \ \ \ 1.350\pm0.675$\,Jy ----------------------} & $^i$\\
   &   850 &  1997-2002 & SCUBA / JCMT        & \multicolumn{3}{r}{-----------------------$\ \ \ \ 0.628\pm0.017$\,Jy ----------------------} & $^h$\\
   &  1100 & 01-12-1994 & OVRO                & $0.397\pm0.035$\,Jy & \multicolumn{2}{r}{----------------- \ \ $< 0.100$\,Jy\ \ ----------}   & $^i$\\
   &  1300 & April 1988 & IRAM                & \multicolumn{3}{r}{-----------------------$\ \ \ \ 0.280\pm0.009$\,Jy ----------------------} & $^h$\\
   &  2800 &  1996-1997 & BIMA                & $0.050\pm0.006$\,Jy & \multicolumn{2}{r}{----------------- \ \ $< 0.009$\,Jy\ \ ----------}   & $^j$\\

\noalign{\smallskip}
\hline
\noalign{\smallskip}
\multicolumn{4}{l}{$^a$ Melnikov, S. (private communication)}     & \multicolumn{4}{l}{$^f$ \cite{ghez91}: flux of T~Tau~N} \\
\multicolumn{4}{l}{$^b$ this work}                                & \multicolumn{4}{l}{$^g$ \cite{herbst97}}\\		   
\multicolumn{4}{l}{$^c$ \cite{herbst07}}                          & \multicolumn{4}{l}{$^h$ \cite{andrews05} and references therein}\\ 
\multicolumn{4}{l}{$^d$ \cite{duchene05}: flux ratios}            & \multicolumn{4}{l}{$^i$ \cite{hogerheijde97}}\\		   
\multicolumn{4}{l}{$^e$ \cite{beck04}: flux of T~Tau~N}            & \multicolumn{4}{l}{$^j$ \cite{akeson98}}\\		   
\noalign{\smallskip}
\hline

\end{tabular}
\end{table*}

In addition to the N-band spectra obtained with MIDI (Section~\ref{phot}), we derived mid-infrared fluxes from measurements with the ``Infrared Space Observatory'' (ISO). For the reduction of the raw data the ``ISOPHOT Interactive Analysis'' software (PIA) version 10.0.0 was used. The long wavelength observations were small rasters, producing at 60 and 100\,$\mu$m final maps of $9\times 3$ pixels with a pixel scale of $43''\times43''$ and at 150 and 170\,$\mu$m maps of $6\times 2$ pixels with a pixel scale of $89''\times89''$. The flux distribution of these small maps was fitted by a sum of a theoretical PSF and a small extended clump (assumed in Gaussian shape). This approach was chosen to eliminate the contributions from the nebulae that are present in the T~Tau system. Details of the method are described in \cite{abraham00}. Nonetheless, due to the large beam size of the telescope the colour-corrected ISO fluxes listed in Table~\ref{table2} will be considered as upper limits only.

In the visual, we took images in four bands by using MPIA's KING telescope on the K{\"o}nigstuhl in Heidelberg with its 0.7\,m primary mirror. These images have been reduced and analysed with standard IRAF routines. For the near-infrared, the sub-mm, and the mm-regime, fluxes from the literature have been adopted, chosing as a rule data corresponding to the low state of T~Tau~Sa. For a discussion of the Q-band fluxes see Section~\ref{dis}. All photometric measurements are listed in Table~\ref{table2}.

The derived spectral energy distributions (SEDs) are displayed in Figures~\ref{Fig_n1}, \ref{Fig_sb1}, and \ref{Fig_sa1}. The ISO fluxes therein are distributed to T~Tau~N and T~Tau~S according to the flux ratio found in the upper N-band between 12\,$\mu$m and 13\,$\mu$m. For the flux ratio of T~Tau~Sb and T~Tau~Sa we assume a value of 0.5. The latter is also applied to the Q-band flux of T~Tau~S.

%

\section{Data Reduction\label{dr}}

\subsection{Visiblities\label{drint}}

\begin{figure*}
\centering
\includegraphics[height=18cm,angle=90]{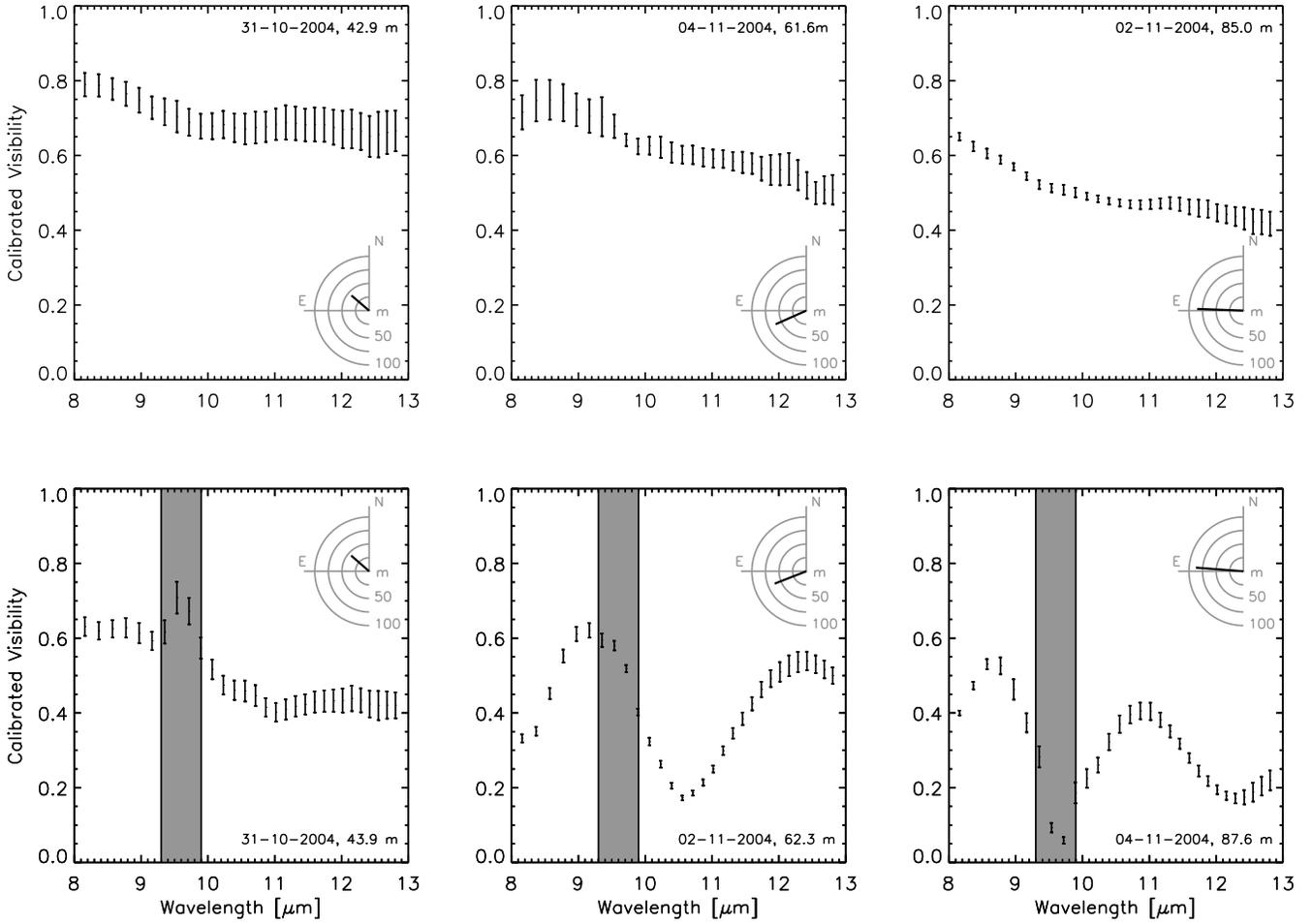}
\caption{The calibrated visibilities for T~Tau~N (upper row) and T~Tau~S (lower row). The errors are the calibration errors. The thick lines within the concentric rings on the right side of the individual panels represent the length and position angle of the projected baseline. Between 9.4\,$\mu$m and 9.9\,$\mu$m the calibrated visibilities for T~Tau~S are not used for analysis. This wavelength range is strongly affected by the atmospheric ozone band due to the low flux of the source in the silicate band.}  
\label{Fig2}
\end{figure*}

For the data reduction, a custom software called {\it MIA+EWS}$^{3,4}$, written in IDL and the C language, was used, where we chose the first branch based on power spectrum analysis (called {\it MIA}). This package works best when OPD scans are centred on the white-light fringe (zero OPD), as in most of our observations, but can also handle data tracked at non-zero OPD. The data reduction steps are described in a tutorial on the cited web page, as well as in \cite{leinert04} and \cite{ratzka05}. The results are raw values for the 8-13\,$\mu$m spectrum $F_{\rm raw}\left(\lambda\right)$ and the 8-13\,$\mu$m correlated flux $F_{\rm raw}^{\rm cor}\left(\lambda\right)$.
\footnotetext[3]{http://www.strw.leidenuniv.nl/$\sim$nevec/MIDI/}
\footnotetext[4]{http://www.mpia-hd.mpg.de/MIDISOFT/}

To correct for the bias introduced by signal fluctuations not related to the fringe signal, the off-fringe power spectrum, determined on the source, but far from zero OPD, is subtracted from the signal. The fringe amplitude (or correlated flux) as a function of wavelength is the square root of the fringe power spectrum after this "background'' subtraction. Division of this final fringe amplitude by the photometric flux gives the raw visibility of the object as a function of wavelength. To allow at least approximately for the influence of unequal fluxes in the two interfering beams, the raw visibilities $V_{\rm raw}$ are calculated by 
\begin{equation}
V_{\rm raw}\left(\lambda\right)={F_{\rm raw}^{\rm cor}\over \sqrt{F_{\rm raw}^A\left(\lambda\right)F_{\rm raw}^B\left(\lambda\right)}}\mathrm{\,,}\label{eq1}
\end{equation}
where A and B are the fluxes determined from the photometric datasets of the two incoming telescope beams. A precise correction, which would imply determining the fluxes in the two beams for each individual interferometric scan or even each exposed frame, is not possible with the instrument mode used. 

Calibrated visibilities for an object are obtained by dividing its raw visibility by the instrumental visibility derived from one or more calibrators with known diameter observed within the same night:
\begin{equation}
V\left(\lambda\right) = {V_{\rm raw}\left(\lambda\right)\over V_{\rm instr}\left(\lambda\right)}
\mathrm{\, .}\label{eq2}
\end{equation}

The second branch of the data reduction software, called {\it EWS}, uses a clever shift-and-add algorithm in the complex plane, averaging suitably modified individual exposures to obtain the complex visibility A($\lambda$)$\cdot$e$^{i\phi(\lambda)}$ (see \cite{jaffe04} and the documentation on the web page just referenced). Thus, this algorithm is ``coherent''. It works best if the OPD scan of the interferometric measurements is offset from zero OPD. For the low spectral resolution of the prism ($\lambda/\Delta\lambda$ $\approx$~30) an offset of about $5\lambda$ is suitable. However, EWS can also be used safely without an offset from zero OPD when the observed sources are reasonably bright so that the background in no major source of error. We thus used EWS to confirm the visibility results obtained with the MIA power spectrum analysis, but in particular analysed the phases computed by EWS to identify the brighter source in the close binary T~Tau~S (see Section~\ref{dis} and Appendices~\ref{phases}\,\&\,\ref{etavir} for details).

\subsection{Photometry\label{drphot}}

While the determination of the normalised and calibrated visibilities requires the special software package, the absolute spectra $F\left(\lambda\right)$ can be derived from the photometric measurements $F_{\rm raw}^A\left(\lambda\right)$ and $F_{\rm raw}^B\left(\lambda\right)$ by following the standard procedures applied when reducing mid-infrared spectra. To do this, however, at least a few of the interferometric calibrators have to also be spectrophotometric standard stars with well-known spectra. Furthermore, the different airmasses of the scientific targets and the calibrators as well as the temporal variations in seeing conditions have to be taken more seriously than for the visibility determination.

\subsection{Correlated Flux\label{drcorr}}

The wavelength-dependent correlated flux $F^{\rm cor}\left(\lambda\right)$ can now be derived by a simple multiplication of the above determined quantities:
\begin{equation}
F^{\rm cor}\left(\lambda\right) = V\left(\lambda\right)\cdot F\left(\lambda\right) 
\mathrm{\, .}\label{eq3}
\end{equation}
Although one may think of directly calibrating $F_{\rm raw}^{\rm cor}\left(\lambda\right)$ instead of using Equation~(\ref{eq3}), we determine in this paper the correlated flux of our comparatively bright sources with the already calibrated quantities $V\left(\lambda\right)$ and $F\left(\lambda\right)$.

Since the correlated spectra reflect to a first appromixation the emission from the region not resolved by the interferometric instrument, they carry valuable information about the dependence of the spectrum on the distance from the star. 

%

\section{Direct Results\label{results}}

In the following, we present the visibilities and photometry of T~Tau~N and T~Tau~S. These results can be directly derived from the observations with MIDI without making assumptions or using simulations.

\subsection{Visibilities\label{visi}}

Separate interferometric observations were possible for T~Tau~N and the T~Tau~S subsystem. The results are shown in Figure~\ref{Fig2}. Here, the calibration is performed with all calibrators of one night that show a reasonable visibility. Calibrator measurements that have not been used are indicated in Table~\ref{table1}. The errors represent the standard deviation of the ensemble of calibrator measurements. For the two measurements of the scientific targets with corrupted photometry (see Table~\ref{table1}) we used the photometry for the object in the two other nights and averaged the resulting visibilities. In this case, the error bars of the average visibility span the whole range that was originally covered by the error bars of the visibilities averaged, i.e., the new error bars form an ``envelope'' around the original visibilty errors. 

Both T~Tau~N and T~Tau~S are well resolved  with visibility values corresponding to a FWHM of a Gaussian brightness distribution of $\sim 10\pm5$\,mas at 8\,$\mu$m. The FWHM roughly increases with wavelength to $\sim 20\pm5$\,mas at 13\,$\mu$m. However, the unequal appearance of the visibilities of T~Tau~N and T~Tau~S calls for a different treatment of the two measurements (Sections~\ref{ttaun}\,\&\,\ref{ttaus}). While the measurements of T Tau N show a wavelength dependence of the visibility typical for emission from a circumstellar disc \citep{leinert04}, the visibilities for T~Tau~S show sinusoidal oscillations. These have to be interpreted as the signature of the binary components T~Tau~Sa and T~Tau~Sb. The deep minima in the modulations already indicate that the brightness of the two sources is similar in the N-band.  

%

\subsection{Photometry\label{phot}}

\begin{figure}
\centering
\includegraphics[height=8.5cm,angle=90]{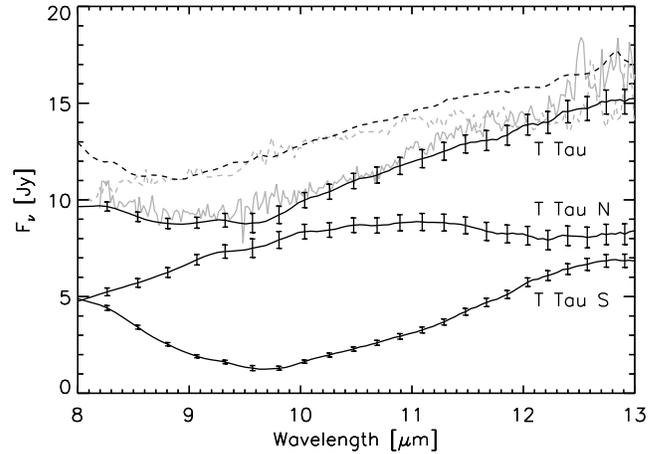}
\caption{The spectra for the northern and the southern component together with a combined spectrum of the whole source. A spectrum obtained with Spitzer in February 2004 is drawn as dashed black line. Underlaid in grey are two spectra taken with TIMMI\,2 in February 2002 (solid line) and December 2002 (dashed line).}
\label{Fig1}
\end{figure}

We obtained separate spectrophotometric observations for T~Tau~N and the T~Tau~S subsystem (Figure~\ref{Fig1}). From the comparison with TIMMI\,2 spectra taken by \cite{przygodda04} and a Spitzer spectrum, we tentatively conclude that no large brightness changes occured in the T~Tau system between 2002 and our measurements. This is similar to what was observed in the K-band \citep{duchene05}, where T~Tau~S appears to be in a low brightness state. However, it is worth mentioning that the spectra deviate significantly from each other, i.e., we witness a continuous change of the contributions of the southern components to the overall spectrum.

%

\section{The northern component\label{ttaun}}

%

\subsection{Radiative Transfer Model}

We used the radiative transfer code MC3D \citep{wolf99, wolf03a, schegerer08} to model simultaneously the spectral energy distribution and the visibilities of T~Tau~N by manually adjusting the parameters. Manual adjustments have the advantage to allow even in the paramater space of our radiative transfer model with its many dimensions the determination of the best fitting model. On the other hand, this approach always leaves a small chance that only a local instead of a global minimum has been found. But we carefully checked this possibilty and are confident that we determined the best parameters.  

The used MC3D code includes the effects of heating of a disc and an envelope by the central star as well as heating by accretion. The photospheric spectrum of T~Tau~N is represented by a Kurucz model \citep{kurucz92} of a star with temperature $T_{\star}=5250$\,K, surface gravity $\log g=3.7$\,[log cm/s$^2$], and solar metallicity (Table~\ref{MC3D_N}). 

\paragraph{Density Model} The density profile of our radial-symmetric passively-heated (dust) disc is described in cylindrical coordinates $\left(r,z\right)$ that refer to the distance from the star in the midplane and the height above it as
\begin{equation}
\rho_{\rm disc}\left(r,z\right)=\rho_0\left(R_{\star}\over r\right)^{3\left(\beta-{1\over 2}\right)}\exp\left(-{1\over 2} \left[z\over h\left(r\right)\right]\right)\rm{\ ,}\label{rt1}
\end{equation}
where $R_{\star}$ is the stellar radius and $h\left(r\right)$ the disc scale height \citep{wolf03a}. The latter is defined as
\begin{equation}
h\left(r\right)=h_{100}\left(r\over 100\,\rm{AU}\right)^\beta\label{rt2}
\end{equation}
with $h_{100}$ being the scale height of the disc at $r=100$\,AU. Since the Equations (\ref{rt1}) and (\ref{rt2}) fully describe the structure of the disc, $\rho_0$ can be adjusted to reach the assumed disc mass.

As a representation of a potential ``envelope'', we add a simple spherical dust configuration to the disc model. This spherical envelope allows us to bring additional material close to the star to reprocess light more efficiently. It may represent remnant material of the star formation process. The envelope is geometrically constrained by the inner radius $R_{\rm in}$ and the outer radius $R_{\rm out}$ of the disc. With the density distribution of the disc $\rho_{\rm disc}\left(r\right)$ and the position vector $\vec{r}$, the density profile of the envelope is given by  
\begin{equation}
\rho_{\rm env}(\vec{r}) = c_{\rm 1} \cdot \rho_{\rm disc}(R_{\rm in},0) 
\cdot \left(\frac{|\vec{r}|}{R_{\rm in}}\right)^{c_{\rm 2}} \mbox{\ for\ \ } R_{\rm in} \leq |\vec{r}| \leq R_{\rm out}\rm{\ ,}
\label{rt3}
\end{equation} 
where $c_{\rm 1}\ll 1$ and $c_{\rm 2} < 0$. While the constant $c_\mathrm{2}$ determines the concentration of the dust towards the star, the quantity $c_\mathrm{1}$ is adjusted to guarantee a low optical depth of the envelope and the possibility to observe the innermost region of the disc. The disc and envelope are combined by 
\begin{equation}
\rho_{\rm env}(\vec{r}) = \left\{\begin{array}{lcl}
\rho_{\rm disc}(\vec{r}) & \mbox{for} & \rho_{\rm env}(\vec{r})  \leq \rho_{\rm disc}(\vec{r})\\
\rho_{\rm env}(\vec{r}) & \mbox{for} & \rho_{\rm env}(\vec{r}) > \rho_{\rm disc}(\vec{r})
\end{array}\right.\rm{,}
\label{rt4}
\end{equation} 
ensuring a smooth transition from the disc to the envelope.  

\paragraph{Accretion Effects} We extend the passive model further for the implementation of accretion effects \citep{schegerer08}. Apart from the parameters of the disc and the star, this accretion model requires three additional parameters: the accretion rate $\dot{M}$, the radius $R_\mathrm{bnd}$ at which the disc is magnetically truncated, and the boundary temperature $T_{\rm bnd}$ of the accreting regions on the surface of the star. Following the approach of \cite{koenigl91} we assume here that the truncation radius is half the corotation radius. A stellar magnetic field of only about 0.4\,kG is then required \citep{johns07}. For such a steady-state scenario and under the assumption that the shock front is located close to the stellar surface, $T_{\rm bnd}$ can be easily calculated when the stellar mass, the stellar radius, the stellar angular velocity, and the accretion rate are known \citep{koenigl91}. With a rotation period of 2.8\,days \citep{herbst86}, we find for the boundary temperature $T_{\rm bnd}$ of T~Tau~N the value listed in Table~\ref{MC3D_N}. Interestingly, the magnetic truncation occurs at a radius of $R_{\rm bnd}=5.4\,R_{\odot}$ in this accretion model. The boundary radius is thus smaller than $R_{\rm in}=23\,R_{\odot}$ and the magnetic truncation has to take place in a gaseous inner disc. 

\begin{figure}[b]
\centering
\includegraphics[width=8.5cm,angle=0]{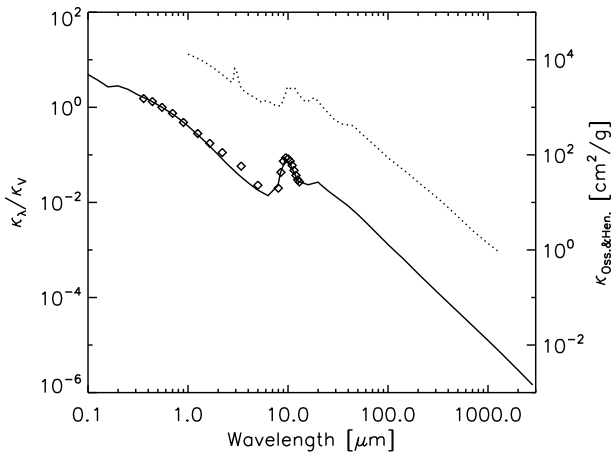}
\caption{Comparison of the mass absorption coefficients of the dust in the disc atmopshere (solid) with those of the insterstellar extinction law (diamonds) from \cite{rieke85}. The values are given relative to the V-band. Plotted as dotted curve are the opacities provided by \cite{ossenkopf94} for a gas density of $10^6\,{\rm cm}^{-3}$, thin ice mantles, and a coagulation over $10^5$\,years.}
\label{Kappa}
\end{figure}

\paragraph{Dust Model} Following \cite{schegerer08}, we assume in our modelling approach a gas-to-dust mass ratio of 100:1 and a dust mixture of ``astronomical silicate'' and graphite with relative abundances of $62.5$\% for astronomical silicate and $37.5$\% for graphite \citep{draine93}. We use the optical data of ``smoothed astronomical silicate'' and graphite published by \cite{draine84} and \cite{weingartner01} and consider a grain size power law $n(a) \propto a^{-3.5}$ with $a_\mathrm{min} \leqq a \leqq a_\mathrm{max}$, where $n(a)$ is the number of dust particles with radius $a$. This power law was found by \cite{mathis77}, hereafter MRN, in their study of the extinction of interstellar carbon and silicate with typical sizes between $0.005 \mathrm{\mu m} - 0.01 \mathrm{\mu m}$ and $0.025\ \mathrm{\mu m} - 0.25 \mathrm{\mu m}$, respectively. This grain size power law has already been used in former modelling aproaches of discs around young stars, e.g., \cite{wolf03a}. We use a minimum particle size of $a_\mathrm{min} = 0.005\ \mathrm{\mu m}$.

Furthermore, we implement a two-layer dust model. The disc interior contains a maximum dust grain size of $1\ \mathrm{mm}$ while the MRN grain size distribution with $a_\mathrm{max} = 0.25\ \mathrm{\mu m}$ is used in the disc atmosphere. These upper disc layers are defined by an optical depth $\tau_\mathrm{N}$ in the N-band, measured vertical to the disc midplane, below unity. Such a division of the disc is based on the idea of the favoured settling of larger dust grains, e.g., \cite{schraepler04}. Similar models with different grain sizes have been investigated for other discs by \cite{dalessio97}, \cite{menshchikov99}, and \cite{wolf03a}.

In order to avoid determining the temperature distribution of each single dust component and to accelerate the radiative transfer simulations, we average the optical properties of carbon and astronomical silicate for  different grain sizes in each dust layer to derive the optical constants. Such an approach was justified by \cite{wolf03c}.

\begin{table}[b]
\caption{The parameters of the radiative transfer model of T~Tau~N.}
\label{MC3D_N}
\centering
\begin{tabular}{llccrcc}
\hline\hline
\noalign{\smallskip}
               & Parameter & Unit & & Value & & Ref.\\
\noalign{\smallskip}
\hline
\noalign{\smallskip}
\multicolumn{2}{l}{{\bf Stellar Parameters}}\\
\noalign{\smallskip}
\hspace{0.3cm} & Mass ($M_{\star}$)	 	& M$_{\odot}$	   & & 2.1             & & $^a$ \\
               & Temperature ($T_{\star}$)	& K$_{\ \ }$	   & & 5250            & & $^a$ \\
               & Luminosity ($L_{\star}$)	& L$_{\odot}$	   & & 7.3             & & $^a$ \\
               & Radius ($R_{\star}$)	& R$_{\odot}$	   & & 3.3             & & calc.\\
\noalign{\smallskip}
\noalign{\smallskip}
\multicolumn{2}{l}{  {\bf Accretion}}\\
\noalign{\smallskip}
               & Accretion Rate ($\dot{M}$)	& M$_{\odot}\rm{yr}^{-1}$  & & $3\cdot10^{-8}$ & & $^a$ \\
               & Boundary Temp. ($T_{\rm bnd}$) & K$_{\ \ }$	   & & 9700            & & calc.\\
               & Boundary Rad. ($R_{\rm bnd}$)  & R$_{\star}$	   & & 1.6             & & calc. \\
\noalign{\smallskip}
\noalign{\smallskip}
\multicolumn{2}{l}{  {\bf Circumstellar Disc / Envelope}}\\
\noalign{\smallskip}
               & Inner Radius ($R_{\rm in}$)	& AU		   & & 0.1             & & $^b$ \\
               & Outer Radius ($R_{\rm out}$)	& AU		   & & 80	           & & $^b$ \\
               & $\beta$	 	&		   & & 1.25            & & $^b$ \\
               & $h_{100}$  	 	& AU		   & & 18	           & & $^b$ \\
               & $c_1$		 	&		   & & $1\cdot10^{-5}$ & & $^b$ \\
               & $c_2$		 	&		   & & -5.0            & & $^b$ \\
               & Inclination	 	& deg		   & & $<30$           & & $^b$ \\
               & Disc Mass ($M_{\rm disc}$)	& M$_{\odot}$	   & & $4\cdot10^{-2}$ & & $^c$ \\
\noalign{\smallskip}
\noalign{\smallskip}
\multicolumn{2}{l}{  {\bf Interstellar Extinction}}\\
\noalign{\smallskip}
               & Foreground Extinction $A_V$	& mag		   & & 1.5             & & $^a$ \\
\noalign{\smallskip}
\hline
\noalign{\smallskip}
\multicolumn{7}{l}{\parbox{8.5cm}{$^a$ \cite{white01}; $^b$ this work; $^c$ \cite{hogerheijde97}; calc. = calculated}}\\
\noalign{\smallskip}
\hline             
\end{tabular}
\end{table}

\begin{figure*}
\centering
\includegraphics[width=9.0cm,angle=0]{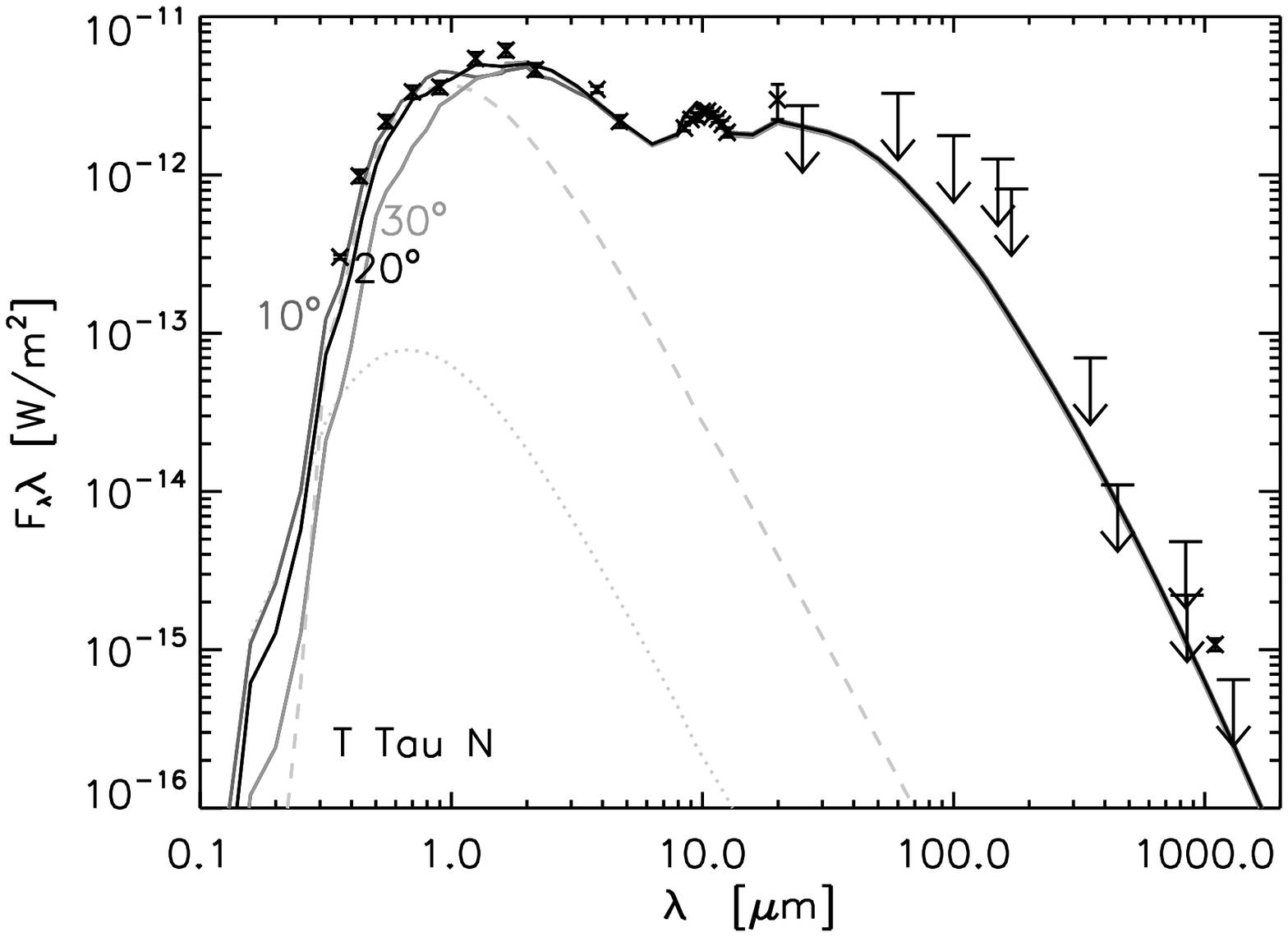}
\includegraphics[width=9.0cm,angle=0]{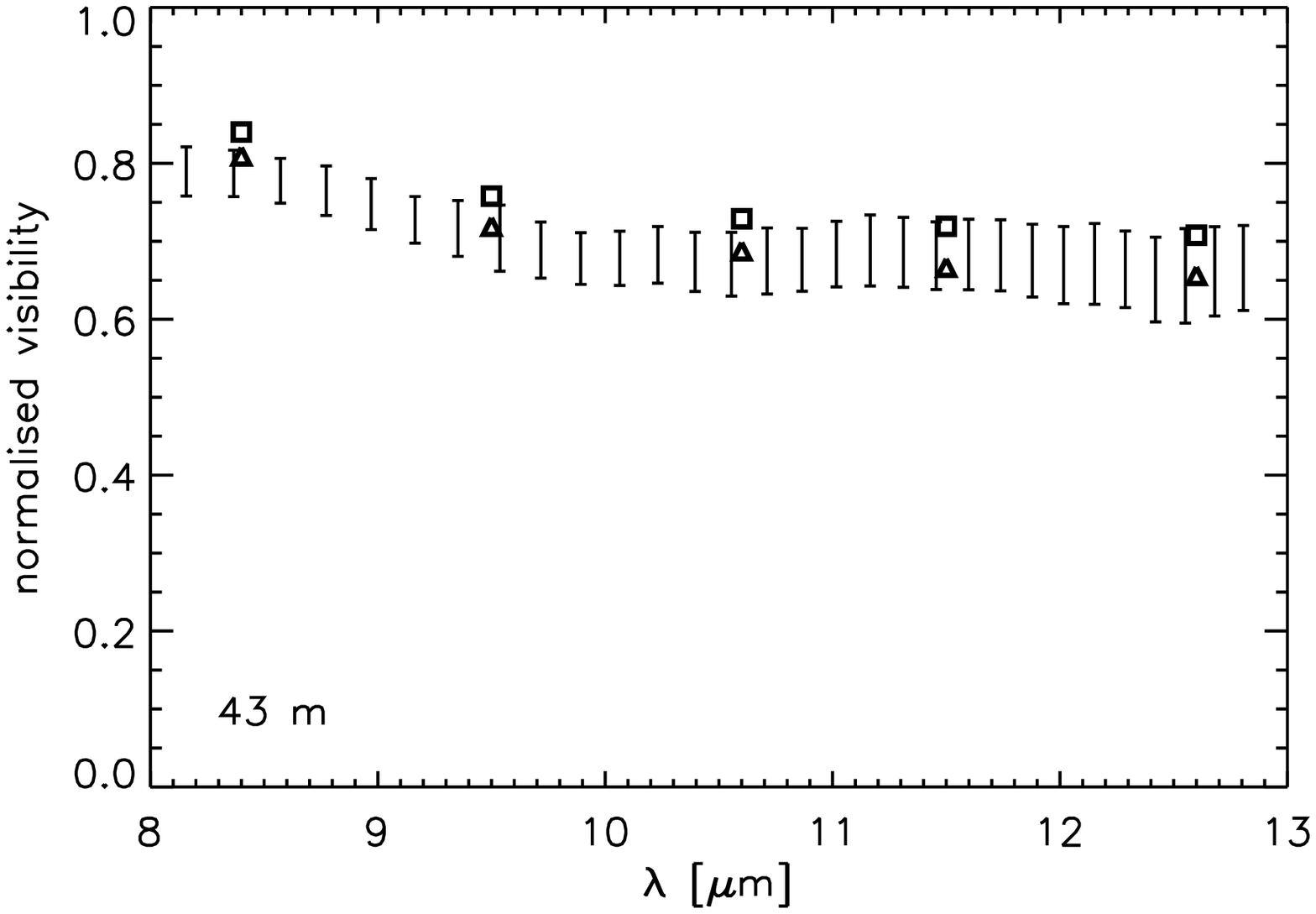}\\
\includegraphics[width=9.0cm,angle=0]{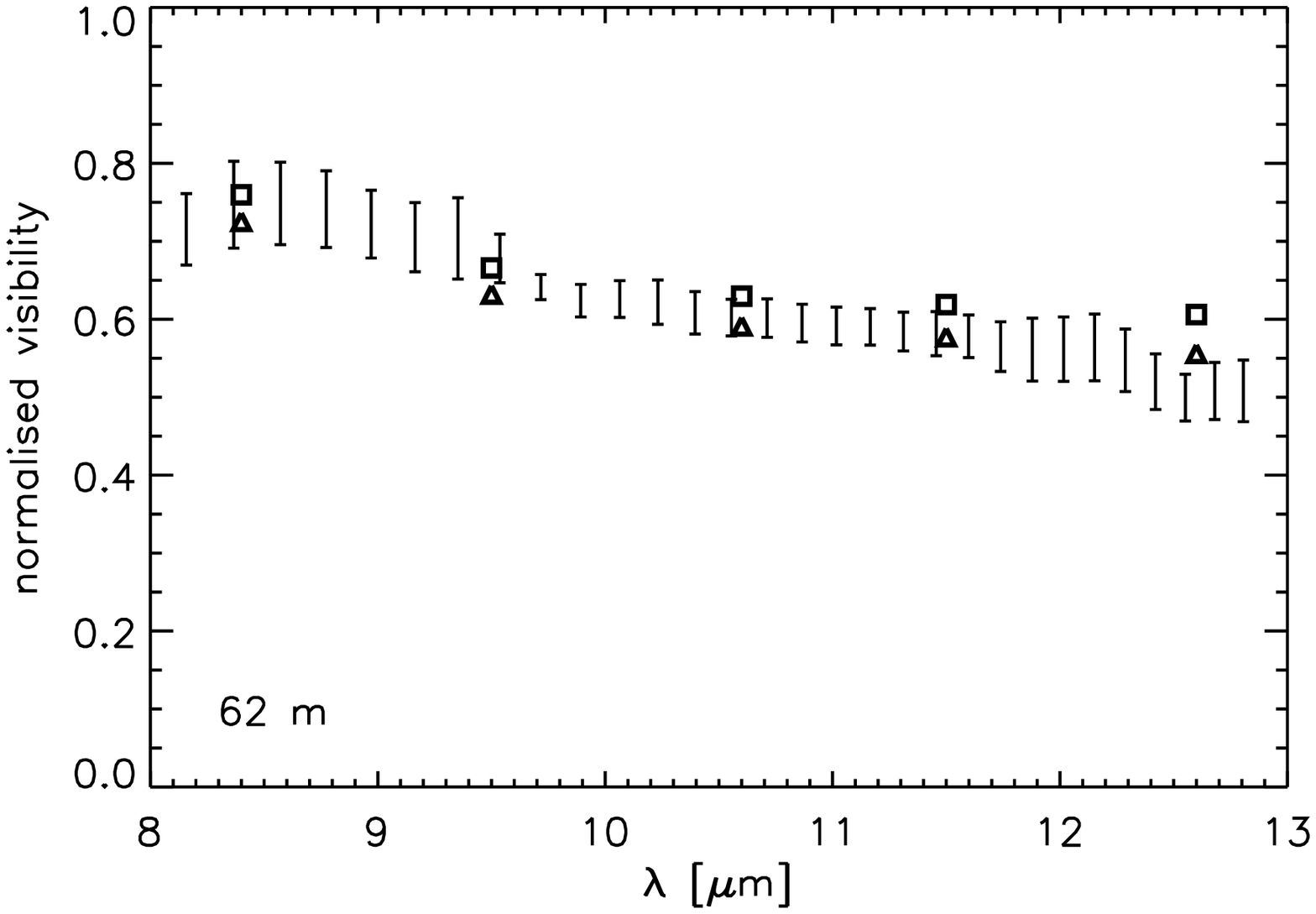}
\includegraphics[width=9.0cm,angle=0]{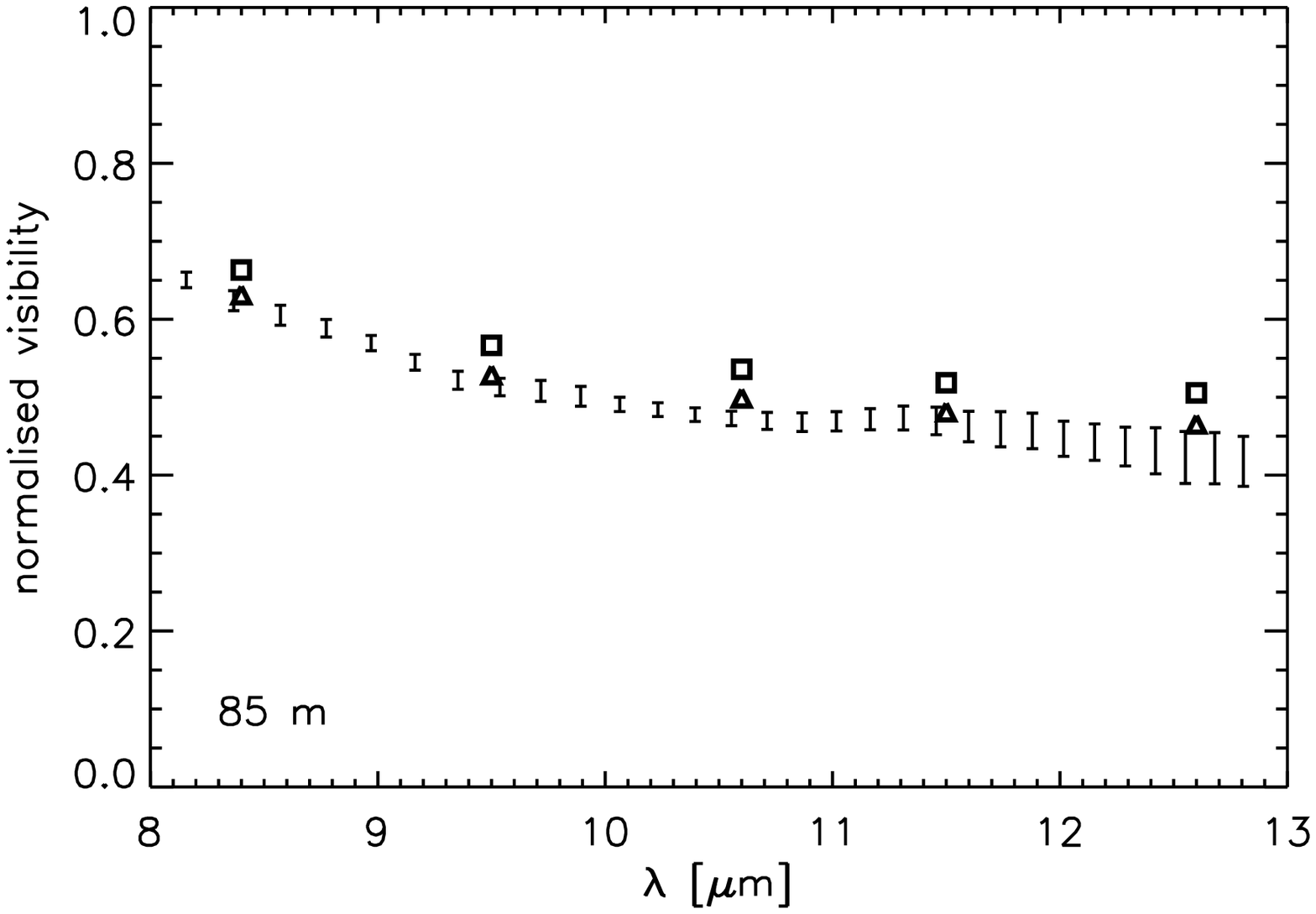}\\
\caption{Results of the simultaneous fit to the SED and the visibilities of T~Tau~N. {\it Top Left:} The model (Table~\ref{MC3D_N}) is plotted for inclinations of $10^{\circ}$, $20^{\circ}$, and $30^{\circ}$. The contributions from the stellar photosphere based on a Kurucz model (dashed) and the accretion (dotted) take foreground extinction into account. Measured photometric data are indicated by black crosses. Arrows indicate upper limits. For references, see Table~\ref{table2}. {\it Top Right and Lower Row:}  The measured visibilities (error bars) are displayed together with the upper and lower limits (squares and triangles) derived from the model with an inclination of $20^{\circ}$ by varying its orientation by $360^{\circ}$.} 
\label{Fig_n1}
\end{figure*}

\begin{figure*}
\centering
\includegraphics[width=5.5cm,angle=0]{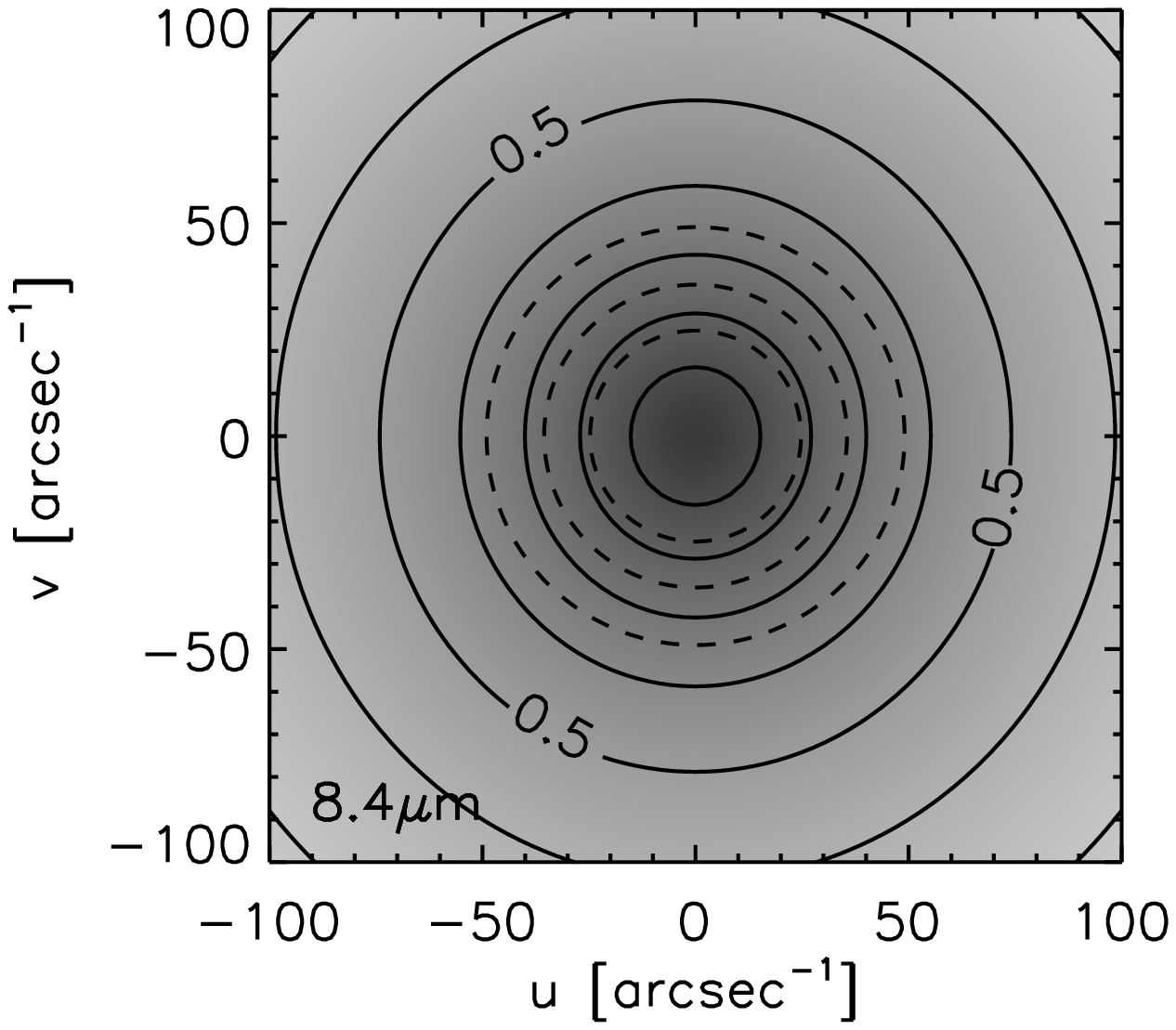}\hspace{0.6cm}
\includegraphics[width=5.5cm,angle=0]{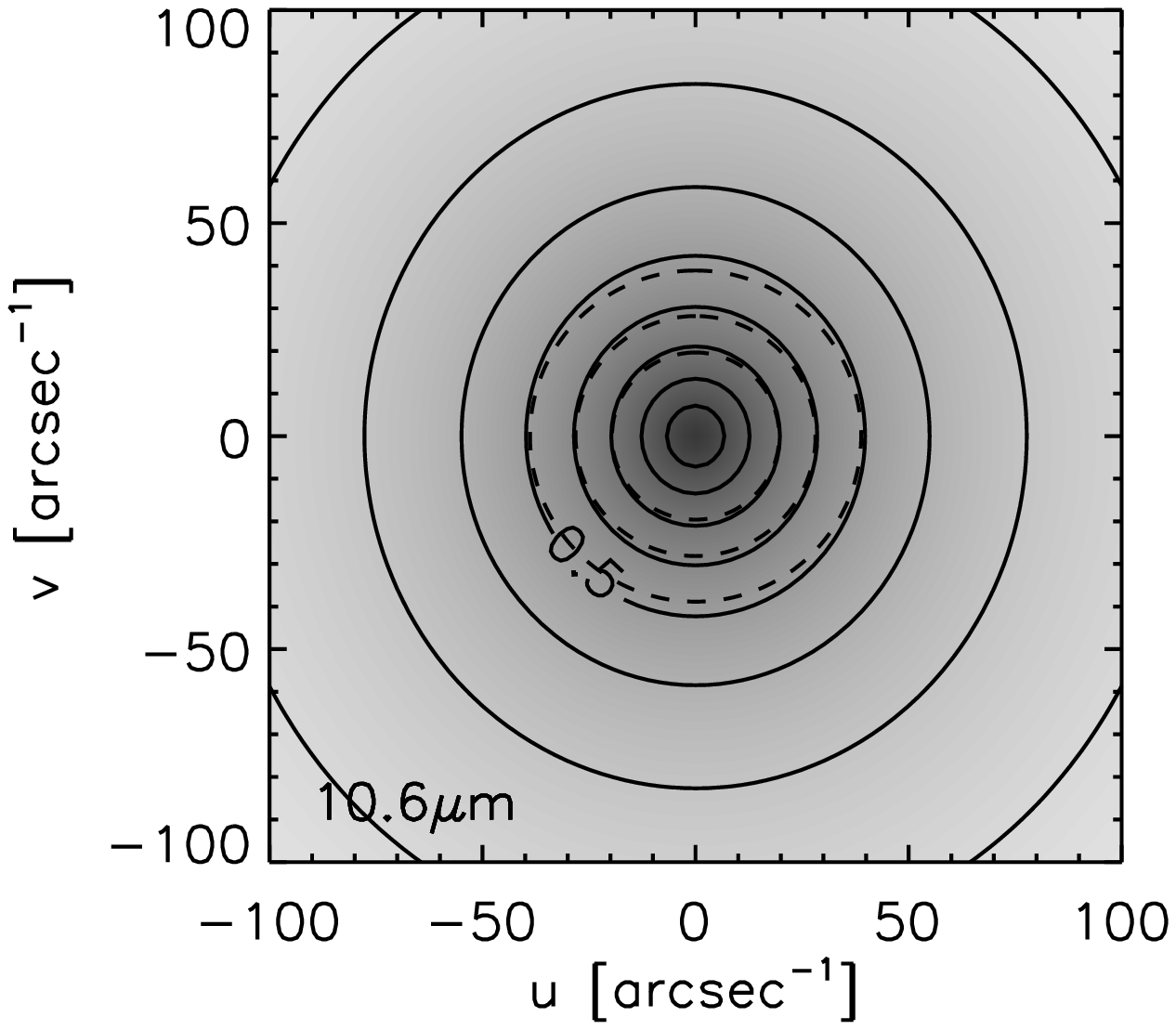}\hspace{0.6cm}
\includegraphics[width=5.5cm,angle=0]{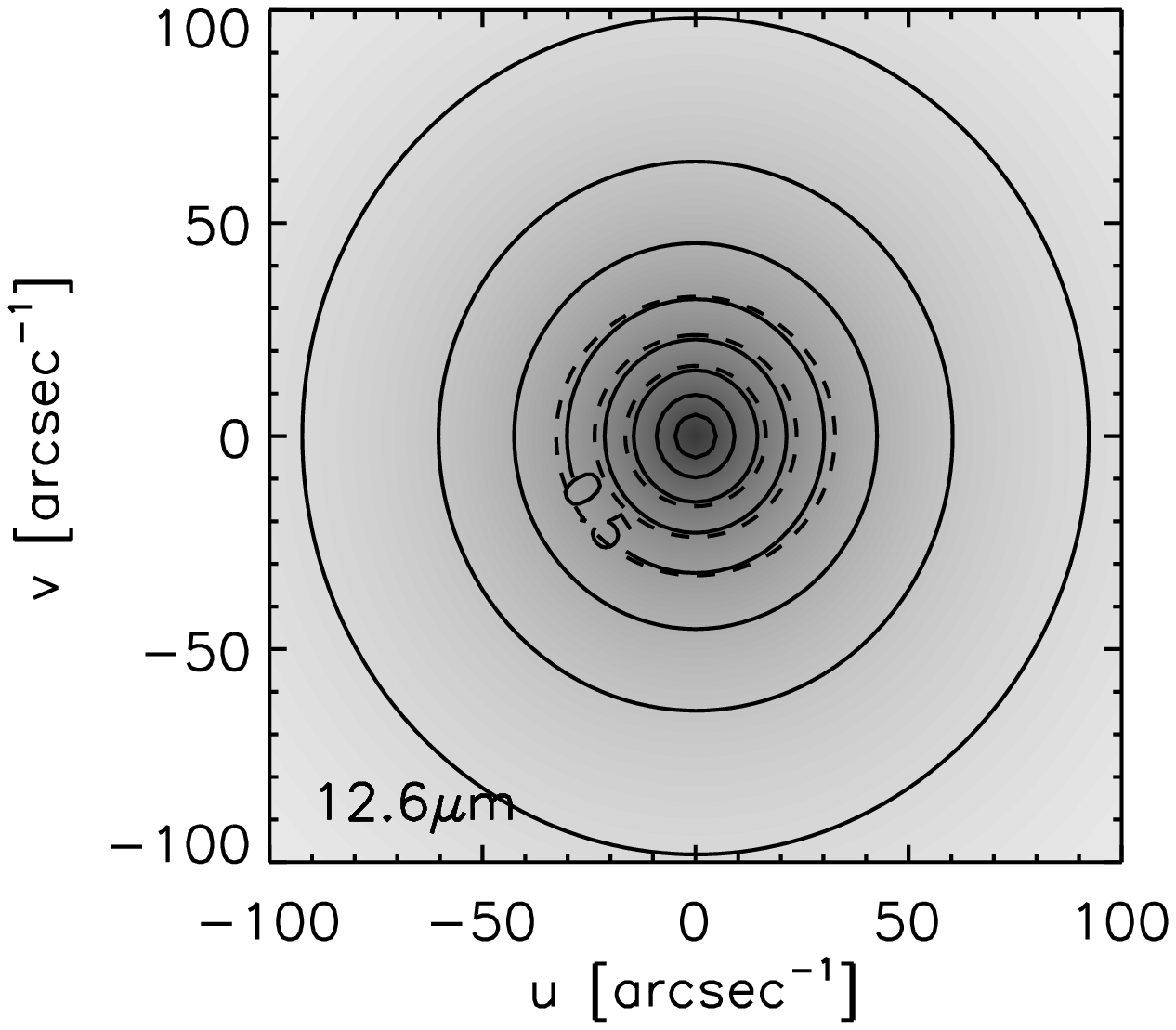}
\caption{The modelled visibilities of T~Tau~N for 8.4\,$\mu$m, 10.6\,$\mu$m, and 12.6\,$\mu$m and a disc inclination of $20^{\circ}$ (from left to right). The solid contours are drawn for steps of 0.1 and the associated grey background is underlaid for easier comparison. The radii of the overplotted dashed circles correspond to the spatial frequencies for which the baselines are sensitive at the various wavelengths.}  
\label{Fig_n2}
\end{figure*}

\paragraph{Interstellar Extinction} Since the interstellar extinction is not taken into account by the model itself, we ``redden'' the fluxes resulting from the simulation before comparing them with the measured values. Starting with the visual extinction $A_{\rm V}$ the extinctions for the other wavelengths are calculated by applying the wavelength-dependent mass absorption coefficients. These or more presisely the cross sections are provided by the model of the dust in the disc atmosphere described in the last paragraph. A comparison with the interstellar extinction law by \cite{rieke85} is shown in Figure~\ref{Kappa}. The visual extinctions $A_{\rm V}$ of the models are given together with the other parameters in Tables~\ref{MC3D_N}\,\&\,\ref{MC3D_S}.

%

\begin{figure}[h!]
\centering
\includegraphics[width=8.5cm,angle=0]{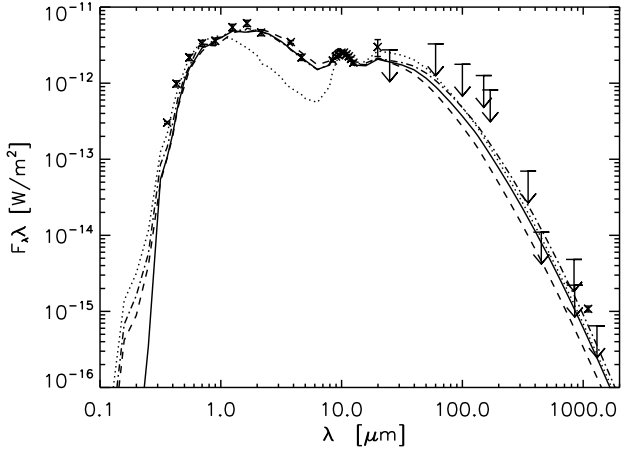}
\caption{The spectral energy distribution for the model of T Tau~N with a disc inclination of 20$^{\circ}$ when neglecting the accretion (solid), removing the ``envelope'' (dotted), and changing the outer radius of the disc to 50\,AU (dashed), 110\,AU (dashed-dotted), respectively.}
\label{CompN}
\end{figure}

\subsection{Results of Model Fits\label{mc3d_n_mir}}

\setcounter{footnote}{4}

The parameters of our best model for T~Tau~N are listed in Table~\ref{MC3D_N}. In Figure~\ref{Fig_n1}, the resulting SEDs and normalised visibilities are plotted together with the measured data. A comparison shows that the model and the data are in good agreement, although the stellar parameters, the disc mass\footnote{The opacities used by \cite{hogerheijde97} are plotted for comparison in Figure~\ref{Kappa} as dotted curve (M. Hogerheijde, priv. communication). Since MC3D only provides cross sections no absolute opacities can be given.}, and the accretion rate were taken as constrained by complementary studies.

However, there are differences between the measured SED and the model. The slight deviations in the N-band are consequences of the simplistic dust model. The ``underestimation'' of the measured SED in the wavelength range from 25\,$\mu$m to 170\,$\mu$m can be attributed to the large beam size of the ISO satellite. We already treated the measured fluxes as upper limits, because the nebulae around T~Tau~N contribute significantly at those wavelengths. The same might be applicable for the Q-band. Here, the resolution of the telescope was limited and the southern binary could only marginally be resolved \citep{herbst97}. 

Due to the uncertainties at wavelengths longwards of the N-band, models with various combinations of the parameters $\beta$ and $h_{100}$ fit the SED and the visibilities. Also $R_{\rm out}$ is not well constrained. Figure~\ref{CompN} shows the changes in the SED when varying the outer radius of the dics $R_{\rm out}$ by 30\,AU while keeping all other model parameters constant. The visibilities both in the near- and the mid-infrared are only marginally affected. In a recent study, based on the extent of the H$_2$ emission and models of the SED, an outer radius of the disc of about 85-100\,AU was found \citep{gustafsson08}. The value listed in Table~\ref{MC3D_N} thus probably underestimates the true size of the disc. 

Nonetheless, our model puts strong constraints on the inclination of the disc. Large inclination angles can be safely excluded on the basis of both the SED and the visibilities. The different levels of the measured visibilities only reflect the spatial resolution reached by the different baseline configurations and are not an indication for an elongated disc (Figures~\ref{Fig_n1}\,\&\,\ref{Fig_n2}). On the other hand, since the variations of the visibilities due to the variation of the position angle are small, the determination of the actual position angle of the major axis of the disc around T~Tau~N is not possible. 

Another constraint of the model is the presence of a thin envelope with a steep density gradient, where the material is highly concentrated towards the central star. Without this aditional material the flux between 3\,$\mu$m and 8\,$\mu$m would be too low (Figure~\ref{CompN}) and the visibilities represent no longer the measurements. The visibilities are reduced by about 50\% when compared to our best model. However, it has to be mentioned that models with a curved inner rim or other special geometries of the inner disc have not been tested, although such models might not require the steep density gradient of the envelope found above.

Although accretion may lead to similar effects as an extended, spherical dust distribution \citep{schegerer08}, according to the measurements the accretion rate of T~Tau~N is not high enough. However, our model is not very sensitive to the accretion rate. Neither the visibilities nor the SED change significantly (Figure~\ref{CompN}).

\begin{figure*}
\centering
\includegraphics[width=9.0cm,angle=0]{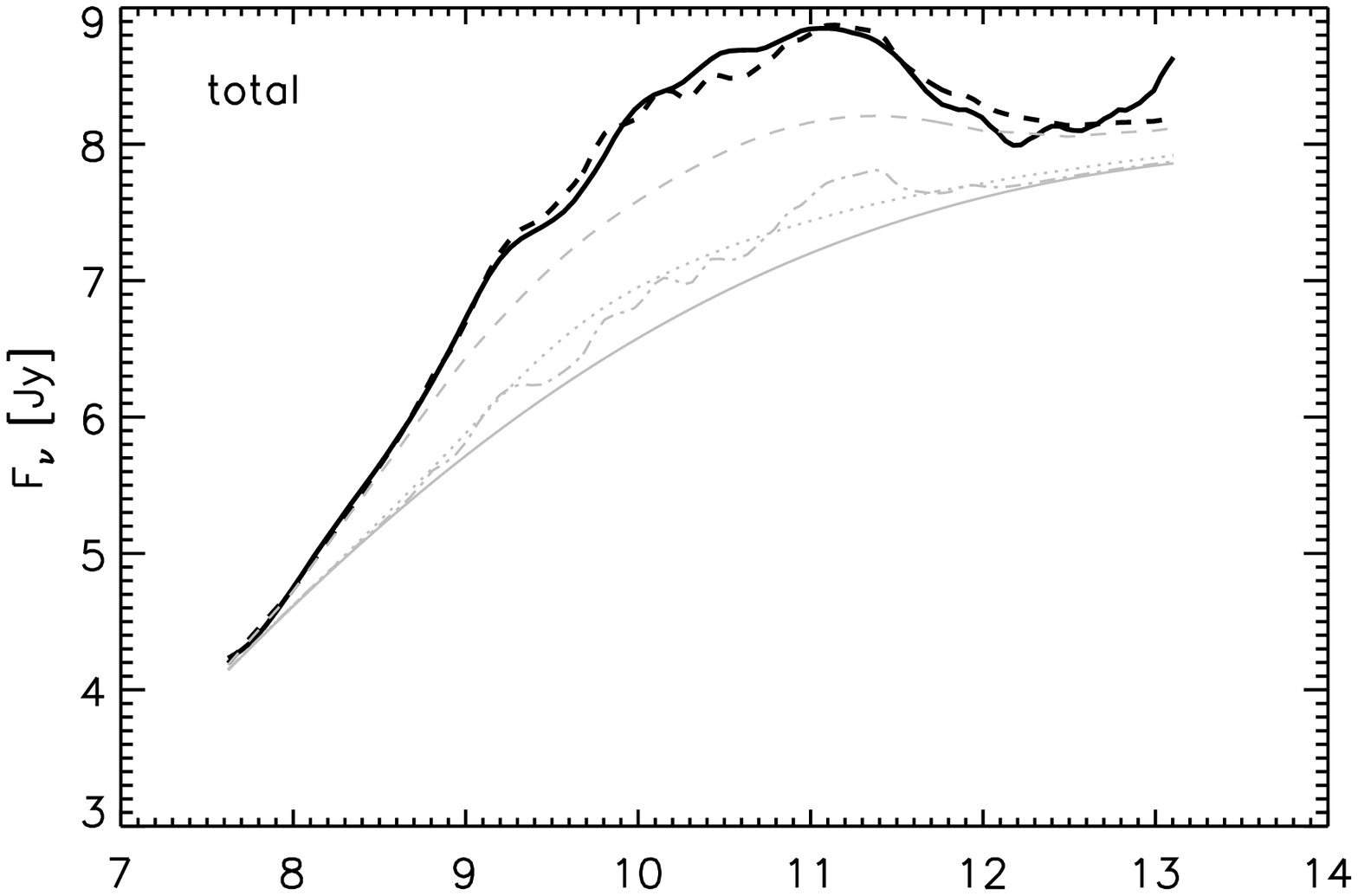}
\includegraphics[width=9.0cm,angle=0]{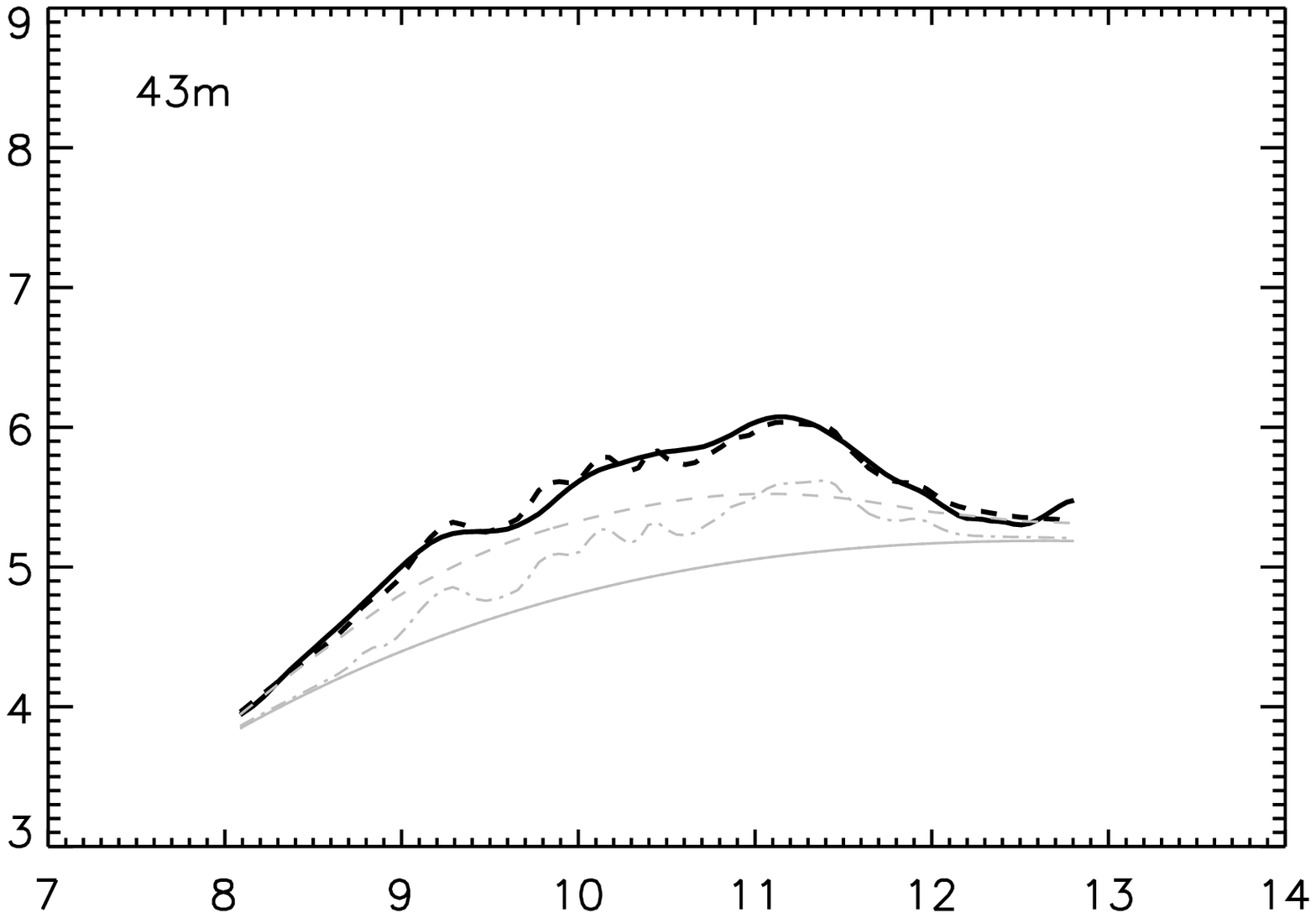}\\
\includegraphics[width=9.0cm,angle=0]{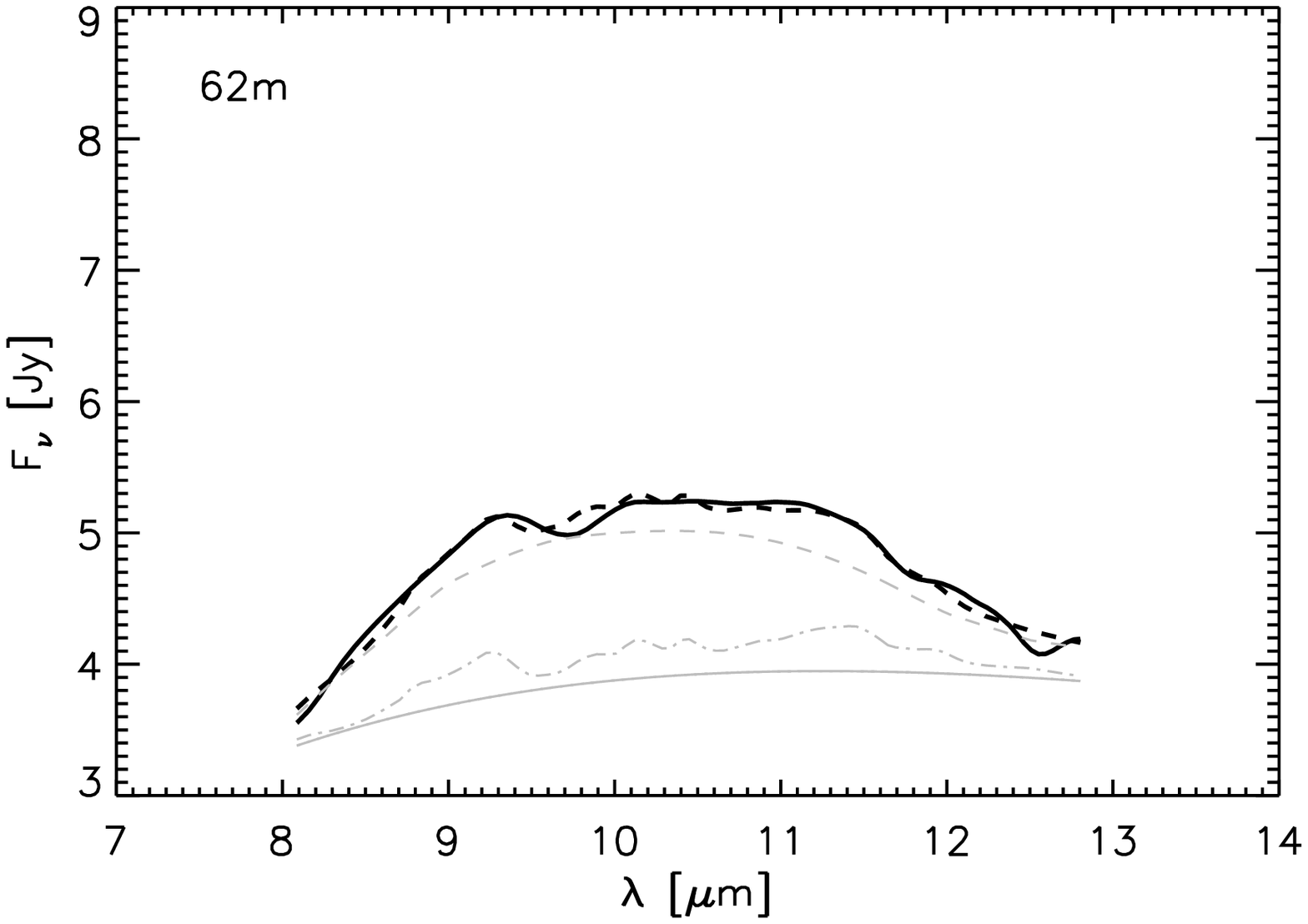}
\includegraphics[width=9.0cm,angle=0]{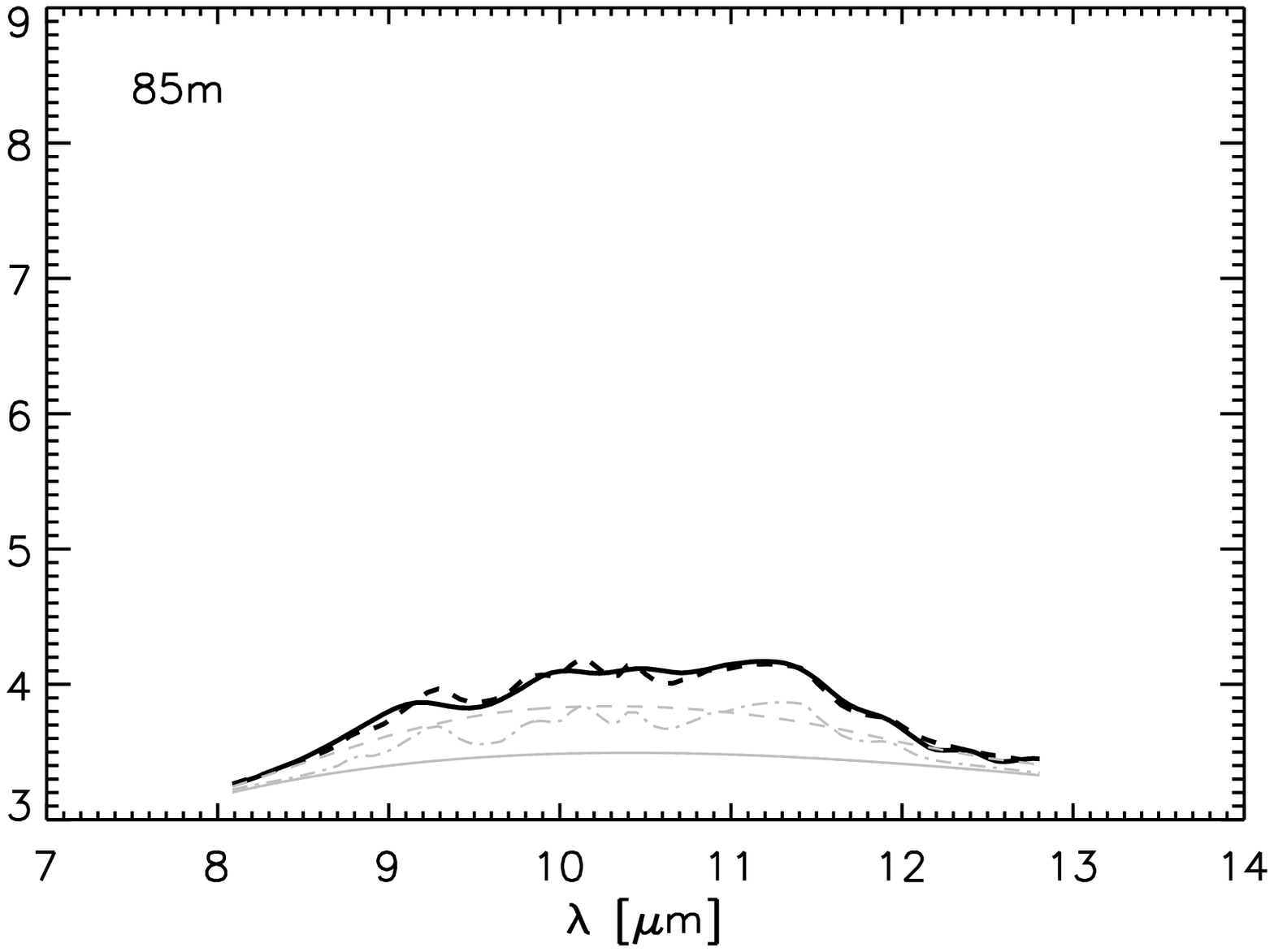}
\caption{Contributions of the dust species to the total flux of T~Tau~N (upper left panel) and the correlated fluxes. The three different baselines are indicated. In each panel the best fit (dashed black) is plotted together with the measured spectrum (solid black). The lines above the fitted continuum (solid grey) represent the various contributions: small amorphous (dotted grey), large amorphous (dashed grey), and crystalline grains (dashed-dotted grey).}  
\label{DustFig}
\end{figure*}

%

\subsection{K-band visibilities\label{mc3d_n_nir}}

The interferometric measurements in the mid-infrared presented in this paper constrain mainly the structure of the warm inner parts of a disc, whereas for the study of the hot innermost regions near-infrared visibilities are required. T~Tau~N was measured with the ``Palomar Testbed Interferometer'' (PTI) at 2.2\,$\mu$m both with the north-south (103-108\,m) and the east-south (83-86\,m) baseline \citep{akeson02}. After correcting for the incoherent contribution of T~Tau~S \citep{akeson00} the squared visibilities are roughly between 0.6 and 0.7 for the short and 0.4 and 0.5 for the long baseline. For our best-fit model (Section~\ref{mc3d_n_mir}) we find at 2.2\,$\mu$m squared visiblities of $0.59\pm0.01$ for a 85\,m baseline and $0.45\pm0.01$ for a 105\,m baseline. The errors given here reflect the uncertainty due to the unknown position angle of the disc. These predictions of our model are in good agreement with the PTI measurements, although no special attempt has been made to fit the near-infrared results.

%

\subsection{Dust Distribution and Evolution}

For the observations on the VLTI, T Tau N is well separated from T~Tau~S in the 10 $\mu$m range, so we can study its uncontaminated mid-infrared spectrum. The modest extinction (Table~\ref{MC3D_N}) and the fact that the silicate band appears in emission allows us to interpret the profiles by a mixture of amorphous and crystalline particles of various sizes. By applying such studies to the total and correlated spectra of selected Herbig~Ae/Be stars, evidence for grain growth and crystallisation particularly in the inner parts of the circumstellar discs was shown \citep{vanBoekel04}. Similar processes in the circumstellar discs of T Tauri stars were expected and recently found for RY~Tau \citep{schegerer08} and even for the less luminous source TW~Hya \citep{ratzka07}. 

Here, we want to analyse the total and correlated spectra of T~Tau~N (Figure~\ref{DustFig}) to determine the silicate composition of its circumstellar disc down to radii of a few AU. For this we use a $\chi^2$-fitting method \citep{bouwman01}. It assumes that the silicate emission feature has its origin in the optically thin surface layer of the circumstellar disc, where it results from a linear combination of mass absorption coefficients $\kappa_{ij}$ of different dust species $i$ of different sizes $j$:
\begin{eqnarray}
F(\nu)=B(\nu,T) \left(C_{0} + \sum_{ij} C_{ij}\ \kappa_{ij}(\nu) \right), 
\label{eq:fit}
\end{eqnarray}
where $C_{0}$ and $C_{ij}$ are fitting parameters, which reflect the underlying continuum and the mass contribution of each component. The quantity $F(\nu)$ is the spectral flux at frequency $\nu$, $\kappa_{ij}(\nu)$ represents the frequency-dependent mass absorption coefficient for a specific component, and $B_{\nu}(T)$ is the Planck function corresponding to a blackbody temperature $T$. 

Observationally, the signature of pristine, interstellar dust particles is a broad bell-shaped emission feature with a maximum near 9.7 $\mu$m. Larger particles are characterised by a flat-topped emission feature, while crystalline particles show narrower emission, most notably one at 11.3\,$\mu$m due to forsterite. As basic dust set for our $\chi^2$-fitting routine we use the same silicate species already used by \cite{schegerer06}: small ($\mathrm{0.1 \ \mu m}$) and large ($\mathrm{1.5 \ \mu m}$) amorphous grains with olivine and pyroxene stoichiometry \citep{dorschner95}, as well as crystalline species such as forsterite \citep{servoin73}, enstatite \citep{jaeger98}, and quartz \citep{henning97}. Carbon is not considered in the study presented here, because its emission profile is monotonic in the 10\,$\mu$m wavelength range \citep{draine84,wolf03b} and thus contributes to the underlying continuum only.

In the case of T~Tau~N, we find from fitting the emission profile of the total flux, corresponding to an emitting region of about 1-20 AU, that small and large amorphous grains dominate with an abundance of about 90\% (Table~\ref{Dust}). On the other hand, the correlated flux spectra clearly show the dominant role of large amorphous and crystalline species (see Figure \ref{DustFig}). These interferometric measurements allow insights into the dust properties at radii close to the central source. The nominal resolutions $\lambda/B$ of the three baselines are 7\,AU, 5\,AU, and 4\,AU at 10\,$\mu$m and the distance of T~Tau. Here, the abundance of small amorphous grains is insignificant. 

However, the trend for an increasing contribution from crystalline species with spatial resolution is broken in the case of the 62\,m baseline. Here, one has to keep in mind that the interferometer resolves the disc only in the direction of the projected baseline, while perpendicular to it, light is also collected from outer disc regions. The determination of the absolute positions and abundances of the dust species thus requires a more detailed study. Nevertheless, the finding that large amorphous and crystalline grains are concentrated towards the central source appears to be robust.

\begin{table}
\caption{Results of the spectral decomposition of the MIDI spectra. Given are the mass fractions of the various components (rows~1, 2, and 3) and the temperature of the continuum (row~4) for the total spectrum and the correlated spectra. The first two rows give the values for the small and large amorphous dust gains. The third row lists the contribution from both the small and large crystalline grains, i.e., enstatite, forsterite, and quartz.}             
\label{Dust}      
\centering                        
\begin{tabular}{lcccc}        
\hline                 
\hline                 
\noalign{\smallskip}
Species & total & 43\,m & 62\,m & 85\,m\\    
\noalign{\smallskip}
\hline
\noalign{\smallskip}
0.1\,$\mu$m & $0.24\pm0.14$ & $0.00\pm0.00$ & $0.00\pm0.02$ & $0.00\pm0.00$\\
1.5\,$\mu$m & $0.64\pm0.10$ & $0.63\pm0.05$ & $0.80\pm0.03$ & $0.59\pm0.24$\\
cryst .     & $0.12\pm0.03$ & $0.37\pm0.07$ & $0.20\pm0.02$ & $0.41\pm0.06$\\
\noalign{\smallskip}
Temp.	& $355\pm5$\,K & $403\pm14$\,K & $453\pm4$\,K & $490\pm25$\,K\\
\noalign{\smallskip}
\hline
\end{tabular}
\end{table}

%

\section{The close binary T~Tau~S\label{ttaus}} 

The observations of T Tau S are complicated by the fact that its two components are not resolved in the N-band by a single VLT telescope, and are therefore superposed both in the photometric and in the interferometric measurements. In the latter the effect of binarity is visible as sinusoidal modulations in the visibilities (Figure~\ref{Fig2}). In the following, inspired by the procedure routinely used in speckle interferometry, we try a decomposition in order to be able to study the two components individually. 

This requires several steps. First of all, we determine from the interferometric measurements the binary parameters, i.e., separation, position angle, and the ratio of the correlated fluxes (Section~\ref{fit}). Then the brighter component is identified by the measured phases. After estimating the visibilities of the fainter component by means of a reasonable radiative transfer model of its SED, we use these visibilities to calculate separate N-band spectra as well as visibilities for T~Tau~Sa (Section~\ref{dis}). Finally, a radiative transfer model for T~Tau~Sa is presented (Section~\ref{ttausa2}).

%

\subsection{Binary Parameters of T~Tau~S\label{fit}}

\begin{figure}[h!]
\centering
\includegraphics[height=8.5cm,angle=90]{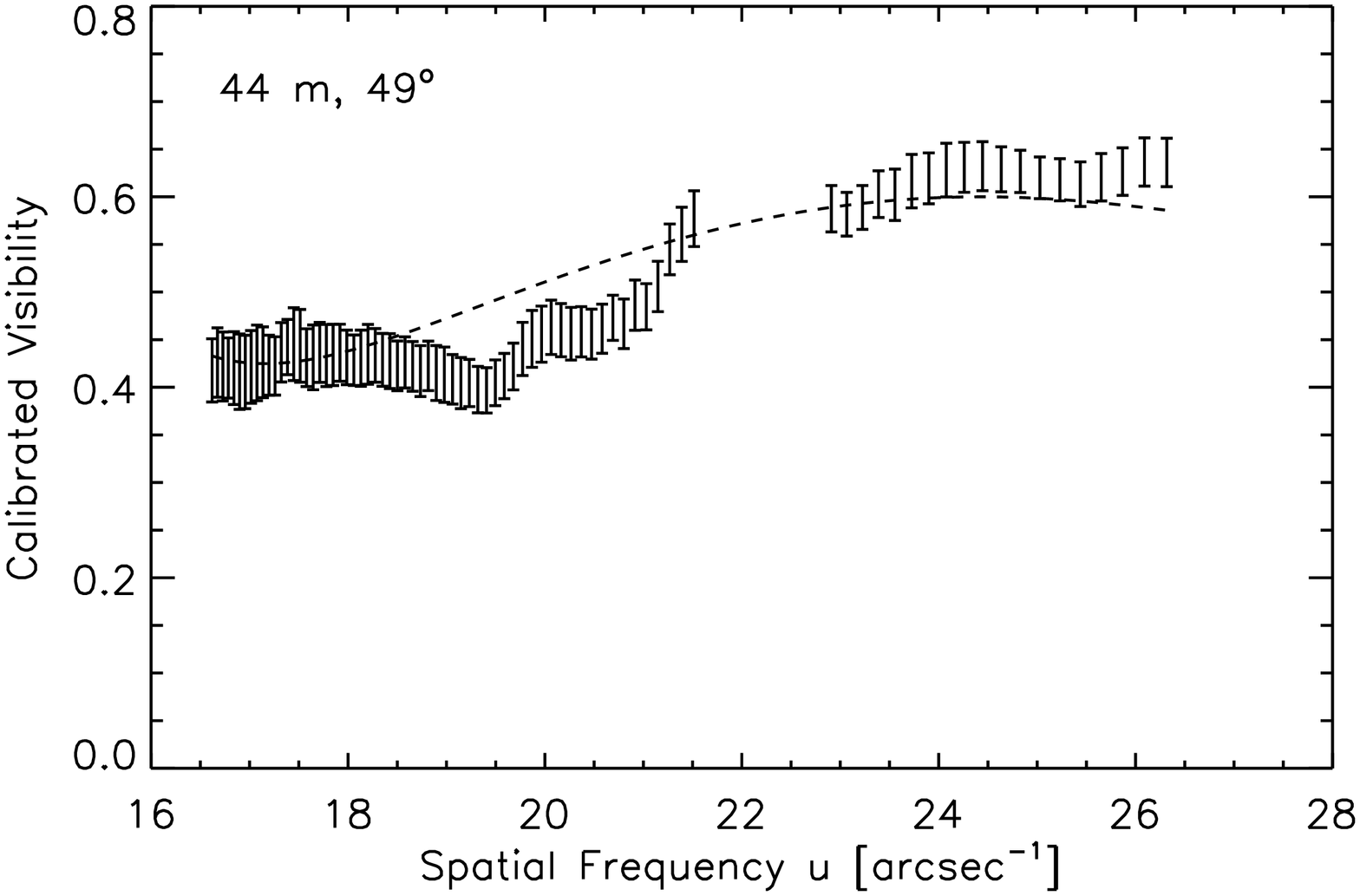}
\includegraphics[height=8.5cm,angle=90]{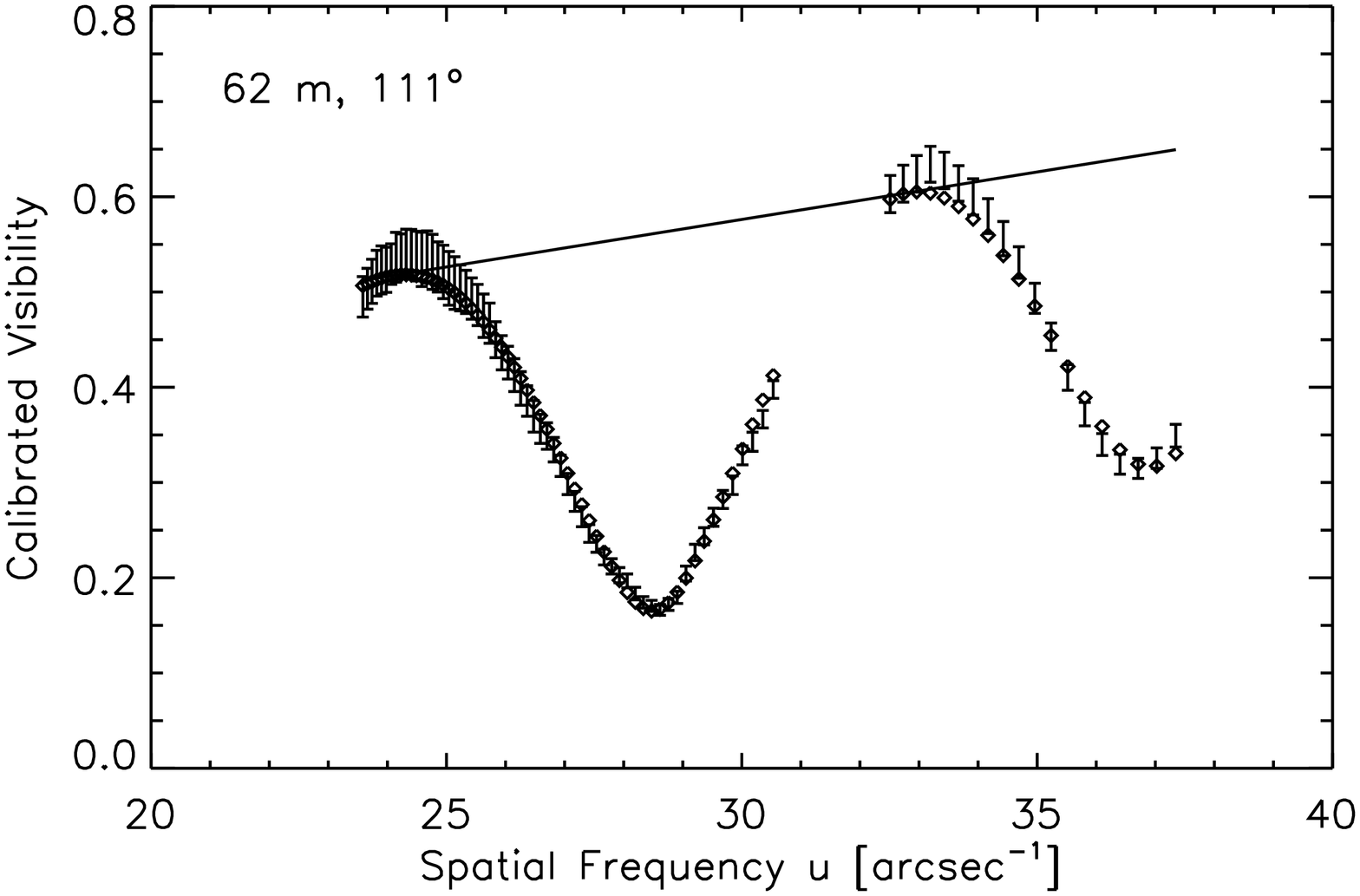}
\includegraphics[height=8.5cm,angle=90]{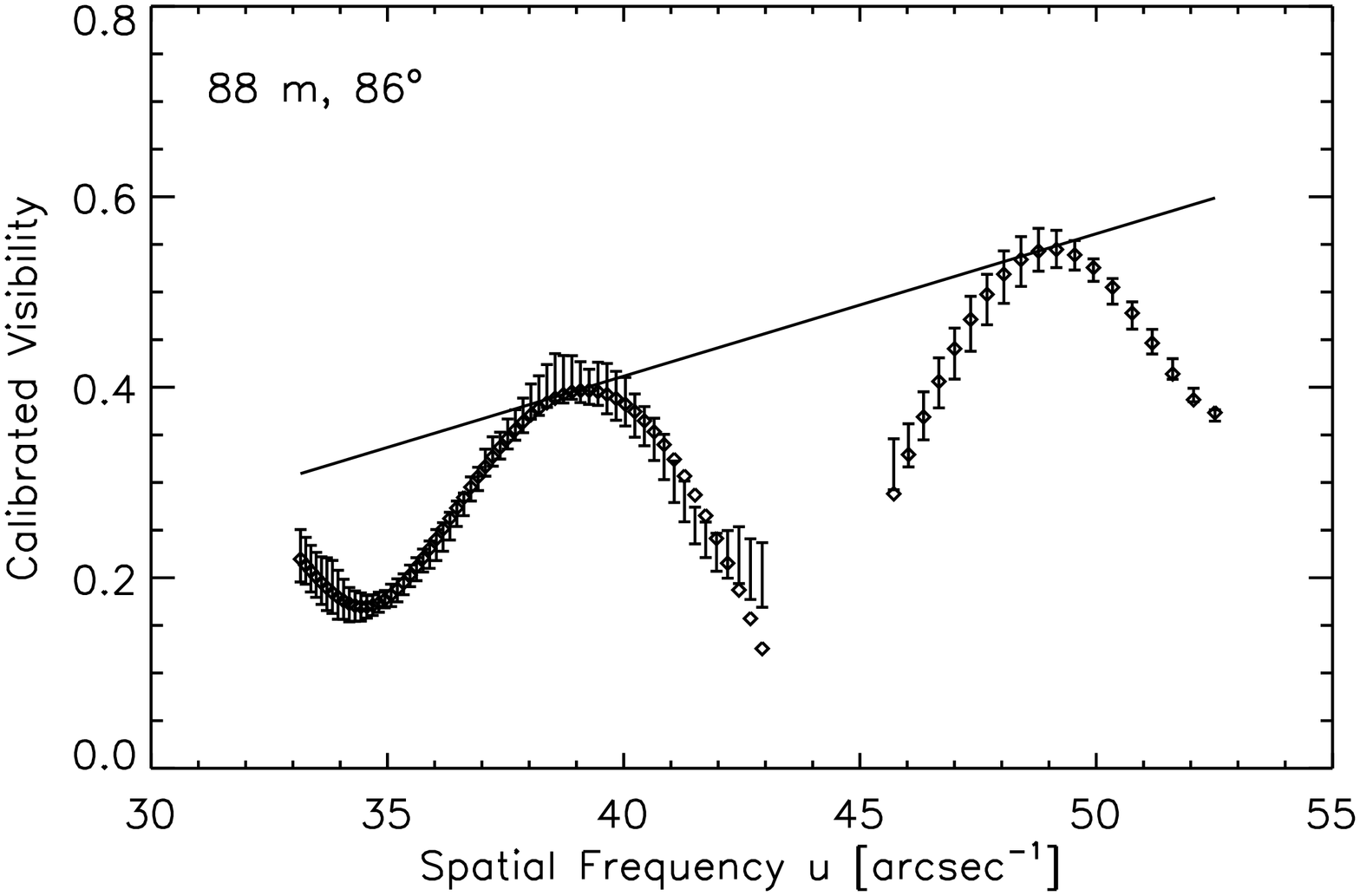}
\caption{The calibrated visibilities vs.\ spatial frequency are given by the error bars. Overplotted are the best fits (diamonds) for the measurements with the two long baselines. The lines are the linear approximations of the extended structures surrounding the components of the southern binary T~Tau~S. A fit for the measurement with the short baseline has not been attempted due to the limited modulation of the signal. The dashed line in the first panel shows the visibility as it is expected from the fitted parameters. A factor 0.6 has been applied.\label{Fig3}}
\end{figure}

All visibilities of T Tau S show that this source is well resolved by the interferometer (Figure~\ref{Fig2}). In addition, a clear sinusoidal binary signal is apparent for the second and third night when the position angle of the projected baseline was fairly close to the position angle of the separation vector of the two components. In Figure~\ref{Fig3} the calibrated visibilities are plotted as functions of the spatial frequency $u=B/\lambda$ for ease of discussion (compare Figure~\ref{Fig2}). Here, $B$ is the length of the projected baseline.

In Appendix~\ref{AppA}, we derive a formula for the visibility of a binary composed of two extended components. Under the assumption that the difference of the Fourier phases is negligible one finds:
\begin{equation}
V(u) = {\sqrt{\left.F_{p}^{\rm{cor}}(u)\right.^2+2F_{p}^{\rm{cor}}(u)F_{c}^{\rm{cor}}(u)\cos\left(2\pi us\right)+ \left.F_{c}^{\rm{cor}}(u)\right.^2}\over F_{p}^{\rm{tot}}(u)+F_{c}^{\rm{tot}}(u)}\mathrm{.}\label{eq6}
\end{equation}
The suffixes $p$ and $c$ refer to the primary and the companion, and $s$ is the separation of the two components along the projected baseline. We use the definition of the flux ratio
\begin{equation}
f(u)={F_{c}^{\rm{cor}}(u)\over F_{p}^{\rm{cor}}(u)} < 1 \mathrm{\ }\label{eq7}
\end{equation}
and obtain
\begin{equation}
V_{\rm{fit}}(u)=V_0(u)\cdot\frac{\sqrt{1+f^2(u)+2f(u)\cos\left(2\pi u s\right)}}{1+f(u)}\mathrm{\ .}\label{eq5}
\end{equation}
Here, $V_0(u)$ represents the visibility as it would be produced in a scan perpendicular to the separation vector, i.e., without the binary modulation. The spatial frequency $u$ is here a function of the wavelength, while the projected baseline length is fixed. The total flux is thus also a function of $u$.  

Fitting the visibility instead of fitting the correlated fluxes has the advantage that we can use the simple linear relation
\begin{equation}
V_0(u) = a_0 + a_1 u\mathrm{\ } \label{eq8}
\end{equation}
as a reasonable approximation. This is supported by the observation of the first night, scanning approximately perpendicular to the separation vector. The derived visibility shows a smooth, almost linear trend with wavelength or spatial frequency, respectively. We expect similar smooth trends to underly on our other observations.

The correlated flux ratio of the two components is modelled by a second-order polynomial in $u$ as
\begin{equation}
f(u) = f_0 + f_1 u + f_2 u^2\mathrm{,}\ \ f(u) < 1\mathrm{\ }\label{eq9}
\end{equation}
to allow for the effects of silicate absorption, while the separation of the components is formally allowed to vary linearly with spatial frequency to take into account that in general the photocentre of an inclined star-disc system can vary with wavelength:
\begin{equation}
s = s(u) = s_0 + s_1 u\mathrm{\ .}\label{eq10}
\end{equation}

A non-linear least squares fit, based on the Levenberg-Marquardt algorithm, of the visibility function to the visibilities measured with the two longer projected baselines has been performed to derive two independent parameter sets (Figure~\ref{Fig3}).\footnote{Since the separation vector of the binary is almost perpendicular to the position angle of the shortest projected baseline, the binary modulation of the visibility measured with this baseline is not pronounced. A fit for the visibility therefore has not been attempted in this case.} We find rather small values for $s_1$ in both fits, which shows that the wavelength-dependence of the separation can be neglected.\footnote{Between 8\,$\mu$m and 13\,$\mu$m, the separation increases from 120\,mas to  125\,mas for the intermediate and from 102\,mas to 103\,mas for the longest baseline.} We take this as indication that our ansatz of Equation~(\ref{eq5}) is  meaningful.


%


The fit to a visibility described above yields $s(u)$, i.e., the separation of the binary projected along the position angle of the baseline projected onto the sky. This means that we can only derive one component of the two-dimensional separation vector from each visibility. Fortunately, we have two fitted measurements at different orientations of the baseline to determine the relative position of the components at the time of the observations.

\begin{figure}[h!]
\centering
\includegraphics[height=8.4cm,angle=0]{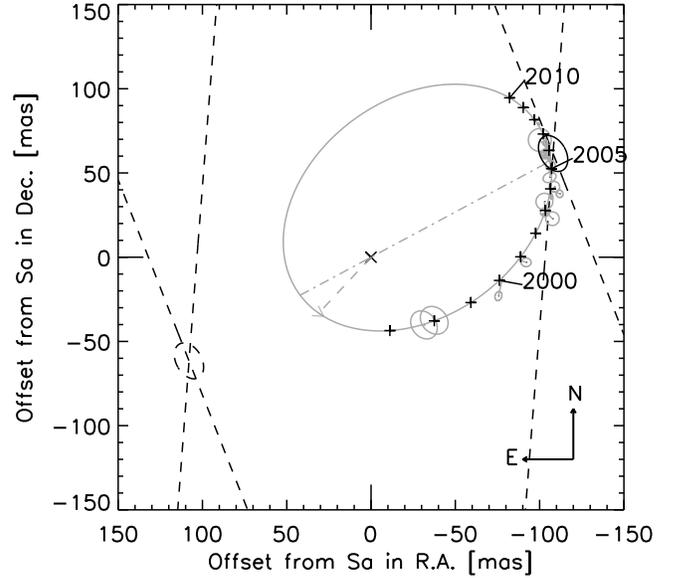}
\caption{In grey the near-infrared measurements of the positions of T~Tau~Sb with respect to T~Tau~Sa are plotted, see \cite{koehler08} and references therein. The ellipses represent the errors. The best-fit orbit derived by \cite{koehler08a} is also shown (dotted: periastron, dashed-dotted: line of nodes). The positions marked by crosses are separated by one year. Overplotted are the projected separations derived from our mid-infrared interferometric measurements and the determined relative position (black ellipse). The dashed, black ellipse is the second possible position due to the 180$^{\circ}$-ambiguity.}
\label{Fig4}
\end{figure}

The two derived projected separations are indicated by the dashed lines in Figure~\ref{Fig4}, where the primary T~Tau~Sa is located at the origin of the coordinate system. The possible positions of the companion (black ellipses) are determined by the intersections of the dashed lines. Two symmetrical solutions exist, because Equation~(\ref{eq5}) does not allow the determination of the sign of $s(u)$. Based on observations of the T~Tau~S binary in the near-infrared we chose the western position (solid ellipse). At a wavelength of 10.5$\,\mu$m we thus find
\begin{eqnarray}
d      & = & \left(124\pm8\right)\mathrm{mas}\nonumber\\
\Theta & = & \left(300\pm5\right)^{\circ}\label{eq13}
\end{eqnarray}
for the separation $d$ and the position angle $\Theta$.

Figure~\ref{Fig4} shows that this result fits reasonably well to earlier measurements and to the orbit of the T Tau S binary determined by \cite{koehler08a}.

%

\begin{figure}[h!]
\centering
\includegraphics[height=8.5cm,angle=90]{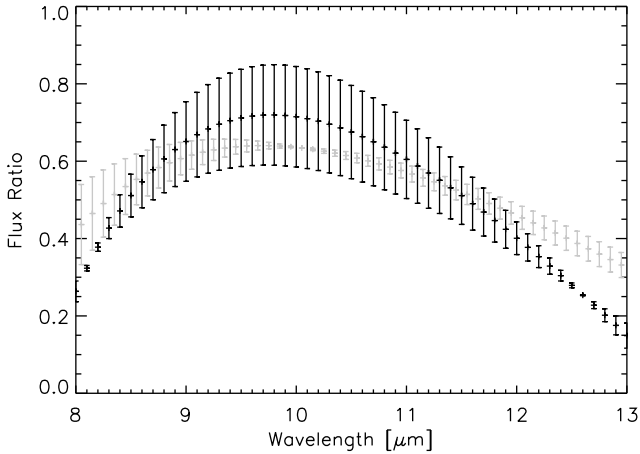}
\caption{The resulting flux ratio of the correlated fluxes when averaging the two polynomials derived from the fits to the two visibilities (black). Also shown is the flux ratio when averaging the two polynomials determined by the fits to the two correlated fluxes (grey). The latter are shifted by 0.05\,$\mu$m to longer wavelengths to allow a better comparison of the results.}
\label{Fig3e}
\end{figure}

Besides the information on the relative position of the companion, the sinusoidal modulation of the visibility as a function of spatial frequency $u$ (Figure~\ref{Fig3}) carries information on the flux ratio of the components. In detail, the observed visibility modulations for T~Tau~S deviate from a sinusoidal shape, and it is this deviation which carries information on the wavelength-dependence of the brightness ratio in the 8-13\,$\mu$m range. The fit procedure actually takes this into account and predicts the brightness ratio as a function of spatial frequency $f(u)$ to have a maximum (in the sense of being close to 1) near 9.5\,$\mu$m and minima of the order of 0.2 at short and long wavelengths (Figure~\ref{Fig3e}).

Although this separation of the fluxes looks elegant, it is an approximation only. What is actually determined and written in Equation~(\ref{eq5}) in form of the polynomial $f(u)$ is not the flux ratio, but rather the ratio of the correlated fluxes. Only with the additional assumption that the correlated fluxes of the two components have a similar dependence on the spatial frequency $u$, i.e., the relative light distributions are the same, this ratio also represents the ratio of the total fluxes. We use this assumption to determine below the separate N-band spectra of T~Tau~Sa and T~Tau~Sb.

To put this assumption on a firmer basis, we repeated the binary fitting for the {\it correlated fluxes} of T~Tau~S for which we describe the ``binary-free'' correlated flux by a second-order polynomial instead of the mere linear pre-factor $V_0(u)$ in Equation~(\ref{eq8}). The second-order polynomial is necessary to fit the silicate absorption feature. The relative position found by these fits is fully consistent with the position determined above from the visibility fits. The ratio of correlated fluxes, independent of the baseline length, is very close to that derived from the visibility fits (Figure~\ref{Fig3e}).

\begin{figure}[h!]
\centering
\includegraphics[height=8.5cm,angle=90]{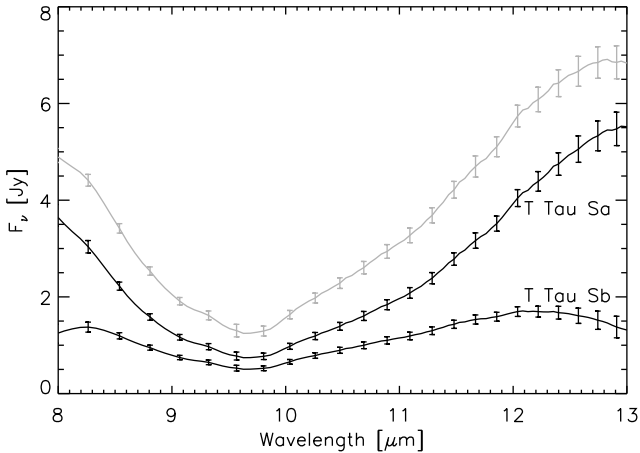}
\caption{The total spectrum of T~Tau~S (grey) as well as the individual fluxes of T~Tau~Sa and T~Tau~Sb under the assumption that the average of the flux ratios given in Figure~\ref{Fig3e} is the flux ratio of the total fluxes. Despite this approximation, the results are close to the final spectral decomposition (Figure~\ref{Fig13}).}
\label{Fig7}
\end{figure}

The average of the flux ratios determined from the fits of the visibilities and the fits of the correlated fluxes has been used to calculate the spectra shown in Figure~\ref{Fig7}. We consider these fluxes as already good estimates, which will be refined in Section~\ref{dis}.

%


Both components of T~Tau~S have the silicate band in absorption, but the optical depth, as usually defined by
\begin{equation}
F_{\nu} = F_{\nu}^{\rm continuum}\cdot e^{-\tau}\label{eq14}
\end{equation}
is quite different. One finds towards T~Tau~Sa an optical depth of $\tau=1.8$, while towards the M1 component T Tau Sb the value is only $\tau=0.9$. For the determination of the optical depth it is assumed here that the flux values at 8\,$\mu$m and 13\,$\mu$m represent the continuum, that this continuum can be linearly interpolated, and that the minimum of the silicate absorption is reached at 9.7\,$\mu$m (Figure~\ref{Fig7}).

Given the small separation between the two components (0.1$''$ or 18 AU), the difference in extinction is hard to explain by variations in the density of foreground material. We conclude that T Tau Sa probably has additional intrinsic extinction in addition to an extinction common to both components. Based on near-infrared observations and after a detailed analysis \cite{duchene05} already found evidence that T~Tau~Sa is surrounded by a small edge-on disc causing the intrinsic extinction. If T~Tau~Sb also possesses a circumstellar disc, it could not be aligned to the disc of T~Tau~Sa, because the extinction then would be much too high. \cite{skemer08} confirm this misalignment observationally based on photometric measurements in the N-band. Perhaps the circumstellar disc around T~Tau~Sb is coplanar with the orbit of this source around T~Tau~Sa, which is inclined by about 35$^{\circ}$ (Section~\ref{3d}).

\subsection{Disentangeling the Visibilities of the Two Southern Components\label{dis}}

We have the following situation: we measured the visibility $V_{\rm meas}$ (Figure~\ref{Fig2}) and the total flux $F^{\rm tot}_{\rm meas}$ (Figure~\ref{Fig1}) of the southern binary. As explained in Appendix~\ref{AppA} the normalised visibility perpendicular to the separation vector of the binary is the flux weighted average of the visibilities of the components, see Equation~(\ref{eqa2}). This visibility must be close to the slowly decreasing visibility measured for the shortest baselines with values between 0.5 and 0.6. By multiplying this ``binary-free'' visibility with the measured total flux one can derive the correlated flux $F^{\rm cor}_{\rm meas}$ and with the flux ratio $f<1$ (Figure~\ref{Fig3e}) the individual correlated fluxes $F^{\rm cor}_P$ and $F^{\rm cor}_C$. To derive the individual visibilities, the following system of equations has to be solved, wherein bold quantities are known\footnote{Mathematically we cannot exclude that one of the components is unresolved. But we expect that this is not the case and that the visibility of the faint component is smooth and monotonous, with no obvious maximum around 9.5 $\mu$m (see the smooth trend of the visibilities observed for T Tau N and of the visibility measured perpendicular to the line connecting the components of T Tau S).} :
\begin{eqnarray}
\begin{array}{ccccccc}
\vspace{0.1cm}
& & {\bf F}^{\rm cor}_{\rm meas}   & = & {\bf V}_{\rm meas}   & \cdot & {\bf F}^{\rm tot}_{\rm meas}\\
\vspace{0.1cm}
&& = &     &       &       &  =\\
\vspace{0.1cm}
{\bf F}^{\rm cor}_{\rm meas}\cdot 1/(1+{\bf f})     & = &{\bf F}^{\rm cor}_P   & = & V_P & \cdot & F^{\rm tot}_P\\
\vspace{0.1cm}
 &&+           &   &     &       &  +\\ 
{\bf F}^{\rm cor}_{\rm meas}\cdot{\bf f}/(1+{\bf f})& =& {\bf F}^{\rm cor}_C   & = & V_C & \cdot & F^{\rm tot}_C\\
\end{array}
\label{syseq}
\end{eqnarray}

To accomplish this, we will constrain $V_C$, but we first have to determine which of the components of T~Tau~S is the primary, i.e., is brighter in the N-band, and whether indeed T~Tau~Sa is the source with the stronger silicate absorption as has been speculated above. Intuitively, we would argue that the eastern, the more deeply embedded source has to be the brighter component in the mid-infrared, but so far we do not know. A phase measurement is needed.

Phase values $\ne$~0 indicate asymmetry of the object. A typical application of phase measurements in a binary is to resolve the right-left ambiguity of the position of the companion with respect to the primary remaining after analysis of the visibility modulus. With the centre of light at the origin, the phase of a binary shows a staircase appearance, rising in the direction towards the companion (if the $+x$ and $+u$ coordinates are going in the same direction). If a linear trend in the phase is removed, corresponding to shifting the primary to the origin, the binary phase will oscillate around zero, rising at the spatial frequency of the visibility maxima if a scan over the binary in the $+x$ direction would hit the companion first. Our data reduction package EWS gives this kind of phase function. We want to use the MIDI observations of T~Tau~S to decide whether it is the eastern component (the infrared companion) or the western component (the M1 star) that appears as primary with the larger correlated flux.

This will only work if the MIDI observing procedure defines an orientation on the sky. And indeed, during the OPD scan performed for interferometric measurements with MIDI, the position corresponding to the white light fringe moves on the sky along the projection of the baseline in a defined direction: closer to that telescope that feeds the right entrance window of MIDI from the viewpoint of the incoming light. This direction can be determined from the header information given with the data.

In Appendix~\ref{phases}, we derive that the phases, as determined with MIDI, indeed should be falling with wavelength (rising with spatial frequency) at the position of the visibility maxima when the companion is offset from the primary component in the direction of negative $x$ or OPD, which is the direction defined above.

In the case of the observations of T Tau S on October 30, 2004 and November 1, 2004, the phase was rising in the visibility maxima, which means that the companion came first when moving along the above mentioned reference directions towards a position angle of 48$^{\circ}$ and 85$^{\circ}$, respectively, in the sky. The opposite was true on November 3, 2004, when the direction on the sky was towards a position angle of 291$^{\circ}$. The conclusion from each of these observations is that at 10\,$\mu$m the eastern source, the deeper embedded companion T~Tau~Sa, dominates the correlated flux and has to be considered as the primary further on in this paper.

%

\begin{figure}
\centering
\includegraphics[width=9.0cm,angle=0]{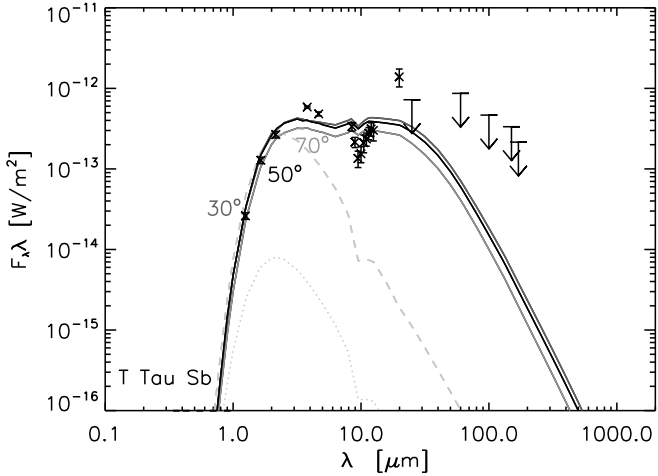}
\caption{Results of the fit to the SED of T~Tau~Sb. The model (see text) is plotted for inclinations of $30^{\circ}$, $50^{\circ}$, and $70^{\circ}$. The contributions from the stellar photosphere (dashed) and the accretion (dotted) take foreground extinction into account. Measured photometric data are indicated by black crosses. Arrows indicate upper limits. For references, see Table~\ref{table2}.}  
\label{Fig_sb1}
\end{figure}


The visibilities of the fainter component in the mid-infrared $V_C$ are now obtained from a ``auxiliary'' radiative transfer model fit to the measured SED of T~Tau~Sb.

Following \cite{duchene02} and \cite{duchene05}, we consider T~Tau~Sb as a T~Tauri star of spectral type M1 behind an extinction screen of A$_V$ = 15 $\pm$ 3 mag. Using the extinction law of \cite{rieke85} and taking the nearby M0.5 giant $\delta$ Oph (HD 146051) as a spectral template, we see an infrared excess in T~Tau~Sb increasing by nearly 3\,mag from H to L$'$. The near-infrared brightness is probably already affected somewhat by additional circumstellar emission. Taking the K brightness of T Tau Sb and extrapolating into the mid-infrared with the spectrum of HD 146051 gives an upper limit to the photospheric contribution of T~Tau~Sb. This upper limit falls from 190\,mJy at 8\,$\mu$m through 140\,mJy at 10\,$\mu$m to 85\,mJy at 13\,$\mu$m. These values are a factor of 2-10 less than the correlated flux attributed to T~Tau~Sb from our measurements. Obviously, there is strong circumstellar emission filling up this otherwise unexplained mid-IR flux, which we model to come from a circumstellar disc at moderate inclination.

\begin{figure}[b!]
\centering
\includegraphics[height=8.5cm,angle=90]{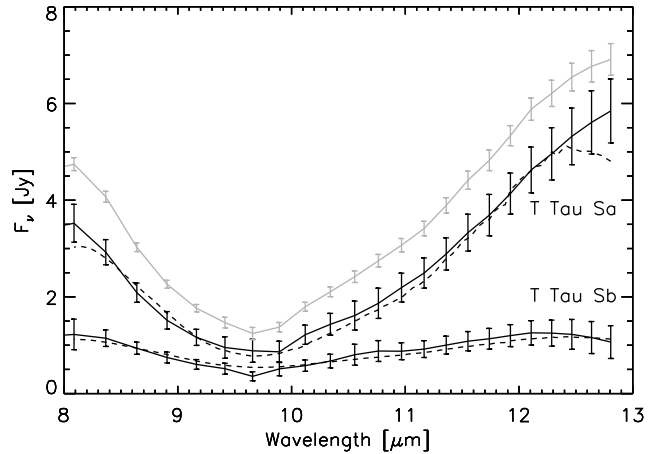}
\caption{The total spectrum of T~Tau~S (grey) as well as the individual fluxes of T~Tau~Sa and T~Tau~Sb as determined with the help of the modelled visibilities of T~Tau~Sb. Overplotted as dashed lines are fits for astronomical silicates with optical depths of $\tau_{\rm Sa}$=1.7 and $\tau_{\rm Sb}$=0.8.}
\label{Fig13}
\end{figure}

\begin{figure*}
\centering
\includegraphics[width=9.0cm,angle=0]{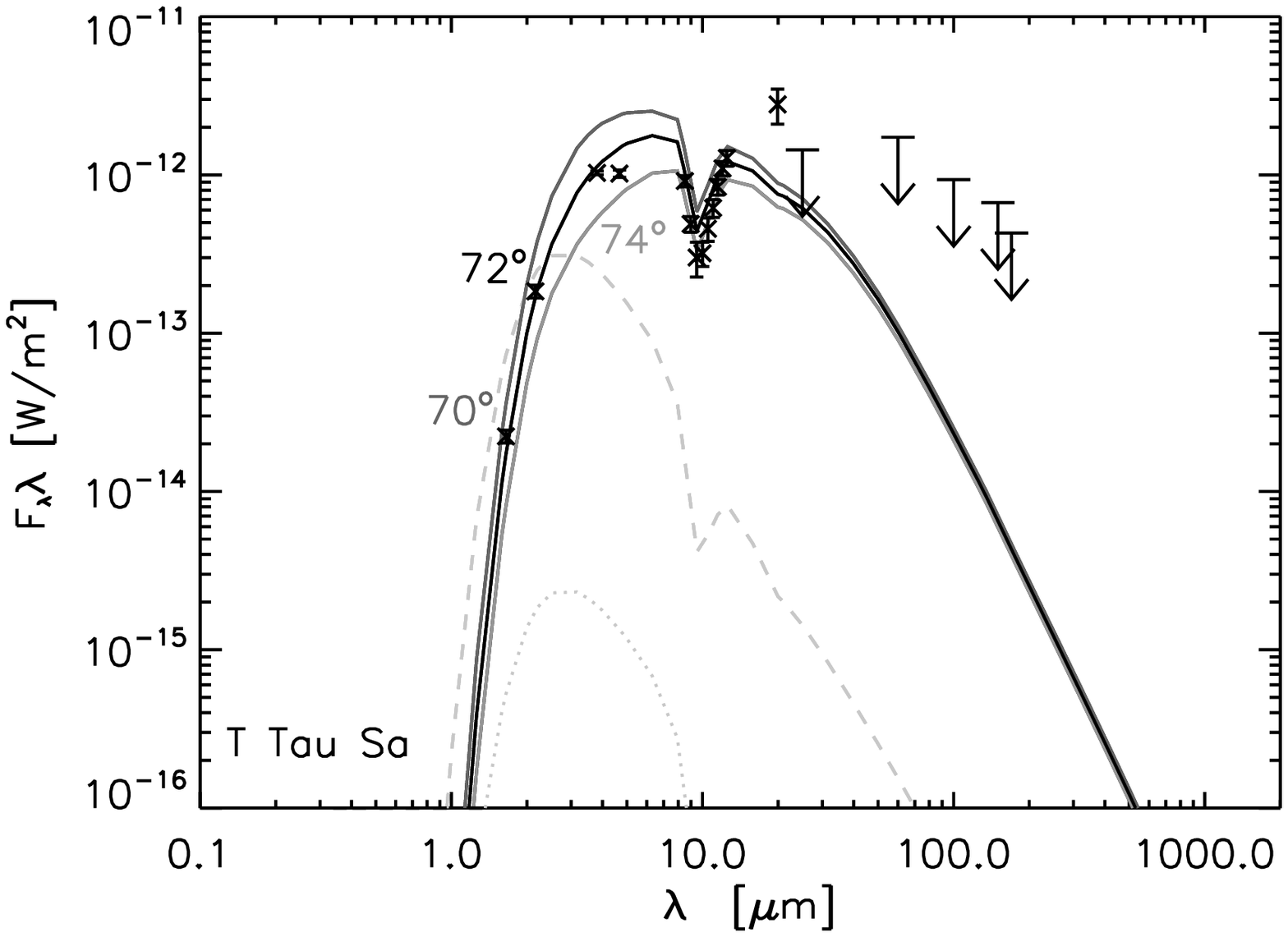}
\includegraphics[width=9.0cm,angle=0]{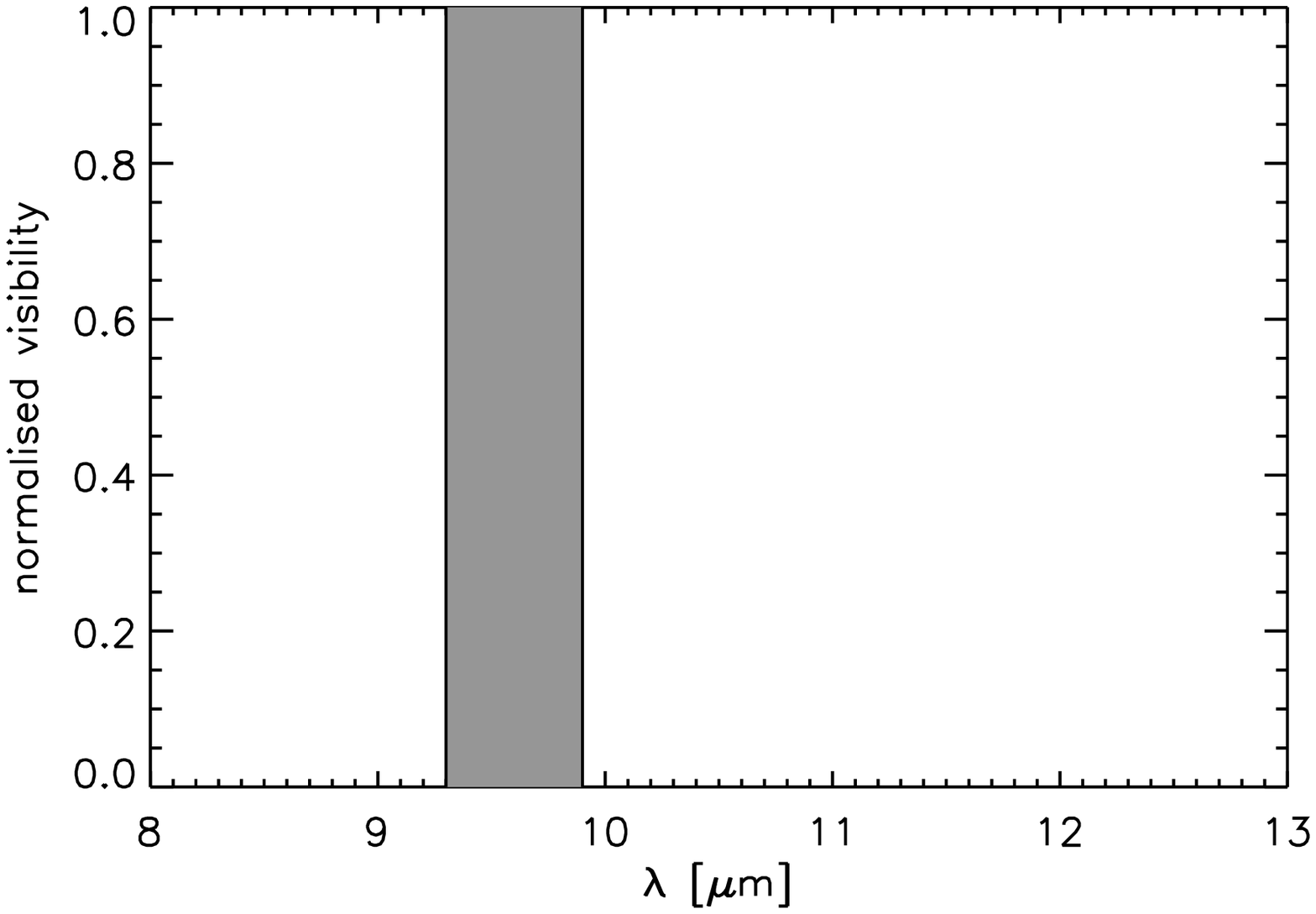}\\
\includegraphics[width=9.0cm,angle=0]{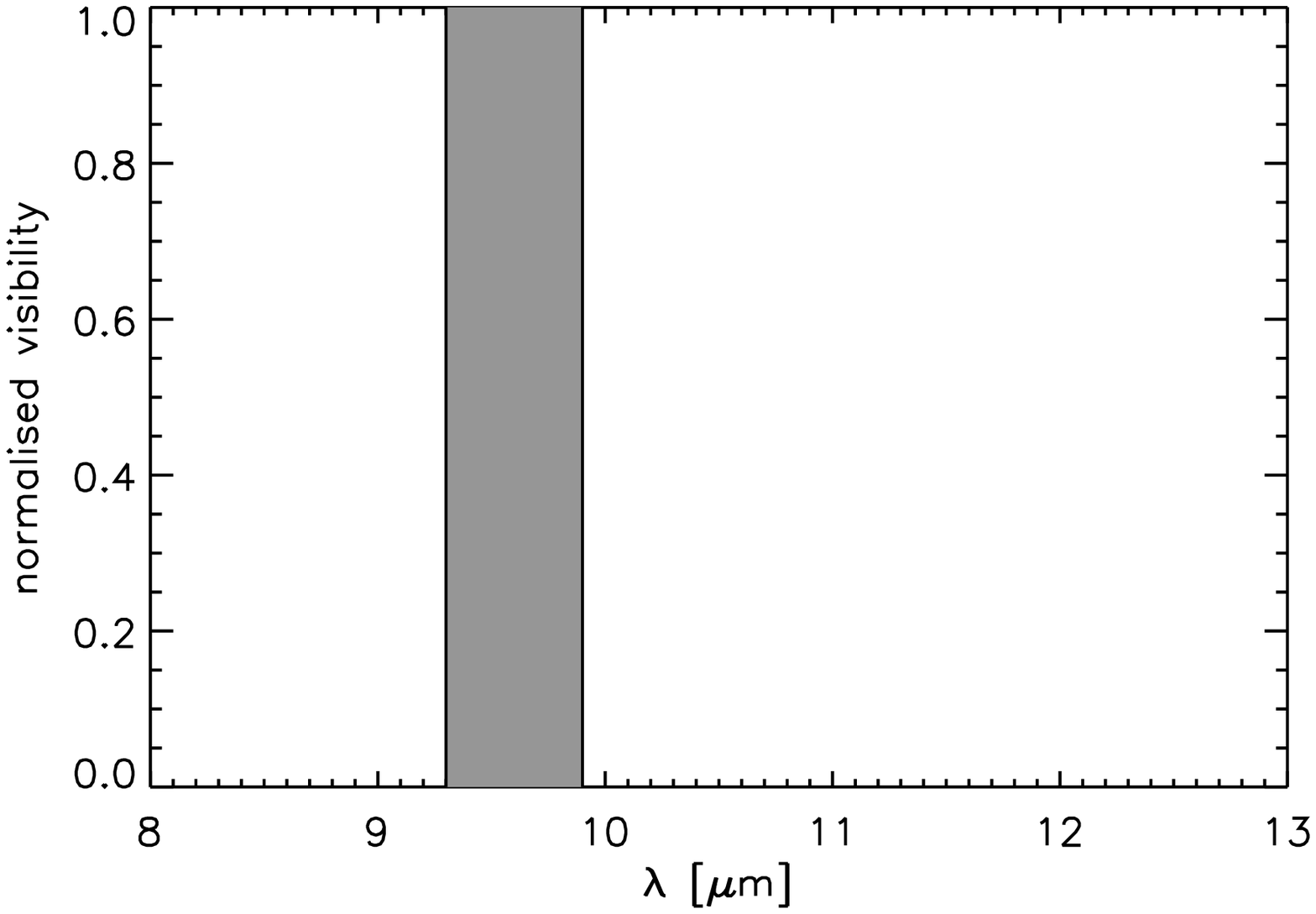}
\includegraphics[width=9.0cm,angle=0]{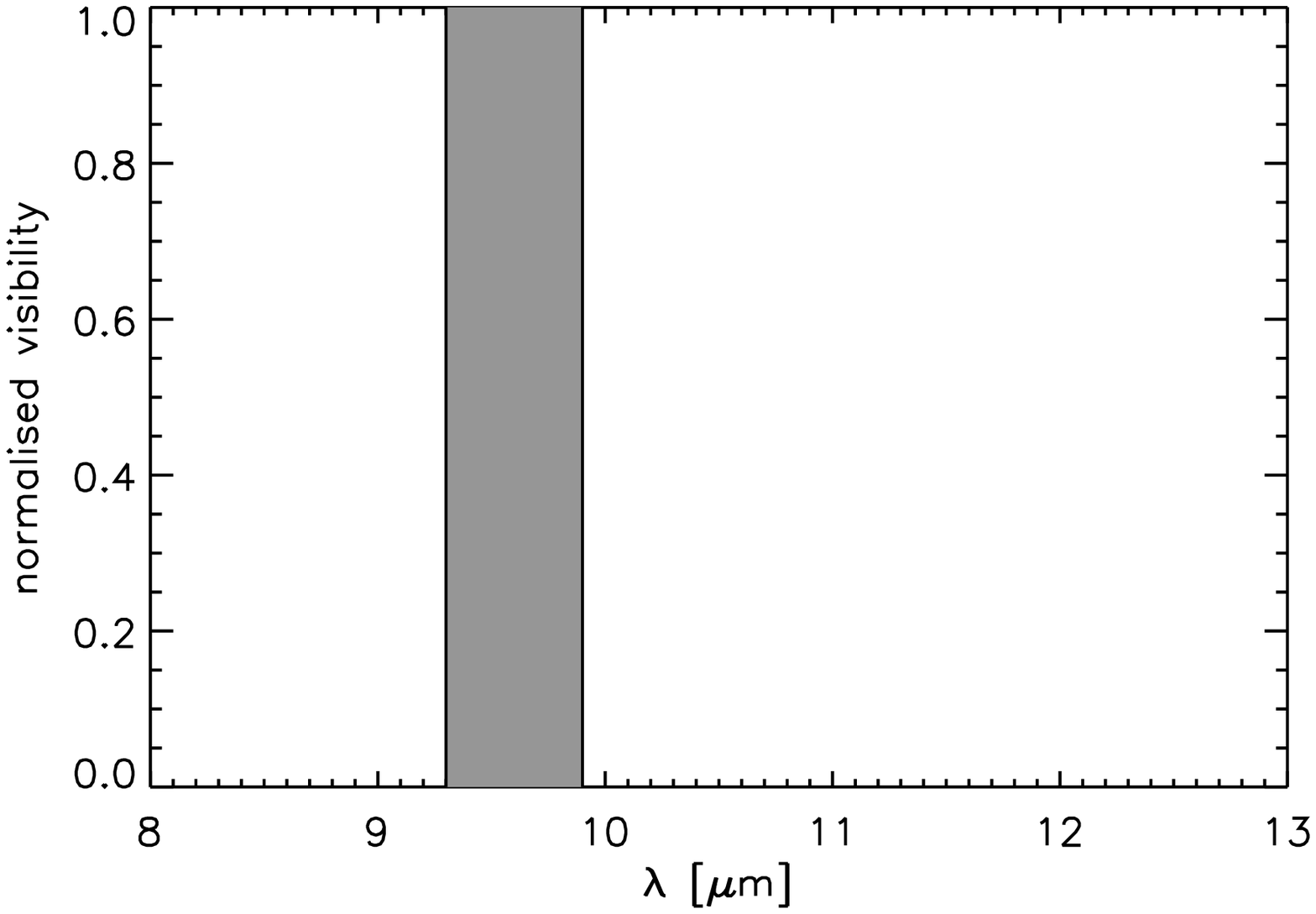}\\
\caption{Results of the simultaneous fit to the SED and the visibilities of T~Tau~Sa. {\it Top Left:} The model (Table~\ref{MC3D_S}) is plotted for inclinations of $70^{\circ}$, $72^{\circ}$, and $74^{\circ}$. The contributions from the stellar photosphere (dashed) and the accretion (dotted) take foreground extinction into account. Measured photometric data are indicated by black crosses. Arrows indicate upper limits. For references, see Table~\ref{table2}. {\it Top Right and Lower Row:} The black squares and triangles represent the upper and lower limits of the normalised visibilities when the orientation of the model for T~Tau~Sa with a disc inclination of $72^{\circ}$ is varied by $360^{\circ}$. The grey squares and triangles are the limits for the modelled visibilities of T~Tau~Sb. The grey dashed lines mark the range of values adopted to derive the visibilities of T~Tau~Sa, which are plotted as solid error bars. The dotted and shifted error bars represent the visibilities derived for T~Tau~Sa, when the visibilities of T~Tau~Sb are fixed to $1.0$, $0.6$, respectively, with an error of 5\%. The grey area between 9.4\,$\mu$m and 9.9\,$\mu$m marks the atmospheric ozone band that affects the measurements due to the low flux of the sources in the silicate band.}  
\label{Fig_sa1}
\end{figure*}

Due to the close periastron passage of the companion, a truncated disc with a size of not more than 5\,AU is expected. Placing such a disc with an inner radius of 0.1\,AU, an outer radius of 5.0\,AU, $\beta=1.00$, $h_{100}=7.0$\,AU, a moderate inclination angle, and a mass of $2.5\cdot10^{-4}$\,M$_{\odot}$ around a M1-type star \citep{duchene02}, whose photosphere is represented by a black body of 3720\,K \citep{lb82}, and applying a foreground extinction of 15\,mag in the visual \citep{duchene05} leads to a reasonable representation of the spectral energy distribution of T~Tau~Sb (Figure~\ref{Fig_sb1}). For this model we adopted a stellar luminosity of 1.8\,L$_{\odot}$ \citep{lb82, bessell88}, a stellar mass of 0.4\,M$_{\odot}$ \citep{koehler08a}, and an accretion rate of $1\cdot10^{-7}$\,M$_{\odot}\rm{yr}^{-1}$ \citep{duchene05}. An ``envelope'' with $c_1=5\cdot10^{-6}$ and $c_2=-1.0$ is also taken into account. It has to be mentioned that according to the model the large foreground extinction is responsible for the silicate absorption. The disc by itself would exhibit the silicate band in emission.

We refer to Section~\ref{mc3d_n_mir} for the explanation of the differences between modelled and measured fluxes longward of the N-band. But we want to return to the underestimation of the Q-band flux. It is not known whether the Q-band flux includes contributions from the northern component or the surrounding nebulae, or whether the flux ratio 0.5 between T~Tau~Sb and T~Tau~Sa as prediced by us based on the N-band fluxes is correct. Nonetheless, the main reason for the discrepancy is that the southern binary was much brighter in the mid-infrared at the time when the Q-band flux was measured. This can be seen, when comparing the integrated fluxes of T~Tau~S reported by \cite{herbst97} with our results presented in Table~\ref{table2}. Depending on the wavelength, the flux was lower by a factor of 3 to 8 in 2004.

Due to the uncertainty what fractions of the Q-band and ISO fluxes originate in the T~Tau~Sb system, the model is based on the near-infrared and N-band fluxes and is thus ambiguous. In the near-infrared the stellar photosphere represents the photometric measurements always well as long as the inclination of the disc is moderate. The N-band fluxes are sensitive to the accretion rate, the disc mass, and the inclination angle, but several combinations lead to the same quality of the modelled SED. The main restriction for a moderate inclination angle comes from the near-infrared. Unfortunately, the model does not well represent the silicate absorption feature, but a higher foreground extinction is excluded again by the near-infrared fluxes.

However, although the SED is not very well known and the model leaves ambiguities, the constraints on size and properties of the source suggest that the object visibilities cannot have values very different from the derived visibilities (Figure~\ref{Fig_sa1}). This was confirmed by comparing the results obtained for several model assumptions.

Finally, the upper and lower limits of the modelled visibilities for a disc inclination of $50^{\circ}$ when changing the position angle of the disc by 360$^{\circ}$ have been used to derive the ``object'' visibilities of T~Tau~Sb. This was done by calculating the average and the standard deviation of the limits and interpolating the results over the whole N-band (Figure~\ref{Fig_sa1}). The visibilities found for T~Ta~Sb are typical for a circumstellar disc.

%


After determining with the radiative transfer model limits for the visibilities of T~Tau~Sb, we can solve Equations~(\ref{syseq}) for the remaining quantities. To this end we divided the measured visibilities $V_{\rm meas}$ by the binary signal and multiplied the results with the measured spectrum of the southern binary $F^{\rm tot}_{\rm meas}$. The resulting, corrected correlated fluxes of the whole system $F^{\rm cor}_{\rm meas}$ can be assigned to the individual components, when taking into account that T~Tau~Sa is the brighter component as has been proven by the phase measurements and that the flux ratio $f$ of the correlated fluxes can be described by the relation plotted in Figure~\ref{Fig3e}. Afterwards, we determine the flux of the companion $F^{\rm tot}_C$ from its correlated fluxes $F^{\rm cor}_C$ with the help of the modelled visibilities $V_C$. This flux then fixes the flux $F^{\rm tot}_P$ and the visibilities $V_P$ of T~Tau~Sa. The final, best estimates of the fluxes for T~Tau~Sa and T~Tau~Sb are given in Figure~\ref{Fig13} and the visibilities for T~Tau~Sa are displayed in Figure~\ref{Fig_sa1}.

Since the determination of the visibilities of T~Tau~Sa is crucial, we checked how much the result is dependent on the model, i.e., the visibilities, of T~Tau~Sb. Even when T~Tau~Sb would be unresolved by the VLTI, i.e., $V_C=1.0$, the derived visibilities $V_P$ for T~Tau~Sa are almost indistinguishable from those that have been derived on the basis of the model for T~Tau~Sb. The same is true, when assuming that $V_C$ is 0.6 and thus close to the measured visibilities of the southern binary. A comparison is made in Figure~\ref{Fig_sa1}.

%

\subsection{A Radiative Transfer Model for T~Tau~Sa\label{ttausa2}}

\begin{table}[t]
\caption{The parameters of the radiative transfer model of T~Tau~Sa.}
\label{MC3D_S}
\centering
\begin{tabular}{llccrcc}
\hline\hline
\noalign{\smallskip}
               & Parameter & Unit & & Value & & Ref.\\
\noalign{\smallskip}
\hline
\noalign{\smallskip}
\multicolumn{2}{l}{{\bf Stellar Parameters}}\\
\noalign{\smallskip}
\hspace{0.3cm} & Mass ($M_{\star}$)  	& M$_{\odot}$	          & & 2.3	          & & $^a$ \\		          
	   & Spectral Type	  	&		          & & Ae	          & & $^b$ \\
               & Temperature ($T_{\star}$)	& K$_{\ \,}$	          & & 9000	          & & $^d$ \\
               & Luminosity ($L_{\star}$)	& L$_{\odot}$	          & & 40	          & & $^g$ \\
               & Radius ($R_{\star}$)	& R$_{\odot}$	          & & 2.6	          & & calc \\
\noalign{\smallskip}
\noalign{\smallskip}
\multicolumn{2}{l}{  {\bf Accretion}}\\
\noalign{\smallskip}
               & Accretion Rate	  	& M$_{\odot}\rm{yr}^{-1}$         & & $1\cdot10^{-8}$     & & $^d$ \\
               & Boundary Temp. ($T_{\rm bnd}$) & K$_{\ \,}$	          & & 8000	          & & est. \\
               & Boundary Rad. ($R_{\rm bnd}$)  & R$_{\star}$	          & & 5.0	          & & est. \\

\noalign{\smallskip}
\noalign{\smallskip}
\multicolumn{2}{l}{  {\bf Circumstellar Disc / Envelope}}\\
\noalign{\smallskip}
               & Inner Radius ($R_{\rm in}$) 	& AU		          & & 0.5	          & & $^g$ \\
               & Outer Radius ($R_{\rm out}$)	& AU		          & & 5.0	          & & $^g$ \\
               & $\beta$	  	&		          & & 1.00	          & & $^g$ \\
               & $h_{100}$ 	  	& AU		          & & 10.0	          & & $^g$ \\
               & $c_1$		  	&		          & & $1\cdot10^{-5}$     & & $^g$ \\
               & $c_2$		  	&		          & & -5.0	          & & $^g$ \\
               & Inclination	  	& deg		          & & 72	          & & $^g$ \\
               & Disc Mass ($M_{\rm disc}$)	& M$_{\odot}$	          & & $3\cdot10^{-3}$     & & $^g$ \\
\noalign{\smallskip}
\noalign{\smallskip}
\multicolumn{2}{l}{{\bf Interstellar Extinction}}\\
\noalign{\smallskip}
               & Foreground Extinction $A_V$	& mag		          & & 30	          & & $^g$ \\
\noalign{\smallskip}
\hline             
\noalign{\smallskip}
\multicolumn{7}{l}{\parbox{8.5cm}{$^a$ \cite{koehler08a}; $^b$ \cite{duchene06}; $^c$ \cite{duchene02}; $^d$ \cite{duchene05}; $^e$ \cite{lb82}; $^f$ \cite{bessell88}; $^g$ this work; calc.=calculated; est.=estimated}}\\
\noalign{\smallskip}
\hline
\end{tabular}
\end{table}

Following the approach of \cite{duchene05}, we assumed that T~Tau~Sa has an effective temperature of 9000\,K and a luminosity of 40\,$L_{\odot}$. The photosphere is represented in our model by a black body. At 2.3\,$M_{\odot}$, T~Tau~Sa is the most massive star in the triple system \citep{koehler08a}. 

Similar to our approach for T~Tau~Sb, we modelled T~Tau~Sa as a source that is surrounded by a small truncated disc with a size of 5\,AU (Table~\ref{MC3D_S}). This circumstellar structure contains a mass of 0.003\,$M_{\odot}$ of gas and dust and thus very efficiently absorbs the photons from the central star due to the high inclination of the disc of around $70^{\circ}$. In addition, an envelope with a steep density gradient seems to be present. On the other hand, the contribution of accretion is negligible. Is has to be mentioned that without the foreground extintintion of 30\,mag in the visual the silicate feature would again appear in emission like it has been found for T~Tau~Sb.

\begin{figure}[h!]
\centering
\includegraphics[width=8.5cm,angle=0]{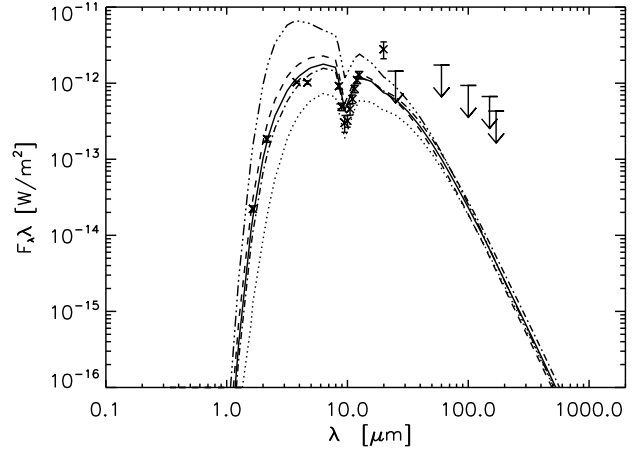}
\caption{The spectral energy distribution for the model of T Tau~Sa with a disc inclination of 72$^{\circ}$ when neglecting the accretion (solid), removing the ``envelope'' (dotted), changing the outer radius of the disc to 4\,AU (dashed), to 6\,AU (dashed-dotted), and using $\beta=1.25$ (dashed-dotted-dotted).}
\label{CompSa}
\end{figure}

In Section~\ref{fit} for the two southern sources a preliminary optical depth of $\tau_{\rm Sa}$=1.8 and $\tau_{\rm Sb}$=0.9 was found. When fitting the spectra plotted in Figure~\ref{Fig13} using the optical depths for astronomical silicates towards the Galactic centre \citep{kemper04}, one finds $\tau_{\rm Sa}$=1.7 and $\tau_{\rm Sb}$=0.8. Conversion of the silicate absorption depth into visual extinction, e.g., \cite{rieke85}, leads to $A_{\rm V}\approx30$\,mag towards T~Tau~Sa and $A_{\rm V}\approx15$\,mag towards T~Tau~Sb. These results are consistent with the values used in our models and are attributed to the foreground screen.

A possible explanation for the additional 15\,mag of visual extinction towards T~Tau~Sa might be a {\it circumbinary} disc currently only extincting this source. Such a scenario has been tested by reducing the foreground extinction towards T~Tau~Sa to 15\,mag, extending its circumstellar disc to 50\,AU, and cutting off therein the area between 5\,AU and 30\,AU. In this model the whole circumstellar structure contained 0.02\,$M_{\odot}$ of gas and dust. It sucessfully reproduced the SED and the visibilities.

The SED derived from our model of a compact circumstellar disc is shown together with the photometric data in Figure~\ref{Fig_sa1}. Also the upper and lower limits of the modelled visibilities of both T~Tau~Sa and T~Tau~Sb are plotted. The visibilities for T~Tau~Sb were used in Section~\ref{dis} to determine the individual spectra of the two components of the southern binary and to separate the measured visibilities of T~Tau~S. The derived visibilities for T~Tau~Sa are indicated in Figure~\ref{Fig_sa1} by the error bars and fall right within the range predicted by the radiative transfer model. The two-dimensional N-band visibilities are shown in Figure~\ref{Fig_sa2} and will be analysed later to determine the orientation of the circumstellar disc (Section~\ref{do}).

In Figure~\ref{CompSa} we show the spectral energy distribution when changing different parameters of the model while keeping the inclination fixed. This demonstrates that the model is to a certain level degenerate. For example adjusting the inclination of the models with a different outer radius leads to similar good agreement with the SED than our best-fit model. Neither the N-band, nor the K-band visibilities can help to overcome this. On the other hand, removing the envelope and especially changing its radial density distribution (not shown) leads to unacceptable SEDs that cannot reproduce the N-band fluxes well. Also changing $\beta$ to a value of $1.25$ is not reasonable.

\begin{figure}[t!]
\centering
\includegraphics[height=8.0cm,angle=0]{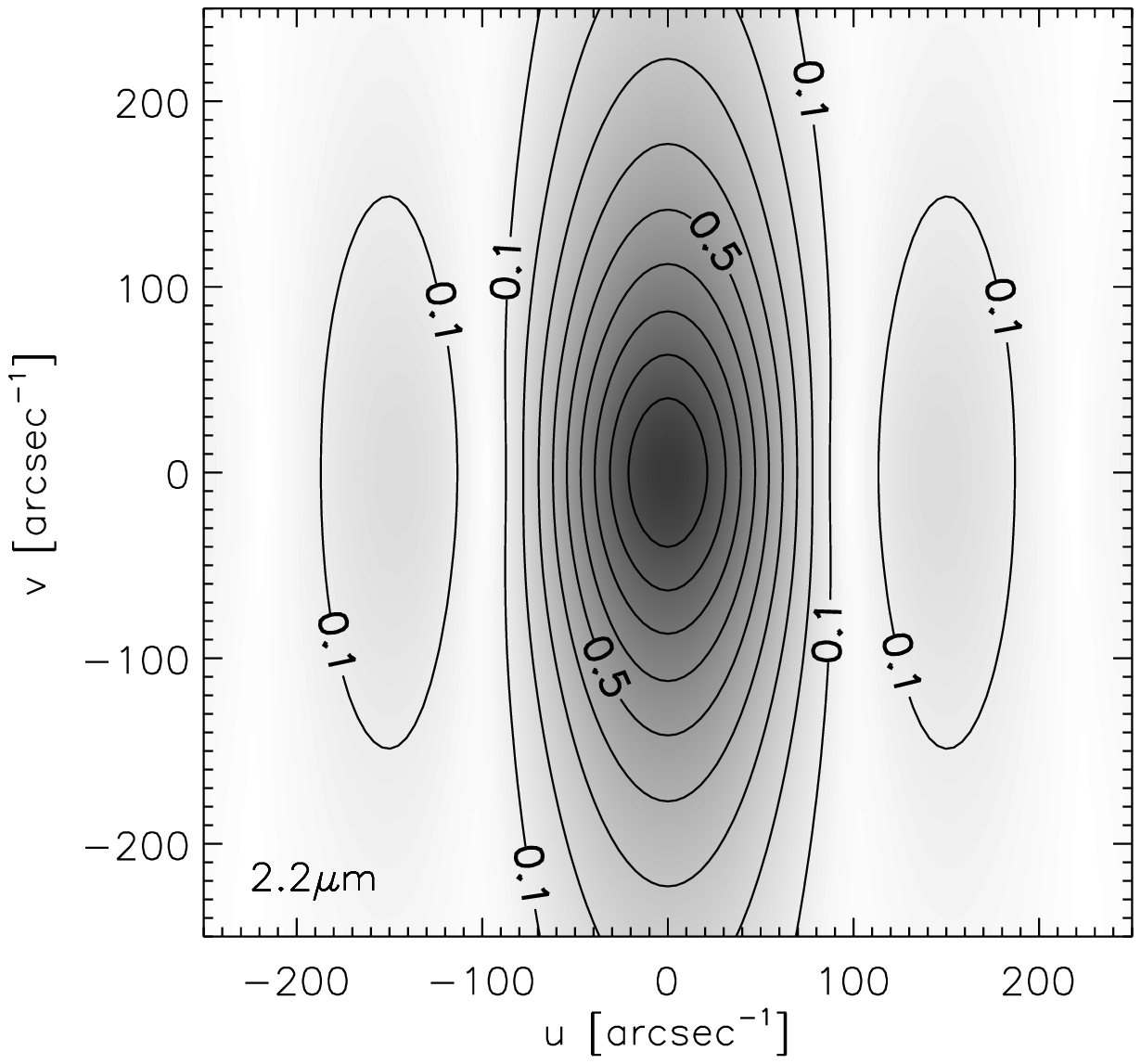}
\caption{The K-band visibility of T~Tau~Sa at 2.2\,$\mu$m as predicted by our radiative transfer model.}
\label{pred}
\end{figure}

Finally, we report on the predictions made by our model for T~Tau~Sa for the K-band visibilities at 2.2\,$\mu$m. We find for the 85\,m baseline used by \cite{akeson02} visibilities between 0.03 and 0.38, while the visibilities for the 105\,m baselines lie between 0.01 and 0.29. The exact values depend on the actual position angles of the baselines. In Figure~\ref{pred} the two-dimensional K-band visibility is presented.

%

\section{The T Tau system\label{3d}}

\begin{figure}[b!]
\centering
\includegraphics[height=8.4cm,angle=0]{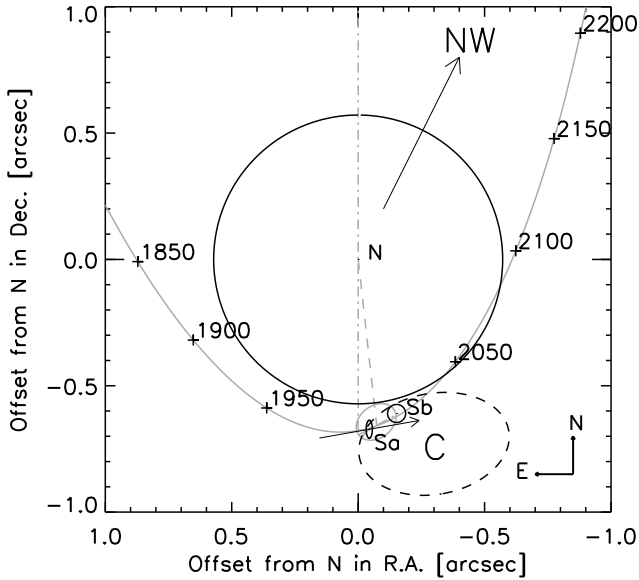}
\caption{A sketch of the T~Tau system combining results of our MIDI measurements and previous observations. The dashed ellipse labelled `C' indicates the emission structure found by \cite{herbst07}. The orbits are taken from \cite{koehler08a}. The dotted line re\-presents the periastron, the dashed-dotted line the line of nodes. See text for further details.}
\label{Fig14}
\end{figure}

%

\subsection{Three-Dimensional Geometry}

Figure~\ref{Fig14} portrays the T~Tau triple system schematically, but to scale. The graph is centred on T~Tau~N. 

The orbit of the centre-of-mass of T~Tau~S around T~Tau~N is drawn in grey and the positions of the centre-of-mass of the southern binary in intervals of 50\,yrs are indicated by crosses. At the time of the inter\-ferometric observations with MIDI, the orbit of the two southern components T~Tau~Sa and T~Tau~Sb around each other is overlaid (see Figure~\ref{Fig4}). The positions of the two southern components are also shown for the time of our measurements. Both orbits are taken from \cite{koehler08a}. While the southern orbit has a period of $28\,[+12,-4]$\,yrs, a semi-major axis of $13\,[+6,-2]$\,AU, an eccentrity of $0.47 \,[+0.16,-0.18]$, and an inclination of $34\fdg8\,[+12\fdg0,-31\fdg0]$, the wide orbit has a period of $17500\,[+7500,-16000]$\,yrs, a semi-major axis of $1200\,[+1200,-600]$\,AU, an eccentricity of $0.92\,[+0.07,-0.08]$, and an inclination of $56\fdg1\,[+3\fdg0,-4\fdg0]$. The wide orbit is much less constrained due to the fact that only a small part of it is covered by the near-infrared observations.

The circumstellar discs around the three components of the T~Tau system appear in black. The sizes and the inclinations of the discs are those that we derived from our simultaneous fits to the spectral energy distributions and the visibilities. The size of the disc around T~Tau~N thus might be similar to the periastron distance of the wide orbit. This indicates that the southern binary has to pass its periastron in the background or in the foreground with respect to the plane of the almost face-on disc around T~Tau~N. Otherwise dynamics would lead to a much smaller truncated disc. Surprisingly, the orbital elements of the wide orbit show that the southern binary is close to the line of nodes. The three stars thus nowadays reside at similar distances from the sun and the distance of the southern binary from the face-on disc has to be indeed quite small. Here it has to be mentioned again that neither the wide orbit, nor the size of the disc around T~Tau~N is well constrained yet. So we have to leave it open, whether refined orbital elements, a smaller outer disc radius or even a larger disc with a gap will help to overcome this inconsistency.

Our simulations show that the two discs of the southern binary are not aligned. While the inclination of the truncated disc around T~Tau~Sb is consistent with the inclination of the orbit of this source around T~Tau~Sa \citep{koehler08a}, the disc around T~Tau~Sa is highly inclined. The inclination of the disc around T~Tau~Sa is again in good agreement with the inclination of the wide orbit. Finally, the face-on disc around T~Tau~N exhibits a third orientation in the T~Tau triple system.

%

\subsection{Formation of the system}

When fragmentation of a rotating cloud is responsible for the formation of a binary, the circumstellar discs are preferentially co-aligned with respect to each other. Also the orientation of the plane of the orbit should be similar to that of the discs. Therefore, one would intuitively argue that the misaligned discs in the T~Tau triple system favour an independent formation of T~Tau~N and T~Tau~S. Consequently, even the misaligned two southern components might have formed independently. However, only in the centre of very dense clusters the required subsequent capturing process is efficient enough \citep{hall96, bonnell01}. Formation by capture is thus very unlikely for T~Tau.

In order to identify a viable formation scenario we have to determine the most fundamental parameters of the system. One finds that about half of the mass of the system is concentrated in the northern, the other half in the southern component. In addition, these two mass concentrations orbit around their centre-of-mass in a highly eccentric orbit. 

A scenario that leads to such a configuration is that of a filamentary collapse \citep{zinnecker89, zinnecker91, bonnell92a}. A gravitationally unstable, elongated fragment tumbling ``end over end'' will subsequently break up into two or even more fragments. Initially, these fragments are heading towards each other due to their gravitational attraction. But for a non-zero impact parameter the angular momentum will prevent the fragments from merging. Already formed discs will evolve and either directly or tidally interact with each other.

The details of the process and the final configuration of the system are very sensitive to the initial conditions \citep{bonnell92a}. For example, a wide range of binary mass ratios can be found in simulations of filamentary fragmentation. But the mass ratio tends to equalise during the evolution of the system, because the accretion onto the secondary is very efficient \citep{bonnell94}. The secondary is able to collect even the high specific angular momentum material, as it resides in a circumstellar structure around the primary, away from the centre-of-mass. Furthermore, the eccentric orbit of the binary is circularised with time, because the specific angular momentum of the material accreted by the companion varies with the distance from the primary.

Another scenario to explain the actual properties of the T~Tau system might be a three-body interaction within a flattened parential envelope \citep{reipurth00}. In a close triple approach a non-hierarchical triple is tranformed into a tight binary and a third component on a long-period or even hyperbolic orbit. Such interactions can also form giant bow-shocks and a circumbinary disc is finally surrounding the binary. Also the IRC phenomenon can be explained by this scenario, when assuming that only the ejected companion has left the large flattened envelope while the binary is still hidden therein.

Whether the T~Tau triple system was really formed by filamentary fragmentation or a three-body interaction is unclear. A detailed investigation is far beyond the scope of this paper. However, only a highly dynamical process can be considered as a reasonable formation scenario \citep{pringle89, whitworth01}.

%

\subsection{Origin of the jets\label{do}}

\begin{figure*}[t!]
\centering
\includegraphics[width=5.5cm,angle=0]{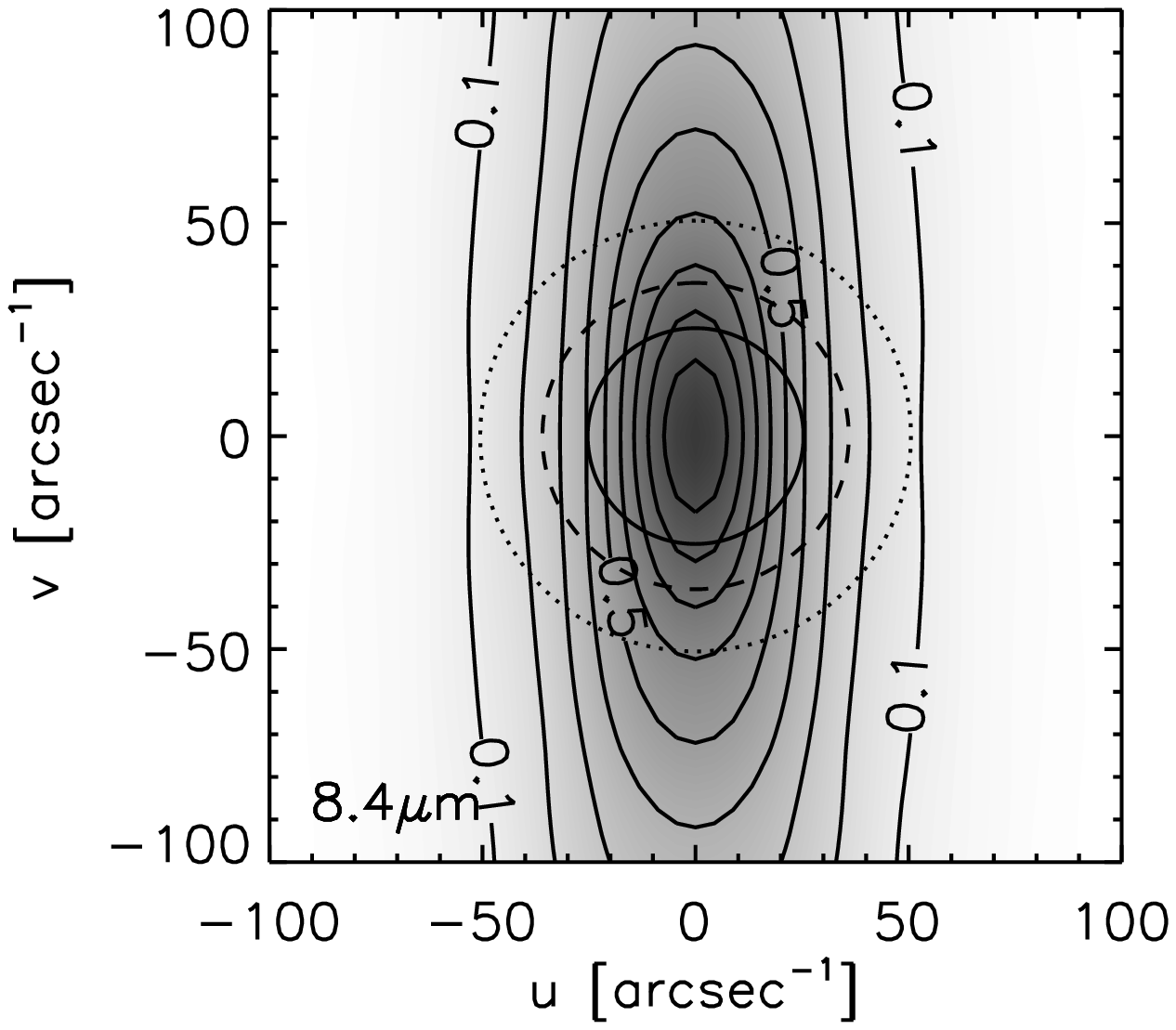}\hspace{0.6cm}
\includegraphics[width=5.5cm,angle=0]{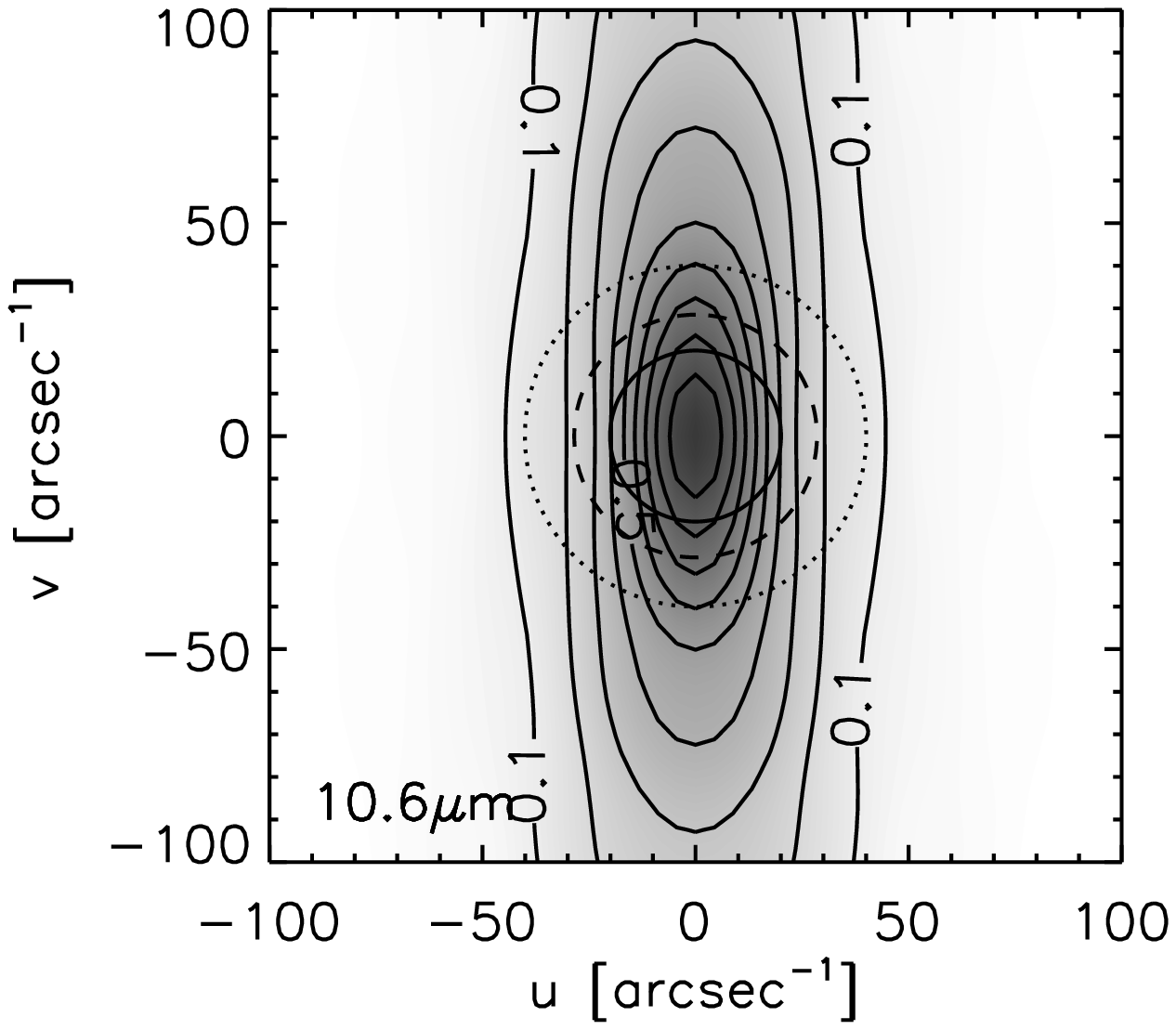}\hspace{0.6cm}
\includegraphics[width=5.5cm,angle=0]{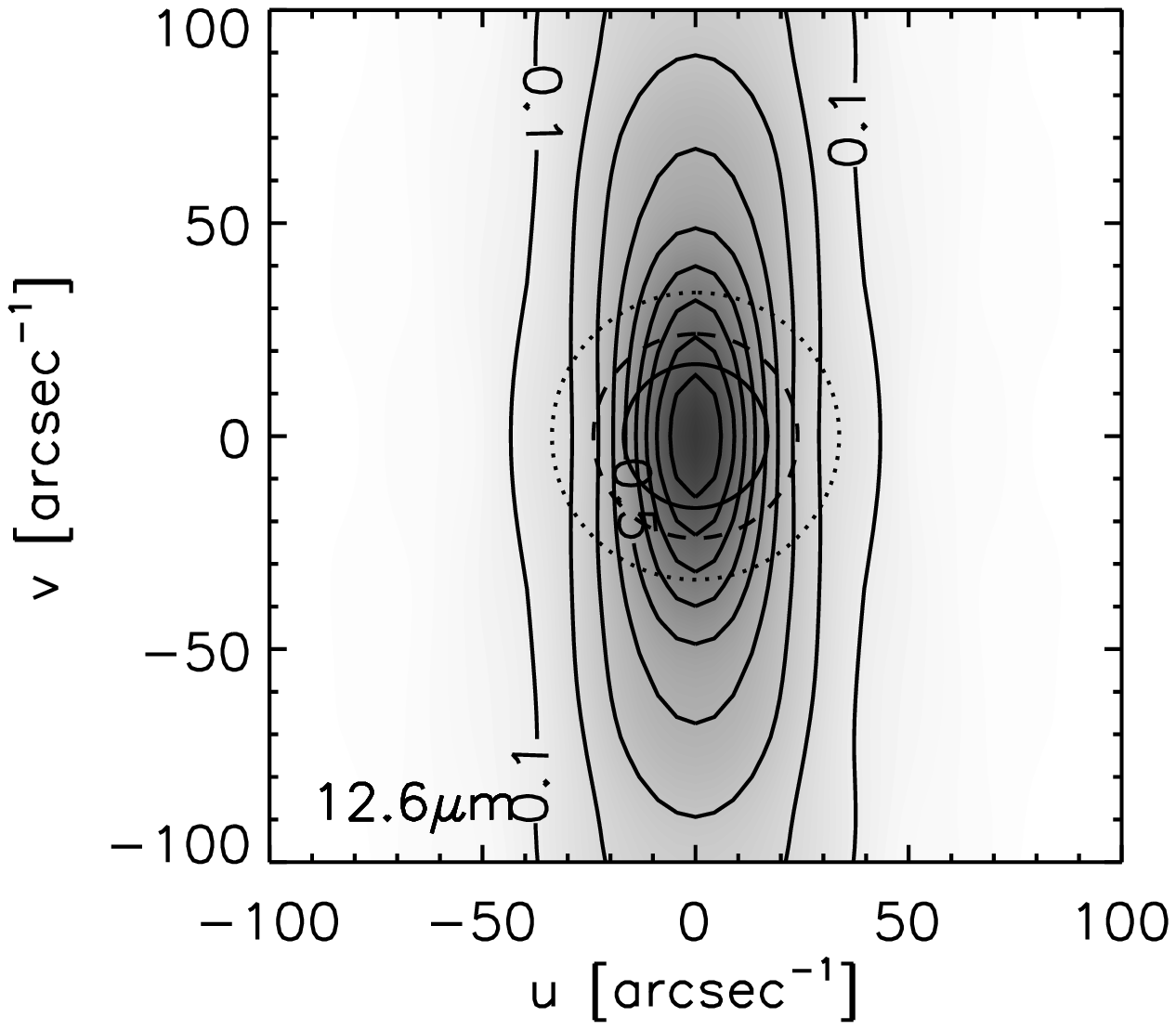}\vspace{0.6cm}
\includegraphics[width=5.5cm,angle=0]{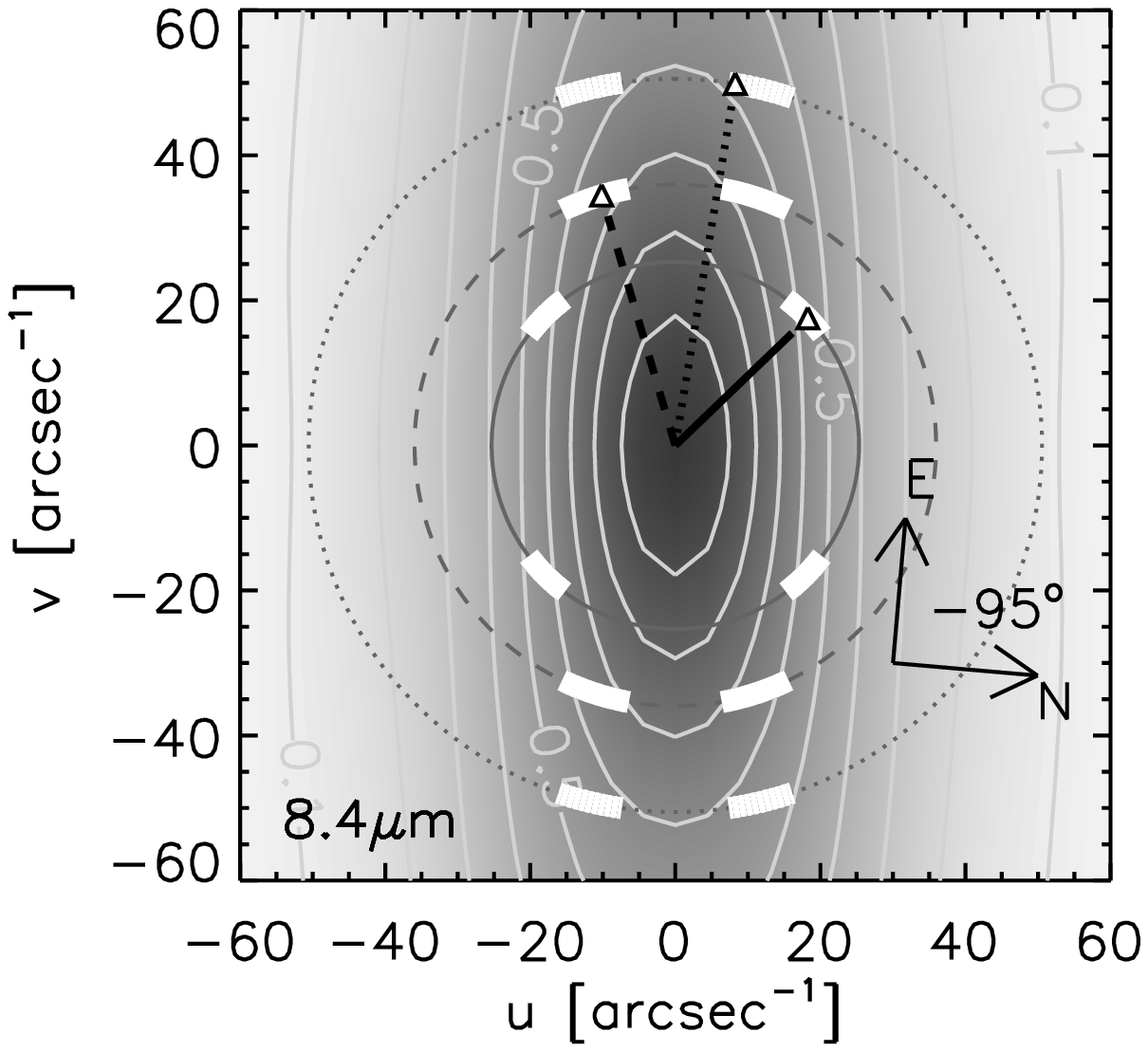}\hspace{0.6cm}
\includegraphics[width=5.5cm,angle=0]{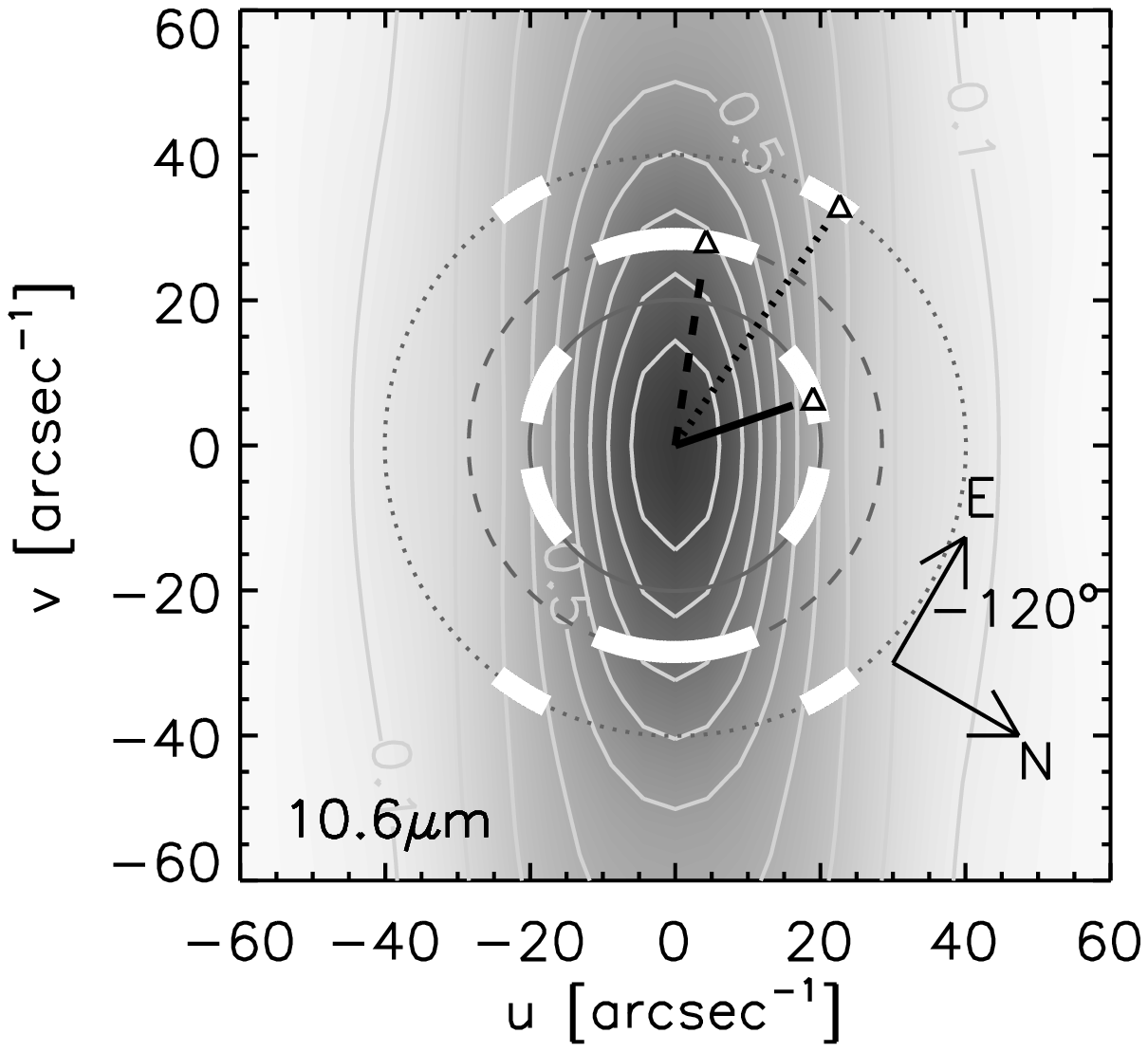}\hspace{0.6cm}
\includegraphics[width=5.5cm,angle=0]{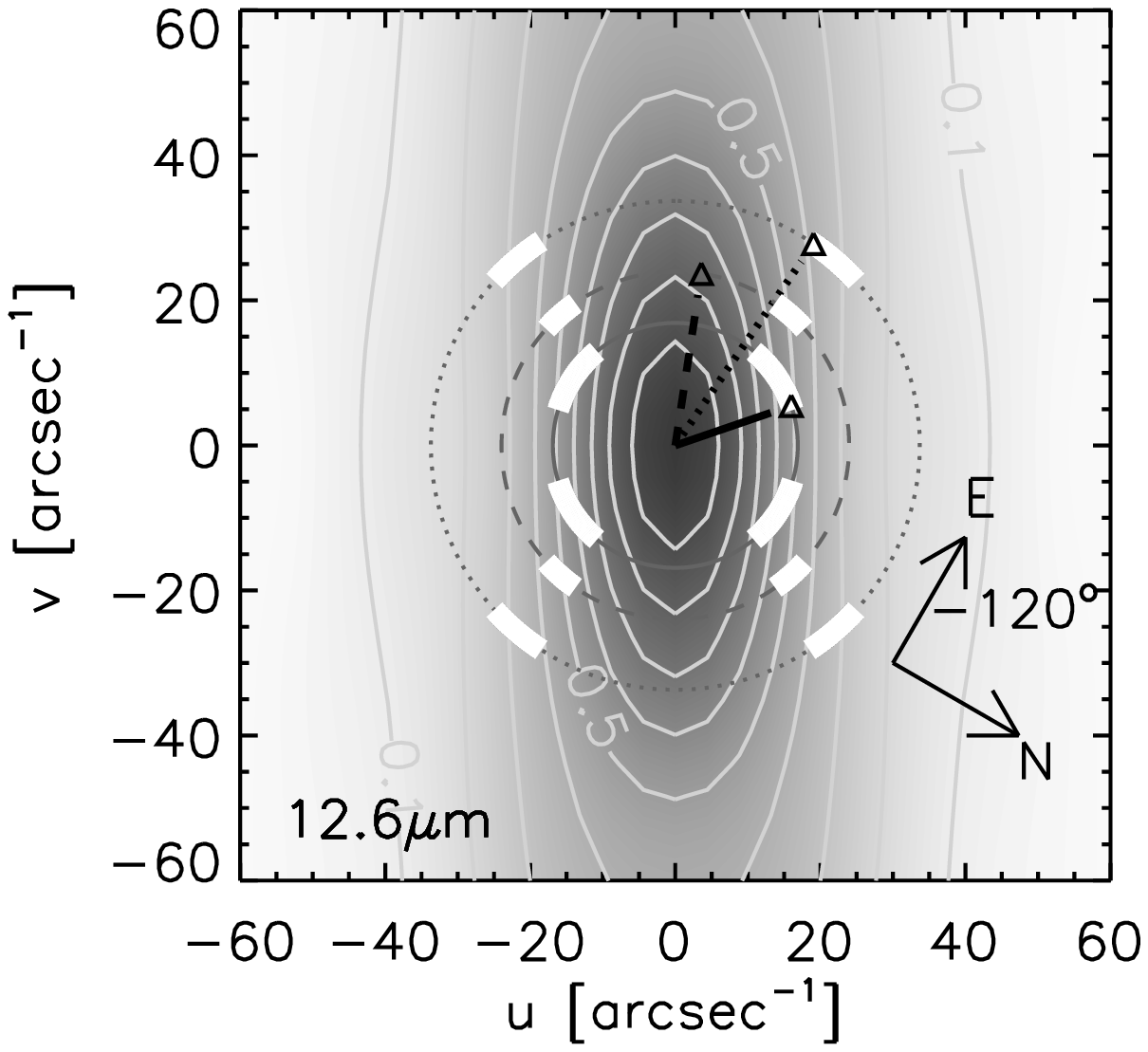}
\caption{{\it Top:} The modelled visibilities of T~Tau~Sa for 8.4\,$\mu$m, 10.6\,$\mu$m, and 12.6\,$\mu$m and a disc inclination of $72^{\circ}$ (from left to right). The solid contours are drawn for steps of 0.1 and the associated grey background is underlaid for easier comparison. The radii of the overplotted circles correspond to the spatial frequencies for which the baselines are sensitive at the various wavelengths (solid: 88\,m projected baseline length, dashed: 62\,m, dotted: 44\,m). {\it Bottom:} Zoom in on the panels in the top row and determination of the orientation of the u-v-plane relative to the sky (see text).}  
\label{Fig_sa2}
\end{figure*}

To cite \cite{boehm94} on the T~Tau system: ``It is intriguing to see two distinct bipolar outflows of different orientation in the same stellar system''. The jet pointing west and ending
in the Herbig-Haro object HH 155 as its working surface \citep{buehrke86}, was attributed by them to T~Tau~N. This association appears convincing: The inclination of the jet axis against the viewing direction was determined from a comparison of radial and tangential velocities to a moderate value of 23$^\circ$ \citep{eisloeffel98} and could be seen in the forbidden lines of [OI] and [SII] well collimated to within 0.3$''$ of the star. The star was found to be also mildly inclined by 19$^\circ$ \citep{herbst97}. Millimetre interferometry of its circumstellar disc found an inclination of i = 41$^\circ$ $\pm$ 3$^\circ$ and a PA of 19$^\circ$ $\pm$ 5$^\circ$ \citep{akeson98}. Near-infrared interferometry \citep{akeson02} confirmed the inclination with i = 29$^\circ$ $\pm$ 12$^\circ$ but disagreed on the PA with 132$^\circ$ $\pm$ 16$^\circ$. Our model above, fitting SED and mid-infrared visibilities, led also to a moderate inclination of $i < 30^\circ$, while the PA could not be determined. With this evidence it was natural that \cite{boehm94} attributed the second outflow, oriented nearly north-south, to the then known second component in the system, T~Tau~S.

A Fabry-Perot map of the T Tau system in the $v=1-0\,{\rm S}(1)$ line of molecular hydrogen led \cite{herbst07} to contradicting conclusions. They found about one arcsecond west of T~Tau~S a bow\,shock-like emission, which the authors named~``C''. Their map clearly shows a huge complex consisting of a bow shock and the walls of the tunnel produced by the jet. This complex is indicated in Figure~\ref{Fig14} by the dashed ellipse. With the high spatial resolution provided by the instrument NACO on the VLT the jet could be unambiguously traced back to T~Tau~S, with the dominating component T~Tau~Sa being the preferred candidate for its origin. The northwest-southeastern (actually NNW-SSE) outflow, which is the inner part of the north-south oriented outflow mentioned above, could be best traced back to T~Tau~N.
 
To increase the difficulty in combining these two sets of measurements and interpretations, \cite{solf99} found a high inclination of $i = 79^{\circ}$ for the north-south oriented outflow on the basis of the high dispersion/average value for its velocity pattern, which makes it hard to imagine the almost face-on circumstellar disc of T Tau N as its source. These complicated geometrical relations led \cite{herbst07} to list four different reasonable combinations of outflows with their stellar sources.

To contribute to this discussion, we also have to derive the position angle of the disc around T~Tau~Sa in addition to the inclination of $72^{\circ}$ resulting from the simultaneous fit to the SED and visibilities. To this end, the modelled, two-dimensional visibilities for T~Tau~Sa are drawn in Figure~\ref{Fig_sa2}. Since the orientation of a baseline with respect to the model, i.e., the object in the sky, is uncertain, the spatial frequencies to which a baseline is in principle sensitive fall onto a circle, whose radius is determined by the projected length of the baseline. However, the relative orientations of the three used baselines are fixed. Or in other words, the three baselines represented in the figure by lines from the origin to the circles can be rotated around the origin, but only together. On the other hand, only at certain locations along the circles the visibilty values are consistent with the visibilities derived from the measurements. These locations are marked by the thick white arcs in Figure~\ref{Fig_sa2}. If it is now possible to find a rotation angle that brings all three baselines into marked locations on the corresponding circles, the orientation of the u-v-plane with respect to the baselines and thus to the sky is known with an ambiguity of $180^{\circ}$.

\begin{figure}[h!]
\centering
\includegraphics[width=9.0cm,angle=0]{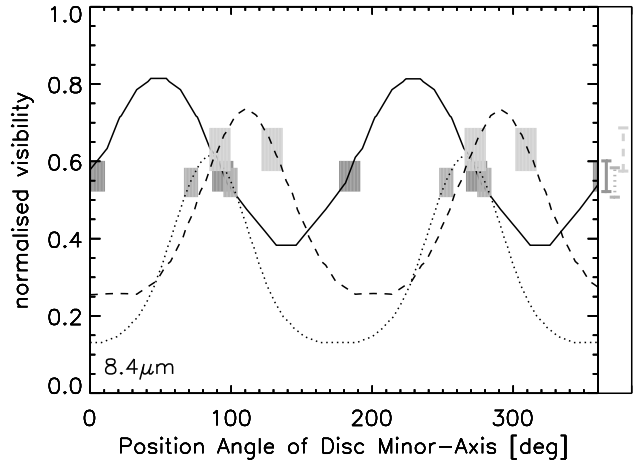}
\caption{The modelled visibilities of T~Tau~Sa drawn as a function of the position angle of the minor axis of the disc (solid: 44\,m projected baseline length, dashed: 62\,m, dotted: 88\,m).  For comparison, the expected visibilities are plotted in the right panel. The corresponding shaded areas in the graph indicate the suitable position angles.}  
\label{Fig_sa3}
\end{figure}

The modelled visibility for T~Tau~Sa at a wavelength of 8.4\,$\mu$m is drawn in Figure~\ref{Fig_sa3} also as function of the position angle of the disc's minor axis. The model is calculated in such a way that without a rotation the minor axis of the disc is oriented north-south, and thus the two-dimensional visibility is elongated along the north-south direction. Therefore, the visibility prediction for the baseline with the longest projected length, which had a position angle of about 90$^\circ$ on the sky, has a minimum without a rotation of the model disc, and a maximum when the model disc is rotated to 90$^\circ$. Similarly, the maxima for the other baselines also are reached when the model disc is rotated by an angle that corresponds to the position angle of that baseline. The visibilities derived from the measurements are drawn as bars in the right panel, and the resulting allowed ranges of position angels are marked in the graph by shaded areas (Figure~\ref{Fig_sa3}). The height of such an area reflects the uncertainty of the corresponding measured visibility.

It appears that agreement between the modelled and the measured visibilities is found when the model disc is rotated by about $100^{\circ}$, i.e., when the disc is elongated almost along the north-south direction. This finding is robust, because the modelled visibilities for the different baselines in Figure~\ref{Fig_sa3} show slopes of opposite sign due to the phase shifts of the maxima. This may qualify T Tau Sa as a launching site for the east-western jet (Figure~\ref{Fig14}). Especially, the position angle of the bow-shock C1 at the western tip of complex C \citep{herbst07} is in excellent agreement with the found position angle of the circumstellar disc around T~Tau~Sa. Also the above discussed scenario of a circumbinary disc only extincting T~Tau~Sa could be realised, because T~Tau~Sb would reside west of the north-south oriented disc plane. Interestingly, if such a circumbinary disc is present, T~Tau~Sb would be regularly extincted by this structure due to its orbital motion.

If the determined north-south orientation of the circumstellar disc around T~Tau~Sa is correct, T~Tau~N would be the origin of the northwest-southeastern outflow with the usual reasoning by exclusion. Unfortunately, the difficulties with picturing the T~Tau system of flows have not been reduced. The found orientation for the disc of T~Tau~Sa indeed perfectly matches the east-western flow of Figure~\ref{Fig14} if it is close to the plane of the sky. But then the difficulty arises to combine this flow with the only moderately inclined jet seen at larger distances out to HH 155. The same inclination discrepancy emerges for the north-south going outflow. To postulate that orientations may have changed strongly in this triple system during the about one million years it may have taken the jet ($v_{\rm rad} = 39$\,km/s) to arrive at the position of HH 155 has no serious foundation at present. On the other hand, the length of a Kozai cycle for the secular evolution of the inclination of the southern binary should be comparable \citep{ford00} and episodic ejections are much shorter \citep{reipurth00}. Perhaps a kinematical determination of the outflow inclination from the H$_2$ emission west of T Tau would be a first clarifying step. Further questions would concern better knowledge of the three-dimensional grouping of outflows around T~Tau and a better understanding of the role of the ``second'' infrared companion T~Tau~Sb. There is no reason why T~Tau~Sb with its circumstellar disc and the highest accretion rate among the three stars in the T~Tau system should not be the origin of one of the two jets.

We presented new information for discussing this puzzle of the origin of the jets, a solution still has to be awaited.

%

\section{Summary\label{sum}}

MIDI at the VLTI allows an interferometric study of the three individual components of the T Tau system and their associated circumstellar material. 

The disc around T~Tau~N was well resolved in our observations. We find a radiative transfer model, including accretion, that fits simultaneously the SED and the visibilities. The physical parameters are in line with earlier measurements. This means that we can confirm the picture of an almost face-on, comparatively massive circumstellar disc around T~Tau~N. An optically thin envelope is needed as an additional component to produce the 3-8\,$\mu$m radiation. The model also fulfils the constraints set by the K-band visibilities measured with the PTI \citep{akeson02}. By fitting the silicate emission feature in the total spectrum and the correlated spectra, we find that dust processing occurs in the inner parts of the circumstellar disc around T~Tau~N.
 
The binary T~Tau~S is also well resolved interferometrically, but not photometrically. The visibilities with the imprinted sinusoidal modulation give the relative position of T~Tau~Sa and T~Tau~Sb and allow us to derive the flux ratio of the (correlated) fluxes. The silicate band is seen in absorption towards both sources and confirms a high foreground extinction of about $A_{\rm V}=15$\,mag. The phases indicate that T~Tau~Sa is the brighter source in the N-band.

For the full use of the data we adopt for T~Tau~Sb the model of \cite{duchene05}, which describes this source as a normal T Tauri star of spectral type M1. The star resides behind the absorbing screen of $A_{\rm V}=15$\,mag and its circumstellar disc is seen not far from face-on. The low-mass disc is truncated because of the proximity of T~Tau~Sa.

With this model at hand, the interferometric data reveal an elongated structure for T~Tau~Sa, interpreted as a disc seen nearly edge-on. The radiative transfer model with a central star as proposed by \cite{duchene05} simultaneously fits SED and visibilities. The model again suggests the presence of an optically thin envelope. The orientation of the disc is almost north-south. T~Tau~Sa might thus be the driving source of the east-west jet. The additional (intrinsic) extinction of about $A_{\rm V}=15$\,mag might be attributed to a circumbinary disc.

The orbits and discs in the T~Tau triple system show complex orientations. While the disc around T~Tau~Sb might be coplanar with its orbit around T~Tau~Sa, the disc around T~Tau~Sa is not aligned. It is almost perpendicular to the disc around T~Tau~N. However, the full picture requires knowledge of the orbital elements of the orbit of the southern binary around T~Tau~N with much better precision than is available now.

%

\appendix

\section{Visibility of a binary with extended components\label{AppA}}

Single-baseline interferometry gives one-dimensional information on the object along the direction of the baseline projected onto the sky. Let P$_{\lambda}$ and C$_{\lambda}$ be the spectrum of the primary and the companion of the binary with projected brightness distributions along this direction of P$_{\lambda}(x)$ and C$_{\lambda}(x)$. The complex visibilities, the Fourier transforms with respect to $x$, are written as $\mathcal{P}_{\lambda}(u)$ and $\mathcal{C}_{\lambda}(u)$. The correlated fluxes of the components then are $F_p^{\rm cor}(u) = \left| \mathcal{P}_{\lambda}(u)\right|$ and $F_c^{\rm cor}(u) = \left|\mathcal{C}_{\lambda}(u)\right|$. The normalised visibilities, also functions of $\lambda$, are $V_p(u) = \left|\mathcal{P}_{\lambda}(u)\right| / P_{\lambda}$ and $V_c(u) = \left|\mathcal{C}_{\lambda}(u)\right| / C_{\lambda}$. Please note that in the case of a measurement with MIDI, $u$ and $\lambda$ are coupled by the relation $u= B/\lambda$, where $B$ is the length of the projected baseline. For extraction of the phase from MIDI measurements, see Appendix~\ref{phases}.

In the special case where the baseline projection is perpendicular to the separation vector of the binary, we have
\begin{enumerate}
\item the flux distribution $O_{\lambda}(x) = P_{\lambda}(x) + C_{\lambda}(x)$\,,
\ \vspace{0.16cm}
\item the correlated flux $F_o^{\rm cor} = \left|\mathcal{P}_{\lambda}(u) + \mathcal{C}_{\lambda}(u)\right|$\,, and
\ \vspace{0.12cm}
\item the normalised visibility V$_o = \frac{\left|\mathcal{P}_{\lambda}(u) + \mathcal{C}_{\lambda}(u)\right|}{P_{\lambda} + C_{\lambda}}$.\\
\end{enumerate}

If $P_{\lambda}(x)$ and $C_{\lambda}(x)$ are symmetric (Fourier transform is real) or the same apart from a constant factor, or if $\mathcal{P}_{\lambda}(u)$ and $\mathcal{C}_{\lambda}(u)$ have the same phase, then we have the naively expected result that the correlated flux of the binary is the sum of the correlated fluxes of the components
\begin{equation}
 F_o^{\rm cor} = \left|\mathcal{P}_{\lambda}(u)\right| + \left|\mathcal{C}_{\lambda}(u)\right|  \label{eqa1} 
\end{equation}
and the normalised visibility of the binary is the flux weighted average of the visibilities of the components
\begin{equation}
V_o(u) = \frac{P_{\lambda}V_p(u) + C_{\lambda}V_c(u)}{P_{\lambda} + C_{\lambda}}\mathrm{\ .}\label{eqa2}
\end{equation}
 
It is common to the three conditions given above that they presume the same phase for $\mathcal{P}_{\lambda}(u)$ and $\mathcal{C}_{\lambda}(u)$. If this does not hold, relations (\ref{eqa1}) and (\ref{eqa2}) are not fulfilled in the strict mathematical sense. But they still may serve as good approximation, to be judged on a case-by-case basis.\footnote{In fact, we will see that for a given wavelength, the phase equality necessary for strict validity of relations (\ref{eqa1}) and (\ref{eqa2}) will be fulfilled at suitable values of the spatial frequencies.}

For all other projected baseline directions, the primary and companion are separated by a projected separation $s$. The correlated flux is then
\begin{equation}
F_o^{\rm cor}(u) = \left|\int_{-\infty}^{\infty}\left[P_{\lambda}(x) + C_{\lambda}(x-s)\right] e^{-2\pi i ux}dx \right|\label{eqa3}
\end{equation}
or after substituting x-s
\begin{eqnarray}
F_o^{\rm cor}(u) & = & \left|\mathcal{P}_{\lambda}(u) +  e^{-2\pi i us} \mathcal{C}_{\lambda}(u)\right|\label{eqa4} \\
& = &\sqrt{(\mathcal{P}_{\lambda}(u) +  e^{-2\pi i us}\mathcal{C}_{\lambda}(u)) (\mathcal{P}_{\lambda}(u) +  e^{-2\pi i us}\mathcal{C}_{\lambda}(u))^{*}}\nonumber
\mathrm{\ ,}
\end{eqnarray}
where the asterisk denotes the complex conjugate of the expression in parentheses.

Writing $\mathcal{P}_{\lambda}(u)$ = $\left|\mathcal{P}_{\lambda}(u)\right|
e^{\phi_p(u)}$ and $\mathcal{C}_{\lambda}(u)$ = $\left|\mathcal{C}_{\lambda}(u)\right|
e^{\phi_c(u)}$, and with $\Delta\phi_{\lambda}(u) = \phi_p(u) - \phi_c(u)$,
the argument of the square root has the value
\begin{equation}
\left|\mathcal{P}_{\lambda}(u)\right|^2
+\left|\mathcal{P}_{\lambda}(u)\right| \left|\mathcal{C}_{\lambda}(u)\right|
\cos(\Delta\phi_{\lambda}(u) + 2\pi us)
+ \left|\mathcal{C}_{\lambda}(u)\right|^2\mathrm{.}\label{eqa5}
\end{equation}
For $s=0$ this is the detailed formulation for the correlated flux observed perpendicular to the separation vector of the binary. For $\Delta\phi(u)$ = 0 and  constant values of $\mathcal{P}_{\lambda}(u)$ and $\mathcal{C}_{\lambda}(u)$, this reduces to the well-known formula for a binary composed of point sources. As in the case of point sources, in this more general case the cosine term is responsible for the cosine-like modulation of the visibility, the correlated flux, respectively.

It is a matter of convenience whether one analyses the correlated flux with the equations given here or the normalised visibility. Then the given equations have to be divided by the combined flux of the binary. In the case of T Tau S, where the visibility varies smoothly with spatial frequency $u$ and hence $\lambda$, while flux and correlated flux show a strong depression in the silicate feature, it is advantageous to discuss and interpret the visibilities.

In the following paragraphs we will now consider two special cases for the product $u\cdot s$.

\paragraph{An integer result for the product $u\cdot s$} The cosine factor equals~1 for $\Delta\phi_{\lambda}(u)$ = 0. This corresponds closely to the maxima of the modulated visibility curve. Here, the correlated flux and hence the visibility are exactly the same as for observations perpendicular to the separation vector of the binary.  For $\Delta\phi_{\lambda}(u)$ = 0, or if $\Delta\phi_{\lambda}(u) + 2\pi us$ is an integer multiple of 2$\pi$, the simple relations (\ref{eqa1}) and (\ref{eqa2}) hold as given above. And, at least for $\Delta\phi(u)=0$, the gradient in these maxima is the same as for the unmodulated curve that would be obtained perpendicular to the separation vector. Therefore, to the extent that linear interpolation of the visibility appears justified, the combined visibility at the given baseline can be reconstructed as a function of wavelength from the two maxima appearing in the modulated visibility curves observed for T~Tau~S.

\paragraph{Case of a half-integer result for the product u$\cdot$s} The cosine factor equals -1 if $\Delta\phi_{\lambda}(u)+2\pi us$ is an odd multiple of $\pi$. This corresponds closely to the minima of the modulated visibility curve, with $F_o^{\rm cor}(u) = \left|\mathcal{P}_{\lambda}(u)\right| - \left|\mathcal{C}_{\lambda}(u)\right|$ for $\Delta\phi_{\lambda}(u)$ = 0. 

Comparing this minimum of the correlated flux or visibility to the maximum value interpolated to the same spatial frequency for the unmodulated visibility curve, we have essentially the same relation as for simple binary stars:
\begin{eqnarray}
\frac{F_o^{\rm cor}({\rm max}) - F_o^{\rm cor}({\rm min})}{F_o^{\rm cor}({\rm max}) + F_o^{\rm cor}({\rm min})}  & = &
\frac{V_o({\rm max}) - V_o({\rm min})}{V_o({\rm max}) + V_o({\rm min})}\nonumber\\ 
& = & \frac{2\cdot \mathcal{C}_{\lambda}(u)}{2\cdot \mathcal{P}_{\lambda}(u)}\mathrm{\ .}\label{eqa6}
\end{eqnarray}

This means that to a good approximation, we can read from the observed modulated visibility curves the {\it ratio of the correlated fluxes of the components} at this spatial frequency. By fitting the shape of the observed visibility modulation for the entire available wavelength range, this ratio can be inferred as a function of $u$. In binaries composed of point sources, the terms in which we are used to think, the equivalent relation gives the brightness ratio of the components, because flux and correlated flux are the same for point sources. With the assumption that the primary and companion show the same visibility, $V_p(u) = V_c(u)$, Equation~(\ref{eqa6}) would also give the brightness ratio.

%

\section{Appendix B. Phase measurements with MIDI\label{phases}}

\paragraph{The basic equations} Consider an object $P_{\lambda}(\alpha)$, where $\alpha$ is an angle on the sky. We write its one-dimensional Fourier transform, i.e., its complex visibility, as 
\begin{eqnarray}
\mathcal{P}_{\lambda}(u) 
& = & \int_{-\infty}^{\infty}P_{\lambda}(\alpha)e^{-2\pi i u\alpha}d\alpha\nonumber\\
& = & \left|\mathcal{P}_{\lambda}(u)\right| \cdot  e^{i\phi_{\lambda}(u)} = A_{\lambda}(u)\cdot e^{i\phi_{\lambda}(u)}
\label{eqa7}
\end{eqnarray}
with visibility amplitude A$_{\lambda}(u)$ and phase $\phi_{\lambda}(u)$. Reversing the direction in which  $\alpha$ is increasing is equivalent to going from $P_{\lambda}(\alpha)$ to $P_{\lambda}(-\alpha)$ and would change the sign of $\phi_{\lambda}(u)$. Observing this object interferometrically with a baseline of projected length $B_{\rm proj}$ and a corresponding spatial frequency $u = B_{\rm proj}/\lambda$ on the sky leads for multiaxial beam combination to a visualisation of the visibility component $\mathcal{P}_{\lambda}(u)$ as an instantaneous fringe pattern in the image plane with the modulating part  
\begin{eqnarray}
I_{\lambda}(x) & \propto & 2 \left|\mathcal{P}_{\lambda}\right| \cos\left(2\pi f_mx + \phi_{\lambda}\right),
\label{eqa8}
\end{eqnarray}
where 1/$f_m$ is the spatial fringe spacing and the sign of the phase is positive for  proper choice of the sign of x. Alternatively, in coaxial beam combination, this visibility component shows an OPD-scanned timewise fringe pattern of 
\begin{eqnarray}
I_{\lambda}({\rm OPD}) & \propto & 2 \left|\mathcal{P}_{\lambda}\right| \sin\left(2\pi {\rm OPD}/\lambda + \phi_{\lambda}\right)\label{eqa9},
\end{eqnarray}
where obviously $\Delta$OPD = $\lambda$ is the fringe spacing during an OPD scan. The latter method is used in the MIDI instrument at the VLTI. The factor of two here results from taking the difference of the two beamsplitter output signals of opposite sign. The presence of the sine function in relation (\ref{eqa9}), instead of the cosine function, results from the well-known $\pi$/2 phase shift in the symmetrically used beam splitter, which allows the coaxial recombination. As we will see, the $\pi$/2 phase shift is of no consequence for the phase determination of $\phi_{\lambda}$. The positive directions of x in Equation~(\ref{eqa8}) and OPD in Equation~(\ref{eqa9}) need to correspond to the positive scan direction $\alpha$ over the source on the sky. Otherwise, the phase $\phi_{\lambda}$ would change sign. In other words, defining the signs of x, OPD, respectively, and keeping Equation (\ref{eqa8}), (\ref{eqa9}), respectively, would also define a positive direction $\alpha$ on the sky.

For later use we note that by the formulation of Equation~(\ref{eqa8}) the maximum of the fringe pattern at the origin will move to negative values of $x$ when $\phi_{\lambda}$ is going positive from the value 0 (which a point source at the origin would have). In the coaxial case the zero crossing of the fringe pattern at the origin will, according to Equation~(\ref{eqa9}), move to negative values of OPD for an increasing $\phi_{\lambda}$.

\paragraph{Derivation from more general principles} Equations (\ref{eqa8}) and (\ref{eqa9}), though plausible, deserve some justification in detail, in particular, to verify the sign of the angle $\alpha$, of the OPD, and of the phases. We start from the presentation by \cite{boden00}. A plane wave propagating from the object in direction ${\bf k}$ towards the observer is written as $e^{i{\bf kx}- i\omega t}$ with $\left | {\bf k} \right| = k = 2\pi/\lambda$. Using telescopes A and B we define the baseline vector {\bf B} as ${\rm A} \rightarrow {\rm B}$. Pointing the telescopes to the object, we are viewing in direction ${\bf s} = -{\bf k}$. The total optical path difference counted in the sense B minus A then is ${\rm OPD} = -{\bf sB} + d_B - d_A$, where d are the pathlengths from the telescopes to the locus of beam combination and $-{\bf sB}$ constitutes the external part of the OPD, which decreases for viewing directions towards telescope B. From the superposition of the waves travelling through telescope A, i.e., $e^{i\left({\bf sB}+ d_A\right) - i\omega t}$, and telescope B, i.e., $e^{i\left(d_B - i\omega t\right)}$, the fringe pattern is
\begin{equation}
F_{\lambda}({\rm OPD}) \propto \cos\left[k\left({\bf sB} + d_A - d_B\right)\right]\label{eqa10}.
\end{equation}
To relate positions on the sky to the instrumentally available internal OPD $d_A - d_B$, we consider the viewing direction ${\bf s}$ as the origin on the sky. We take as the zero point of OPD that value of $d_A - d_B$ where a point source at this origin would produce the white light fringe, i.e., where the argument of the cosine in Equation~(\ref{eqa10}) is zero. We also note that with the accepted sign convention for the OPD (OPD = path length ``MIDI beam B'' minus path length ``MIDI beam A''), an interferometric scan with MIDI moves the internal OPD $d_B - d_A$ stepwise towards more positive values, since the piezo motion is shortening d$_A$. In the current MIDI mechanical-geometrical setup, during such an OPD scan the corresponding position of the white light fringe on the sky moves towards the telescope that sends its light beam into the right entrance window of the MIDI dewar, when seen from the viewpoint of the incoming light (``MIDI beam B''). The FITS headers of the stored data files carry the information that allows transposition of this somewhat convolved definition into a well-defined direction on the sky, such that the phases will give well-defined information on the object. It is natural to count $\alpha$ positive on the sky in the direction where the corresponding white light fringe moves during an internal OPD scan. According to Equation~(\ref{eqa10}), this means that, because d$_A$ decreases during the scan, $\alpha$ as ${\bf sB}$ goes positive towards telescope~B. 

For an extended source, we have to add the independent fringe patterns (Equation~\ref{eqa10}) contributed from all its parts, which means integrating over $\alpha$, i.e., in 
the plane of the sky and perpendicular to ${\bf s}$. The integrand then reads
\begin{equation}
P_\lambda(\alpha)\cdot\cos\left[k\left({\bf s}+{\bf\alpha}\right){\bf B} - k\left(d_B - d_A + \Delta{\rm OPD}_{\rm int}\right)\right]{\rm\ .}\label{eqa11}
\end{equation}
Here, $d_B-d_A$ is the value of the internal OPD corresponding to the white light fringe at $\alpha = 0$, and $\Delta{\rm OPD}_{\rm int}$ is an instrumental offset from the white light position (describing, e.g., the piezo scan). Since ${\bf sB} + d_A-d_B = 0$, the integrand simplifies to
\begin{eqnarray}
P_\lambda(\alpha)\cdot\cos\left[k\left({\bf \alpha B} - \Delta{\rm OPD}_{\rm int}\right)\right]=\hspace{3cm}\nonumber\\
P_\lambda(\alpha)\cos(2\pi\alpha B_{\rm proj}/\lambda)\cos(2\pi\Delta {\rm OPD}_{\rm int}/\lambda)+\nonumber\\
P_\lambda(\alpha)\sin(2\pi\alpha B_{\rm proj}/\lambda)\sin(2\pi\Delta {\rm OPD}_{\rm int}/\lambda)\rm{\ .}\hspace{0.3cm} \label{eqa12}
\end{eqnarray}
Keeping in mind  the above definition of visibility $\mathcal{P}_{\lambda}(u)$, and respecting $\phi_{\lambda} = 2\pi\alpha B_{\rm proj}/\lambda$, the integral over $\alpha$ refers to the integrand
\begin{eqnarray}
\left|\mathcal{P}_{\lambda}(u)\right|\cos(\phi_{\lambda})\cos( 2\pi\Delta{\rm OPD}_{\rm int}/\lambda) -\hspace{2cm}\nonumber\\
\left|\mathcal{P}_{\lambda}(u)\right|\sin(\phi_{\lambda})\sin( 2\pi\Delta{\rm OPD}_{\rm int}/\lambda){\rm\ ,}\label{eqa13}
\end{eqnarray}
which directly leads to the form 
\begin{equation}
F_{\lambda}(\Delta {\rm OPD}_{\rm int}) = \,\,\left|\mathcal{P}_{\lambda}(u)\right| 
\cos(2\pi\Delta {\rm OPD}_{\rm int}/\lambda +\phi_{\lambda}){\rm  ,}\label{eqa14}
\end{equation}
from which Equation~(\ref{eqa9}) follows simply by applying the $\pi$/2 phase shift. This verifies the formulation of Equation~(\ref{eqa9}), including the sign of the phase.

\paragraph{A Gedankenexperiment}
We now consider a binary with the centre of light in the origin to confirm by a detailed look at the expected temporal fringe pattern that the phases determined by MIDI will have the correct sign.

{\em On the sky} we are on safe grounds: The phase would show the form of a staircase with width of the individual steps in spatial frequency $u$ of $\Delta u = 1/{\rm separation}$ and the height of the steps depending on the brightness ratio. The phase will decrease with $u$ if the companion is located at more negative $\alpha$ with respect to the primary. On the other hand, the binary's phase with removed linear trend -- as it is determined in the instrument MIDI -- corresponds to a different position, namely with the primary at the origin. The binary phase now oscillates around zero.  For a companion at more negative position $\alpha$ than the primary, the phase will  increase with $u$ (decrease with $\lambda$) at those spatial frequency values where the maxima of the visibility amplitude are found, a simple and useful criterion.

{\em In the instrument} it looks like this: If we have a companion at slightly negative $\alpha$ with respect to the primary, its contribution to the temporal fringe would have a maximum slightly earlier in the OPD scan, at $\Delta {\rm OPD}_{\rm comp} = -\alpha\cdot B$. The maximum of the combined fringe pattern 
(primary plus companion) then will also be shifted to more negative values, by
\begin{equation}
\Delta {\rm OPD}_{\rm binary} = -\alpha\cdot B\cdot\frac{F_{\rm comp}}{F_{\rm prim}+F_{\rm comp}}\mathrm{ .}\label{eqa15}
\end{equation}
Measured in units of $\lambda$, this shift, having a negative value, will get closer to zero, i.e., will be rising with $\lambda$. The phase $\phi_{\lambda}$, corresponding to this shift of maximum, therefore will be decreasing with $\lambda$. This relation has the correct sign, and so the phases determined from MIDI should be correct with the sign as defined by the Fourier transform and as used in Equations (\ref{eqa9}) and (\ref{eqa14}).

\paragraph{The treatment of phases in  MIDI} In MIDI, OPD in Equation~(\ref{eqa9}) is the actual optical path difference for a given individual exposure. One such exposure with the MIDI instrument gives $F_{\lambda}({\rm OPD})$ as function of $\lambda$ for the range 8-13\,$\mu$m. Many of these exposures at a repeated saw-tooth pattern of OPD steps scanning the fringe constitute an interferometric measurement. In each scan, the position of fringe maximum depends on the phase $\phi_{\lambda}$.

For the determination of phases from MIDI measurements we have to recall that the total optical path difference consists of two parts of different behaviour:  ${\rm OPD}_{\rm total} = {\rm OPD}_{\rm int} + {\rm OPD}_{\rm atm}$, where the ``instrumental'' ${\rm OPD}_{\rm int}$ is quickly stepped through a few times the wavelength of about 10 $\mu$m during an interferometric measurement and is known including its sign, while ${\rm OPD}_{\rm atm}$ is mostly determined by the atmosphere and is more slowly varying with a few $\mu{\rm m/s}$, but in an unpredictable way.

The {\it EWS} part of the MIDI data reduction package {\it MIA+EWS} (Section~\ref{dr}) determines the phase by mathematically shifting each exposure as given by Equation~(\ref{eqa9}) to OPD zero and then averaging to obtain the complex visibility -- modulus and phase \citep{jaffe04}. Remembering that a shift in image space (and here also in OPD) corresponds to a phase factor in Fourier space of the form $e^{2\pi i u {\rm OPD}}$, we multiply (\ref{eqa9}) by the factor $e^{-2\pi i  u {\rm OPD}_{\rm int}}$ to ``shift'' each exposure $F_{\lambda}({\rm OPD})$ back to position ${\rm OPD}_{\rm int} = 0$.  Equation~(\ref{eqa9}) now reads -- expressing the sine function by exponentials and apart from factors that do not matter here -- as follows:
\begin{equation}
F'_{\lambda}({\rm OPD}) = 
\frac{A_{\lambda}}{i}e^{2\pi i  u OPD_{\rm atm} +i \phi_{\lambda}} - \frac{A_{\lambda}}{i}e^{-4\pi i u {\rm OPD}_{\rm int}  -2\pi i u  {\rm OPD}_{\rm atm}  - i \phi_{\lambda}}\mathrm{.}\label{eqa16}
\end{equation}

To determine the unkown atmospheric optical path difference ${\rm OPD}_{\rm atm}$, we Fourier transform Equation~(\ref{eqa16}) with respect to $u$ and obtain peaks at $+{\rm OPD}_{\rm atm}$ and $(-{\rm OPD}_{\rm atm} - 2{\rm OPD}_{\rm int})$. Of these, the second term, rapidly changing with ${\rm OPD}_{\rm int}$ during a scan, quickly smears out by averaging over different exposures, so the only remaining peak gives the wanted value of ${\rm OPD}_{\rm atm}$ with the correct sign. We can now perform the ``shifting back'' correction also for the atmospheric part of the OPD by multiplying Equation~(\ref{eqa16}) with $e^{-2\pi i  u {\rm OPD}_{\rm atm}}$. This leads to
\begin{equation}
F''_{\lambda}($OPD-corrected$) =  
A_{\lambda}e^{+i( \phi_{\lambda} \pm \pi/2)}\mathrm{,}\label{eqa17}
\end{equation}
where we have omitted the strongly jittering part with the factor $e^{-4\pi i u {\rm OPD}_{\rm int} -4\pi i  u {\rm OPD}_{\rm atm}  - i \phi_{\lambda}}$. The constant $\pi$/2 results from the factor 1/$i$. Its sign depends on which of the two complementary interferometric signals after the beam splitter is subtracted from the other one. Averaging Equation~(\ref{eqa17}) over the typically several thousand exposures of an interferometric measurement thus will reliably determine -- apart from a constant known in principle -- the desired phase $\phi_{\lambda}(u=B_{\rm proj}/\lambda)$ over the wavelength range 8-13\,$\mu$m\footnote{At the same time, it will also give the visibility amplitude $A_{\lambda}$.}.

We remove any linear trend with $u$ from this phase, since it would simply correspond to a position offset of the source from the origin. In addition, we remove any remaining constant term $\ne$0. Again, apart from our constant, this term could be and appears to be mostly due to differential longitudinal dispersion over the unequally long air paths of the interfering light beams and it thus mostly carries information not related to the object. Of course, subtraction of the constant term also takes care of the added $\pm$ $\pi$/2. This ends the determination of the phase. We note that the data reduction procedure described here preserves the sign of the phase as defined by the Fourier transform and given in Equation~(\ref{eqa9}).

The phases determined with MIDI on the VLTI thus are not fully complete, i.e., the constant and linear terms are set to zero. But they allow to resolve the ambiguity of where the companion in a binary is located with respect to its primary. This property of the MIDI data is needed and used in the discussion of the close binary T~Tau~S (Section~\ref{dis}).

\paragraph{A check by a known binary} We wanted to check the signs of the phase derived above from ``first principles'' with an object where we assume to know the position of the companion. The close binary \object{Z~CMa}, with separation of 0.1$''$, consists of an infrared companion and a FU~Ori component \citep{koresko91}. At 4\,$\mu$m the infrared companion is brighter by a factor of 6. This ratio rises with wavelength. The ratio of correlated fluxes at 10\,$\mu$m, determined as described in Appendix~A, is in the range 5-10. We assumed that with these clear inequalities the component much brighter at 4\,$\mu$m will also be the much brighter component at 10\,$\mu$m and that it also will show the larger correlated flux. Z~CMa then can be used for calibrating the sign of the phase relation, with the result that the relations derived in this appendix appear correct. A final confirmation came from observations of $\eta$~Vir (Appendix~\ref{etavir}).

%

\section{`The \object{$\eta$~Vir} Experiment'\label{etavir}} 

We got permission from ESO to prove experimentally the conclusions derived in Appendix~\ref{phases}. As target we chose the well-studied triple system $\eta$~Vir \citep{hartkopf92, hummel03}. $\eta$~Vir has a composite spectral type  A2IV and is formed by a close spectroscopic binary and a wide companion. The orbit of the spectroscopic binary has a semi-major axis of 7\,mas and can thus only be marginally resolved by means of optical long baseline interferometry in the mid-infrared. On the other hand, the wide companion is in an orbit with a semi-major axis of 134\,mas and can be well detected with a suitable baseline configuration.

The observations were performed on May 17, 2006 with the baseline UT3-UT4. The projected length was 53\,m at a position angle of 113$^{\circ}$. According to the orbital parameters published by \cite{hummel03}, the wide companion was at that time at a position angle of 281$^{\circ}$  and separated by 89\,mas from the primary. The projected separation along the projected baseline was thus 87\,mas. For the calibration of the data, observations of \object{$\epsilon$~Crt} have been obtained right after those of  $\eta$~Vir. The projected baseline length was 45\,m at a position angle of 130$^{\circ}$.

\begin{figure}
\centering
\includegraphics[height=8.5cm,angle=90]{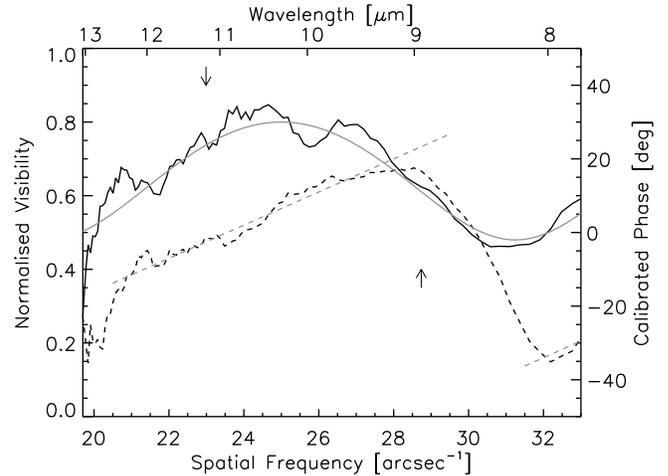}
\caption{The normalised visibility (solid black) and the phase (dashed black) of $\eta$~Vir. The arrows indicate the predicted positions for the maximum and minimum of the visibility, while the grey solid line represents a binary model. The grey dashed line indicates the plateau of the phase. }  
\label{FigC1}
\end{figure}

In Figure~\ref{FigC1}, the normalised visibility (solid black) and the phase (dashed black) are shown. The modulation of the visibility is obvious, although the extremes are shifted to higher spatial frequencies when compared to the predictions by the orbital fit \citep{hummel03}. The latter are indicated by arrows. The MIDI measurements are best represented by a binary with a separation of about 80\,mas and a flux ratio around 0.25. The visibility of such a binary model is shown as solid grey line in Figure~(\ref{FigC1}) after multiplying it by a value of 0.8. The phase, plotted as dashed black line, shows a rising plateau (dashed grey) that is centred around the visibility maximum. The smooth transition to the next plateau occurs as expected around the visibility minimum. The change in phase can be well estimated, although the second plateau can be seen only marginally. We measure a value of $\Delta\phi\approx 75^{\circ}$. On the other hand, the value derived from the flux ratio of 0.25 is $\Delta\phi=2\pi-2\pi/(1+0.25)=2/5\pi=72^{\circ}$, almost identical. We thus conclude that we see in Figure~\ref{FigC1} the signal of the wide binary.

With this knowledge, the conclusion drawn in Appendix~\ref{phases} can now be tested. From Figure~\ref{FigC1}, one sees that the phase increases with spatial frequency (decreases with wavelength) at the visibility maximum. According to Appendix~\ref{phases}, this is a clear sign that the companion was scanned before the primary. Looking at the headers of the files one finds that beam~A was associated with UT3, while beam~B is related to UT4. Since the scan is done towards beam~B, its direction was towards position angle 113$^{\circ}$. This leaves us with the result that the companion was indeed scanned first and that the conclusion drawn in Appendix~\ref{phases} is correct.

%

\begin{acknowledgements}
We thank the ESO staff at Paranal and Garching for their kind support during the preparation and execution of the observations. We are grateful to Benjamin Moster for the reduction of the KING data. We further want to thank Roy van Boekel, Jeroen Bouwman, Michiel Hogerheijde, Gwendolyn Meeus, and Veronica Roccatagliata for discussions and their help. We kindly thank the anonymous referee for his helpful suggestions and comments.

Th.~R.~cordially thanks ``P{\"u}tz'' for confiding the little great wonder to him. May the splendours of heaven always make its path safe and shimmering. Island!
\end{acknowledgements}

\bibliographystyle{bibtex/aa}

\end{document}